\documentclass[a4paper,11pt]{article}
\pdfoutput=1 

\usepackage{jcappub} 
\usepackage{graphicx}
\usepackage{subcaption}                 
\usepackage[T1]{fontenc} 
\usepackage{amsmath}
\usepackage{multirow}
\usepackage[normalem]{ulem}
\usepackage[dvipsnames]{xcolor}
\definecolor{mycolor}{rgb}{1.0, 0., 1.0}

\newcommand{\rose}{Mff$\epsilon_{009}$-MecFB}
\definecolor{colorROSE}{HTML}{DE1562}

\newcommand{\rouge}{Mff$\epsilon_{100}$-MecFB}
\definecolor{colorRED}{HTML}{8B0000}

\newcommand{\blue}{Mff$\epsilon_{009}$-DCool}
\definecolor{colorBLUE}{HTML}{16A9D1}
\newcommand{\verte}{Mff$\epsilon_{100}$-DCool}
\definecolor{colorGREEN}{HTML}{006400}
\newcommand{\jaune}{KSlaw-DCool}
\definecolor{colorYELLOW}{HTML}{CCAA11}

\title{\boldmath Cosmological simulations of the same spiral galaxy: connecting the dark matter distribution of the host halo with the subgrid baryonic physics}

\author[a]{A. Nu\~nez-Casti\~neyra,}
\author[b]{E. Nezri,}
\author[b]{P. Mollitor,}
\author[c]{J. Devriendt,}
\author[d]{R. Teyssier.}

\affiliation[a]{Universit\'{e} Paris Cit\'{e} and Universit\'{e} Paris Saclay, CEA, CNRS, AIM, F-91190 Gif-sur-Yvette, France}
\affiliation[b]{Aix Marseille Univ, CNRS, CNES, LAM, Marseille, France 
}

\affiliation[c]{Sub-department of Astrophysics, University of Oxford, Keble Road, Oxford OX1 3RH, UK}
\affiliation[d]{Department of Astrophysical Sciences, Princeton University, Princeton, NJ 08540, USA}

\emailAdd{arturo.nunez@cea.fr}
\emailAdd{emmanuel.nezri@lam.fr}
\emailAdd{pol.mollitor@lam.fr}
\emailAdd{julien.devriendt@physics.ox.ac.uk}
\emailAdd{romain.teyssier@gmail.com}

\abstract{
The role of baryonic physics, star formation and stellar feedback, in shaping the galaxies and their host halos is an evolving topic. The dark matter aspects are illustrated in this work by showing distribution features in a Milky Way sized halo.
We focus on the halo morphology, geometry, and profile as well as the phase space distribution using one dark matter only and five hydrodynamical cosmological high-resolution simulations of the same halo with different subgrid prescriptions for the baryonic physics (Kennicut versus multi-freefall star formation and delayed cooling versus mechanical supernovae feedback). If some general properties like the relative halo-galaxy orientation are similar, the modifications of the gravitational potential due to the presence of baryons are found to induce different dark matter distributions (rounder and more concentrated halo). The mass density profile as well as the velocity distribution are modified distinctively according to the specific resulting baryonic distribution highlighting the variability of those properties (e.g inner power index from 1.3 to 1.8, broader speed distribution). The uncertainties on those features are of  paramount importance for dark matter phenomenology, particularly when dealing with dark matter dynamics or direct and indirect detection searches. As a consequence, dark matter properties and prospects using cosmological simulations require improvement on baryonic physics description. Modeling such processes is a key issue not only for galaxy formation but also for dark matter investigations.}

\begin{document}
\today
\maketitle

\flushbottom
\section{Introduction}\label{sec:intro}

In a $\Lambda$CDM universe, the formation and dynamics of galaxies are seeded by the massive presence of cold dark matter (DM). 
However, there is no consensus on the expected behavior of DM at galactic scales. Such predictions carry special importance to most, so far failed, DM detection efforts. 
Regardless of the detailed nature of cold DM, its predicted gravitational influence on baryons helps to explain a wide range of observations at the scales of galaxies and clusters of galaxies. On the other hand, the effect that baryons might have on the detailed dynamics and distribution of DM is not well known and could be key information for detection strategy. 
As successful as the $\Lambda$CDM cosmology might be at large scales, the galactic scales remain a very debated ground. At astrophysical scales i.e the scales of the interstellar medium, cosmological simulations rely on ad-hoc simplification of complex baryonic processes to evolve the primordial gas to fully-formed galaxies. These simplifications are known as sub-resolution physics. In particular, the main focus is centered on phenomena like star formation (SF), stellar winds, cosmic ray pressure, and AGN energy injection \cite{Scannapieco2012,Gunn1972,Rosdahl2017, Kimm2015, Teyssier2013, Kretschmer2020,Koudmani2022,FaucherGiguere2022, Dubois2013}.

At early stages in the universe's evolution, while the gravitational potential is dominated by DM, its dynamical evolution can be described by linear theories \cite{Press1974}. However, their dynamics becomes non-linear after the first dark matter halos are formed and, smaller halos and baryons are accreted into the main halos while simultaneously interacting among them. Additionally, at this stage the central gravitational potential of halos starts to be dominated by the baryonic matter. 
Since in cosmological simulations, some baryonic processes evolve through sub-resolution numerical implementations, the galactic DM evolution gains \textbf{baryonic-dependent} complexity. The effect of the central baryonic distribution on the embedding DM halo has been observed in several simulations \cite{Duffy2010,Cui2012, Cui2014,Schaller:2014uwa,Chua2017,Chua2019}. However, what is presented here is a detailed study of the global features of one DM halo hosting a Milky Way size galaxy simulated with different implementation strategies for the sub-resolution physics. 

In parallel with advancements in galactic observations, the resolution of cosmological simulation improves. Some general features of the MW halo (or a  simulated MW analog halo) are no longer out of reach and now present discrepancies between what is inferred from observations and obtained in first principle cosmological simulations. For example, most observations of galactic rotation curves suggest central cores (constant central densities) in the DM density profile \cite{Posti2019b,Li2020} while simulations are almost exclusively obtaining cusps (centrally divergent DM densities). Another discrepancy is found in the presence of stellar bars in disc galaxies: while $\sim 70\%$ of the observed nearby disc galaxies show a central stellar bar \cite{Erwin2018}, cosmological simulations very seldomly yield barred galaxies \cite{Reddish2022}. This hints at the importance of baryonic modelling for the description of galactic central regions inside numerical simulations.

A MW-size halo is expected to extend over a few hundreds of kpc in DM, while the central baryonic component, weighting $\sim 5\%$ of the total mass would extend over a few tens of kpc, thus, dominating the gravitational potential in the center.

Such a DM halo is often considered as a spherical distribution of mass with a steep reduction in density from the centre outwards. Cosmological simulations including only DM create highly triaxial halos \cite{Allgood2006,VeraCiro2011}, however, once baryons are included, the halos become more spherical due to the now deeper central potential \cite{Debattista2008,Chua2019}. On the observational side, the MW halo shape is uncertain and subject of significant discussion. It is argued to be slightly oblate in the center and to become triaxial at large distances \cite{Law2010,Ibata2013,VeraCiro2013}.

 A halo feature that has been extensively discussed is the radial density profile since, as mentioned above, it is the subject of tensions between observations and simulations. In cosmological simulations the full DM distribution is accessible and thus, is commonly used as a tool to derive the shape of dark matter density profiles through curve fitting. However, this is usually done without taking into account the inherent degeneracy between the fitting parameters. On top of that, there is the added uncertainty related to baryonic effects, which will certainly impact the center of the halo \cite{Peirani2017,Mollitor:2014ara}. The solution to the core-cusp issue could help to reveal the nature of the dark matter since the central distribution plays a big role in DM detection. The presence of a core or a cusp can drastically change the prospects to experiments looking for annihilation/decay products of dark matter in (sub)galactic or cluster halos \cite{Gondolo1999} and impact inferences from gravitational lensing observations \cite{MiraldaEscude2002}.

Finally, the phase-space distribution of galactic dark matter is relevant for several fields, from galactic dynamics to dark matter detection, and is easily accessible in simulations. In addition, dynamical approaches like Eddington inversion or action-angles can predict or crosscheck the distribution function of galactic DM \cite{Lacroix2018, Petac2019,Posti2019}. The phase-space distribution is determining for direct detection as well as dark matter capture in the Sun/Earth for neutrino telescope signals. Depending on the dark matter model, it can be also fully relevant  for the indirect detection in satellite galaxies (p-wave annihilating dark matter). 

Cosmological simulations do not have the last word, but as consistent objects evolved from first principles, they give interesting indications on dark matter distribution aspects even if the mass and spatial resolution remain a limiting factor. Furthermore, they are essential for the calibration and validation of semi-analytical works like dynamical approaches before using them on observations \cite{Lacroix2018, Pascale2018, Petac2019}. 
Additionally, simulations give hints and priors for the numerous parameters of halo modelling like density profile, substructure spectrum, mass-concentration relation \cite{Springel2008,Hutten2019,Ando2019,Stref2017,Navarro2010,Bonnivard2016}. Therefore, the conjunction of simulations, semi-analytical models, and observations regarding all those approaches represent a complementary front to address the questions of the dark matter distribution features efficiently.

In that spirit, this paper uses the high-resolution cosmological hydrodynamical simulations published in \cite{Nunez-Castineyra:2020ufe} (paper 1 from now on), where the same spiral galaxy is simulated with different baryonic physics for star formation and supernova feedback. The aim is to illustrate that while the practical ambiguity of baryonic physics in numerical simulations is still important, the uncertainties on the inferred dark matter distribution will remain, limiting our ability to make robust predictions for DM experimental prospects.

The paper is organized as follows: Section \ref{sec:thesimulation} gives an overview on the presented simulations. Section \ref{sec:halos} starts by addressing the static aspects of the halo morphology; the shape and edge of our DM halos, followed by a detailed look at the radial density profile, its properties, and the compression by baryons. This is followed by a study of some dynamical aspects of the DM halo focusing on phase-space distributions in section \ref{sec:localDMPhaseSpace}. Finally, the summary and conclusions are presented in section \ref{sec:summary}. In addition, the appendices contain some explicit complementary calculations and methods to facilitate the comparison with this work.


\section{Simulations}\label{sec:thesimulation}

The following analyses consider the dark matter distribution of the central halo in the simulations presented in paper 1, where the same halo is simulated six times; one run comprised of dark matter only and five high-resolution cosmological hydrodynamics (hydro) runs all resulting in a spiral galaxy, the \emph{Mochima} galaxy. The simulations are evolved with the AMR code \textsc{RAMSES} \cite{Teyssier:2001cp} from the same initial conditions generated with the MUSIC package \cite{Hahn2011} inside a cubic cosmological box of $\sim$36~Mpc of side. In these zoom-in simulations, the initial volume is built as a nested set of 5 convex-hull volumes of increasing (decreasing) DM resolution (particle mass), from a resolution equivalent to $128^3$ particles in the outermost region to $2048^3$ particles in the innermost volume. The inner and most resolved level corresponds to the decontaminated\footnote{The decontamination process is done using the public HAST package \url{https://bitbucket.org/vperret/hast/wiki/Home}} Lagrangian volume of the final galactic halo where DM particles have a mass of  $\sim 1.9\times 10^5$~M$_{\odot}$.

While the five hydro runs share initial conditions, they differ in the numerical models implemented to describe the star formation and supernova feedback evolution. These different prescriptions are described in detail in paper 1. This suite of simulations consists of one control run, using benchmark baryonic physics implementations as in previous simulations \cite{Mollitor:2014ara,Marinacci:2013mha}, and four combinations of two recently introduced SF and SN feedback models. The labels and main prescriptions are as follows; The control run, labelled KSlaw-DCool, uses a SF based on the Kennicutt-Schmidt law (KSlaw) \cite{Kennicutt1998} and for the SN feedback uses the so-called Delayed Cooling prescription (DCool) \cite{Teyssier2013}. This numerical set-up is confronted with two other recent numerical implementations. The first modification refers to the star formation strategy by changing to the so-called multi-freefall (Mff) version \cite{Federrath2012} of a star formation model which is based on turbulent magnetized molecular clouds \cite{Krumholz2005}. This model has one free parameter, $\epsilon$, multiplying the total star formation efficiency. It corresponds to proto-stellar feedback coming from stellar winds for which two extreme cases were tested: a strong proto-stellar FB where $\epsilon = 0.09$ and a weak proto-stellar FB where $\epsilon=1$, leading to the runs labelled Mff$\epsilon_{009}$-DCool and Mff$\epsilon_{100}$-DCool respectively. In the last two runs the SN feedback prescriptions are changed to consider a feedback prescription based on the Sedov-Taylor stages of the supernova explosion called Mechanical feedback (MecFB) \cite{Kimm2015}, leading to the runs labelled Mff$\epsilon_{009}$-MecFB and Mff$\epsilon_{100}$-MecFB corresponding to the strong and weak proto-stellar feedback assumptions.

In every case, the final galaxy is a disc galaxy with a heavy central bulge, but the relative bulge mass with respect to the rest of the galaxy varies. At red-shift 0, depending on the run, the mass of the DM halo is between 0.92$\times 10^{12}$~M$_{\odot}$ and 1.13$\times 10^{12}$~M$_{\odot}$ (see table 1 of paper 1). This halo has a quiet merger history, no major merger for $z<2$, and lies in a filament with a massive neighbour of $\sim1.\times 10^{13}$~M$_{\odot}$ located at $\sim6$~Mpc.


\section{Morphology of the halo}\label{sec:halos}
The dark matter halos have been studied in detail mainly in cosmological simulations with only dark matter. Since its non-collisional evolution should be dominated in the centre by the baryonic counterpart, subsequent studies have looked into the effect that the mere baryonic presence has on the DM halo. 
This section addresses  the main features of the dark matter halo, notably its shape, its outer border, and the dark matter halo density profile. 
\subsection{The shape} \label{sec:Shape}
The most common approaches to model the growth of cosmological structures rely on spherical symmetry \cite{Gunn1972,Peebles1980,Padmanabhan1993}. For big matter over-densities, such assumptions  describe well the evolution of the halo at early times \cite{Bernardeau1994}. However, in the late stages of the halo's history, the mass accretion turns into a violent process highly dependent on its environment. Therefore, there is no reason to assume sphericity anymore.
Nevertheless, the final halos are usually described by spherical halo profiles, even though by now it is well established that halos exhibit triaxial shapes, especially in DMO simulations. The shape of DM halos in numerical simulations have been extensively studied with several techniques, using the mass distribution of the DM  \cite{Barnes1987,Dubinski1991,Katz1991,Warren1992, Dubinski1994,Jing1995,Tormen1997,Thomas1998, Jing2002,Bailin2005,Kasun2005,Hopkins2005,Bryan2013,Kazantzidis2004,Zhu2016}, or the gravitational potential induced by all the components of the simulations \cite{Kazantzidis2010,Abadi2010}, the latter having the advantage of reducing the fluctuations caused by the presence of DM sub-halos (see appendix \ref{appe:shape} for a detailed discussion). It is often observed that once baryons are included in the simulation, the central presence of the  baryonic component of the galaxy can turn a triaxial DMO halo into a more spherical halo \cite{Allgood2006,Hayashi2007,Chua2019}.

\begin{figure}[!ht]
  \centering
  \begin{subfigure}[b]{0.45\linewidth}
    \includegraphics[width=\linewidth]{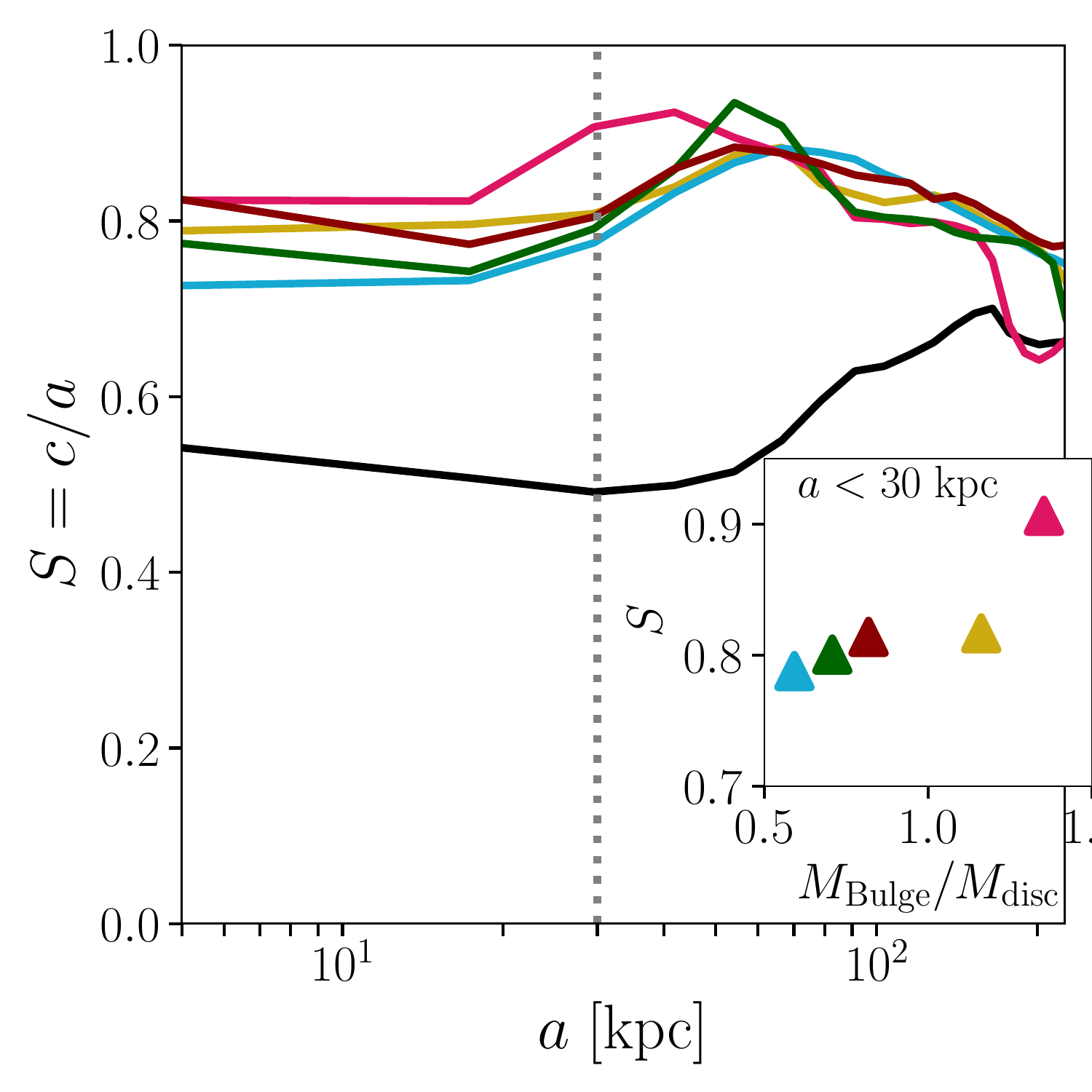}
  
  \end{subfigure}
    \begin{subfigure}[b]{0.45\linewidth}
       \includegraphics[width=\linewidth]{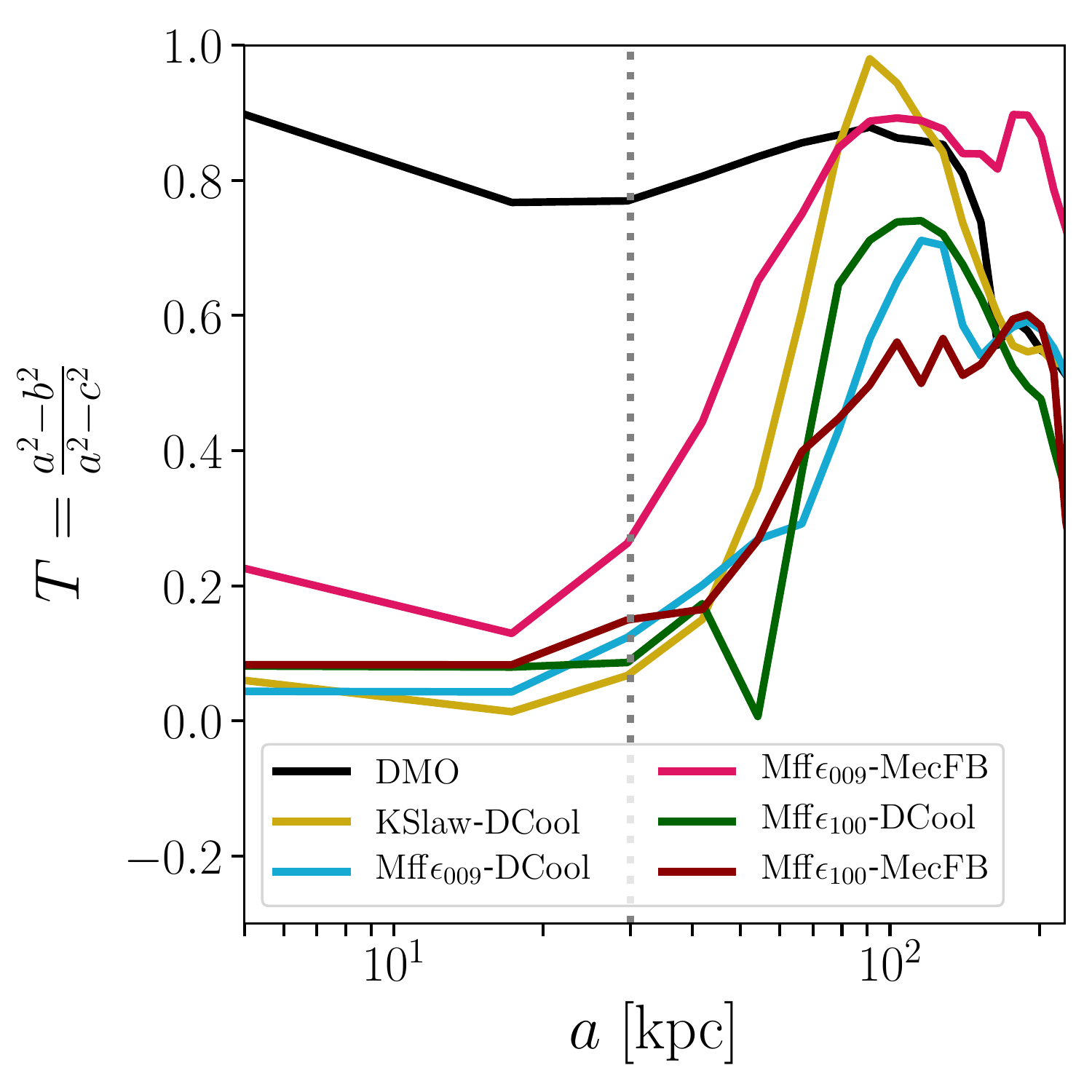}

    \end{subfigure}

  \caption[Sphericity and Triaxiallity]{Sphericity, $S=c/a$, and triaxiallity, $T=(a^2- b^2)/(a^2- c^2)$, of the six runs as a function of the resulting major semi-axis. The  $S$ and $T$ parameters are calculated using the gravitational potential measured at the position of the DM particles (see appendix \ref{appe:shape}). The vertical dotted line separates the two regions of interest, the inner halo region (a<30 kpc) and the outer halo region (a>30 kpc) as described in the text. The inner panel of the sphericity shows the relation of the sphericity of the halo calculated using all DM particles in the inner halo with respect to the bulge-to-disc mass ratio.}\label{fig:SandT}
  
\bigskip
\end{figure}

The investigation of the shape of the halo starts with the computation  of the semi-principal axis of a distribution of particles using the shape tensor (see appendix \ref{appe:shape}), where these axes are $a\geq b\geq c$. This calculation is sensitive to the observable and the volumetric selections used. The results presented here are computed using the gravitational potential  at the position of each DM particle inside iterative ellipsoids. The main advantage of these choices are discussed in Appendix \ref{appe:shape} and compared with other techniques in figure \ref{fig:SandTallTech}. The sphericity (S) and the triaxiality (T) parameters are then defined as:
\begin{equation}
S=c/a \;\;\; \mathrm{and}\;\;\;T=\frac{(a^2-b^2)}{a^2-c^2}\;\;\;.
\end{equation}  
Figure \ref{fig:SandT} shows  the resulting S and T parameters as a function of the biggest semi-principal axis, $a$, for the DMO run and the hydro runs. As expected, in the hydro runs the central presence of the baryonic disc has a drastic effect on the shape of the halo when compared with the DMO run. This highlights that DMO simulations should not be used to address the DM distribution in the inner halo.

When comparing the hydro runs, there are two main regions, the inner halo region ($a<30$ kpc) where the baryonic content  is expected to dominate the gravitational potential, and the outer halo region ($a>30$ kpc), where the presence of massive subhalos, often with non-negligible baryonic content, will drive the shape calculation. In the inner halo the sphericity follows the hierarchy of the bulge-to-disc ratio, i.e. the bigger the stellar bulge with respect to the stellar disc is, the more spherical the inner DM halo becomes. The bulge-to-disc ratio is shown in the inner panel of the sphericity plot in figure \ref{fig:SandT} where the mass of the bulge is defined as the stellar mass inside 3 kpc. This effect can also be seen on the variation of the black ellipsis on figure \ref{fig:maps-pot} (Appendix \ref{appe:shape}). The more prominent the bulge with respect to the disc is, the less flattened the ellipsis is in the edge-on view.

In the case of the outer halo the sphericity becomes degenerated (figure \ref{fig:SandT}) so the focus is turned to the triaxiality. A higher degree of variability in $T$ amongst the different hydro runs reflects the resilience of massive substructures against disruption from the central potential. A detailed analysis of the distribution and evolution of substructures in the Mochima runs is the subject of upcoming work, nevertheless, regarding the resilience of sub-halos some observations can still be done using $T$. The presence of dense DM substructures in the outer part of the halo will increase the value of $T$ at r$> 60$~kpc. However, in the presence of a baryonic disc, sub-halos will be disrupted more efficiently than in the DMO run resulting in cases with $T\simeq 0.6$. Such reduction in $T$ due to the destruction of substructures is observed in three out of the five hydro runs. The remaining two constitute extreme star formation cases and recover a triaxiality value comparable to that of the DMO run ($T>0.8$): the KSlaw-DCool run, which is the least star forming case and thus has the shallower central potential, is less efficient in disrupting satellites which means that the excess in surviving satellites increases $T$. Secondly, the Mff$_{\epsilon099}$-MecFB run has a central stellar content that is too massive and that induces a deeper central gravitational potential. On one side, the enhanced potential should have a destructive effect on the sub-halo population. On the other side, such an efficient star formation also allows satellites to form an important stellar mass fraction (see figure 2a of paper 1). This extra mass fraction ultimately helps sub-halos to resist the harassment from the central potential and keep $T$ for being reduced.



\begin{figure}[!htb]
  \centering
  \begin{subfigure}[b]{0.98\textwidth}
    \includegraphics[width=\textwidth]{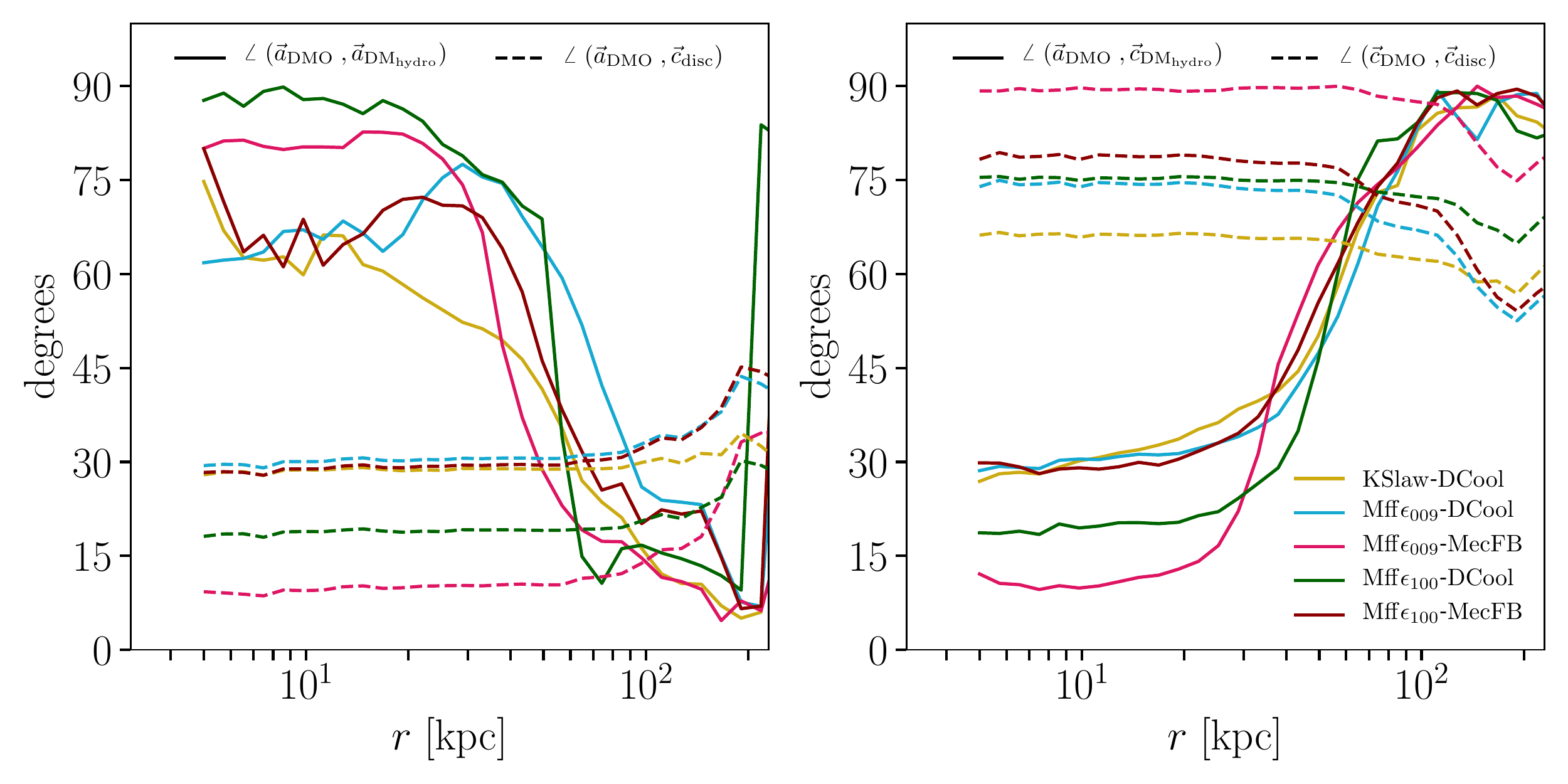}
  \end{subfigure}

\caption[Alignment]{Angles between semi-axes with respect to halo radius, comparing the DM distributions of the dark-mater-only (DMO) runs versus the DM in the hydro runs in solid lines and the DM in the DMO runs versus the semi-axis of the baryonic disc in dashed lines. }
\label{fig:anglesVS}

\end{figure}



The orientation of the halos with a baryonic component is not similar to the original triaxiality in the DMO halo. The halo in the DMO run is a prolate spheroid with $|\vec{a}_{\rm DMO}|>|\vec{b}_{\rm DMO}|\simeq|\vec{c}_{\rm DMO}|$ while the halos in the hydro runs exhibit a rather oblate shape with two comparable major semi-axes i.e. $|\vec{a}_{\rm DM_{\rm hydro}}|\simeq|\vec{b}_{\rm DM_{\rm hydro}}|>|\vec{c}_{\rm DM_{\rm hydro}}|$, which are unsurprisingly aligned with the baryonic disc. This can be understood by looking at the angles between the different semi-axes of the DM distributions (from DMO or hydro runs) and of the baryonic disc. Figure \ref{fig:anglesVS} shows the angles between the different semi-axes of DM distributions from the DMO run and hydro runs in solid lines, and between semi-axis of the DM distribution of the DMO run and the baryonic disc in dashed lines. While there is a variation of $\sim \pm 30$ degrees, the galactic disc is formed on the plane perpendicular to the major semi-axis of the DMO halo, as evidenced by the comparisons between the semi-axis of the DMO inner halo and the baryonic disc shown in dashed lines in figure \ref{fig:anglesVS}. The major semi-axis of the DMO halo, $\vec{a}_{\rm DMO}$, is roughly parallel to the minor semi-axis of the baryonic disc, $\vec{c}_{\rm disc}$, and therefore roughly perpendicular to the minor semi-axis of the DMO halo, $\vec{c}_{\rm DMO}$. The presence of the baryonic disc in this plane induces the oblate shape of the inner DM halo in the hydro runs. An oblate inner DM halo, in this case, means that the two major semi-axes, $\vec{a}_{\rm DM_{\rm Hydro}}$ and $\vec{b}_{\rm DM_{\rm Hydro}}$, lie roughly in the same plane of the galactic disc, i.e. in the same plane as $\vec{b}_{\rm DMO}$ and $\vec{c}_{\rm DMO}$, which explain the fact that $0 <\angle (\vec{a}_{\rm DMO}, \vec{c}_{\rm DM_{\rm Hydro}}) < 30 $ and therefore $60 <\angle (\vec{a}_{\rm DMO}, \vec{a}_{\rm DM_{\rm Hydro}}) < 90 $ degrees as shown by the solid lines in  figure \ref{fig:anglesVS}. The observed variability of the angles between the DMO run and the different hydro runs could be the result of the inherent stochasticity of the simulations and of the highly non-linear evolution of the halo.

One should notice that this remaining difference between the galactic orientation and the halo principal axes of $\sim 30$ degrees can have some consequences regarding analytical modelling based on symmetry assumptions and set limitations on such formalism \cite{Petac2019}.


\begin{figure}[!htb]
  \centering
  \begin{subfigure}[b]{0.92\textwidth}
    \includegraphics[width=\textwidth]{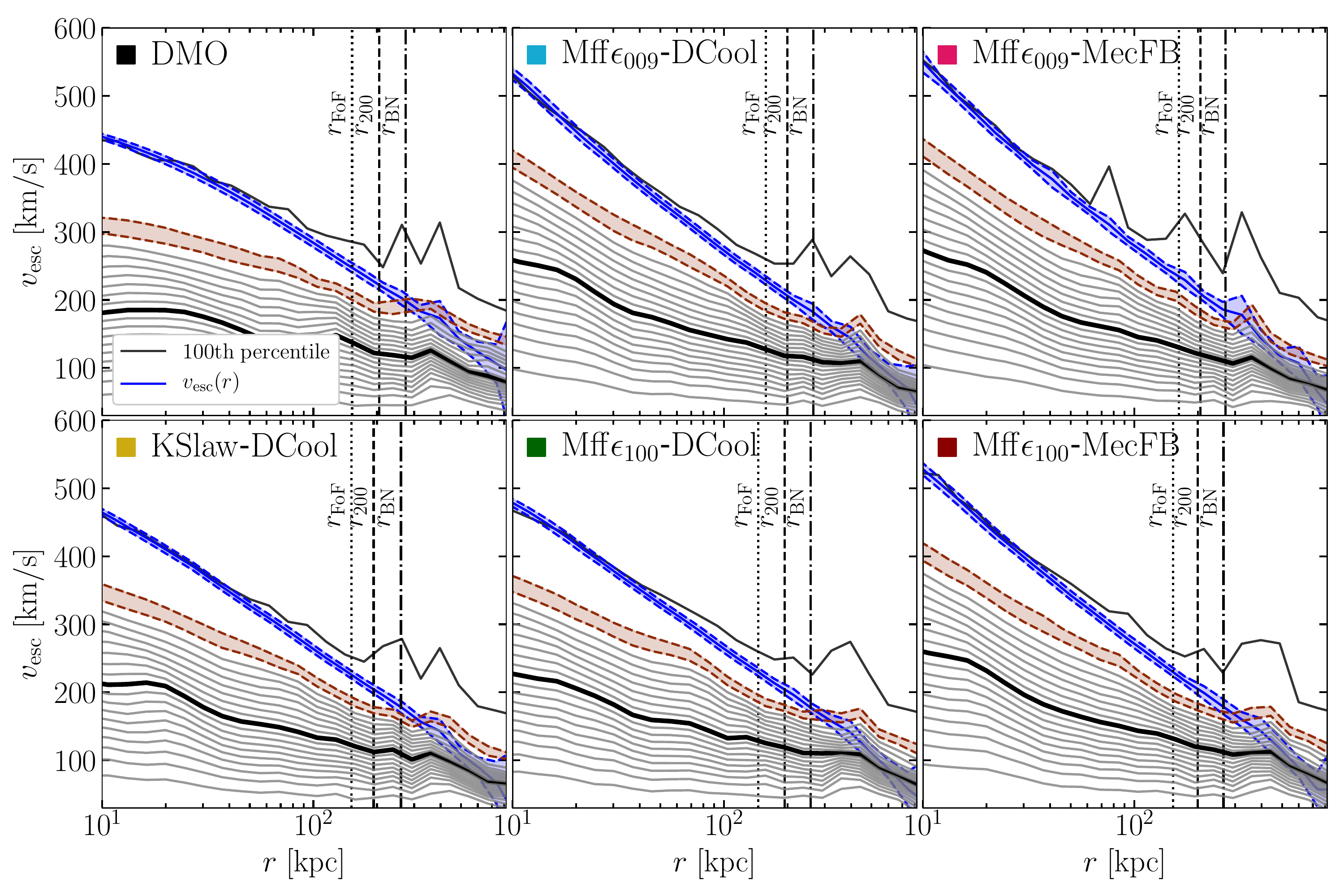}
  \end{subfigure}

\caption[Escape velocity spherical]{The escape speed with respect to the galactocentric radius with its $1 \sigma $ band (blue). The grey lines represent every 5th percentile of the local velocity distribution. The 90th and 95th percentiles are shown in red dashed lines and a filling red band. The vertical lines are $r_{FoF}$, $r_{200}$ and $r_{97}$. }
\label{fig:vesc-vmax-spherical}

\end{figure}

\subsection{The edge} \label{sec:Edge}

The extension of a self-gravitating structure could be considered as the limit where the central object no longer exerts any gravitational influence. In spherical symmetry such radius is defined as the distance, $r_{\mathrm{max}}$, where the gravitational potential reaches its maximum value in the line direction connecting the involved system and its closest massive neighbour (see \cite{Lacroix2020} for details). This radius is environment-dependent and is typically much bigger than the most common definitions  to characterize the size of halos found in the literature, such as the $r_{200}$ based on spherical collapse models \cite{Press1974}, the $r_{BN}$ based in a cosmological extension of the $r_{200}$ simplification \cite{Bryan:1997dn} or models based on particle distributions and clustering algorithms like the friend-of-friend radius $r_{FoF}$. The resulting halos size coming from such models could be rather different. Thus an additional test, using the escape speed of particles in the self-gravitating structure, to investigate the extension of the cloud of gravitationally-bound particles is proposed. In spherical symmetry, the escape speed is defined as 
\begin{equation}
 v_{\mathrm{esc}}(r)=\sqrt{2(\Phi(r_{\mathrm{max}})-\Phi(r))}\;\;,
 \end{equation}

\noindent where $\Phi(r)$ is the spherically averaged gravitational potential at a radius $r$. The $r_{\mathrm{max}}$ and other typical halo size values are given in table \ref{tab:halo_features}.

Even if formally, $r_{\mathrm{max}}$ represents the end of a halo's region of influence, more and more unbound particles, i.e particles going faster than the local escape speed, can be found further from the halo center. Therefore, it might be more precise to define the true outer limit of a halo as the last region that is almost exclusively populated by bounded bodies. Figure \ref{fig:vesc-vmax-spherical} illustrates  the radial evolution of the DM velocity distribution and how it compares with  the local escape speed (shown in blue  with its uncertainty band). Every 5th percentile of the velocity distribution is shown as gray lines with the 50th percentile in solid black. For emphasis, the 90th and 95th percentiles are shown as band bordered by dashed red lines. The vertical lines indicate the different calculations of the halo sizes mentioned above.


Typically $r_{200}$ is a fairly good approximation of the halo boundary but $r_{BN}$ is more precise as it coincides almost systematically with the region where at least 95\% of the contained particles have velocities below the local escape speed. On the other hand, $r_{FoF}$ appears too constraining. Nevertheless, this definition can be tuned as it depends on one parameter, the linking length, which was fixed here at 0.2 as it is usually done for cosmological simulations \cite{Huchra1982}. 

However, it is important to beware of the spherical symmetry assumption, particularly in non-fully-virialized structures. To consider the extension of a halo in such a situation one could calculate the so-called ``splash-back radius'', which is built by locating the orbital apocenter of accreted matter \cite{More2015}, while this method requires several snapshots of the simulation it has the advantage of not relying on spherical symmetry or virialization. The Mochima halo has been chosen to be a fairly isolated halo. Therefore it is not expected to present excessive departures between these definitions of the outer radius, nevertheless, it could present significant discrepancies for halos undergoing mergers.

A step beyond spherical symmetry consists in considering isopotential selections instead of spherical shells, taking into account the geometry and the morphology. The results are shown in figure \ref{fig:vesc-vmax-iso} of Appendix \ref{appe:vesc-isopot}. Naturally,  $v_{\rm esc}$ and 
 $v_{\rm max}$ follow each other and the crossing  or the separation is less sharp. This suggests a smaller edge/extension of the halos than in the spherical approach even if the comparison is not trivial since it is based on different morphological considerations.



\begin{table}[ht!]
  \centering
  \caption[Features of the halos]{Caracteristic radii and concentration values for the 6 simulations. 
  The resolution limit $r_\textrm{lim}$ corresponds to $r_{P03}$ in the case of DMO-S, and to $r_{\rm 3hsml}$ in the case of the hydro runs.}
  \label{tab:halo_features}
\begin{tabular}{|c|c|c|c|c|c|c|c|}
\hline
              run          & $r_{200}$ [kpc] & $r_{BN}$ [kpc] & $r_{FoF}$ [kpc] & $r_{-2}$ [kpc] & $c$ & $r_\textrm{lim}$ [pc] & $r_{\textrm{max}}$ [kpc] \\ \hline
DMO                       & 204.6     & 273.7    &  152.4    & 20.5     &  9.9 & 400 & 1824.71 \\ \hline
\textcolor{colorYELLOW}{KSlaw-DCool}               & 192.8     & 259.6    &  151.2    & 9.5     &  20.4 & 105 & 1029.36 \\ \hline
\textcolor{colorBLUE}{Mff$\epsilon_{009}$-DCool} & 199.8     & 265.4    &  158.3    & 7.7      &  26.0 & 105 & 1015.23 \\ \hline
\textcolor{colorGREEN}{Mff$\epsilon_{100}$-DCool} & 193.9     & 257.2    &  145.6    & 11.9     &  16.3 & 105 & 1062.90 \\ \hline
\textcolor{colorROSE}{Mff$\epsilon_{009}$-MecFB} & 205.7     & 270.1    &  162.7    & 6.9     &  29.5 & 105 & 1038.59 \\ \hline
\textcolor{colorRED}{Mff$\epsilon_{100}$-MecFB} & 199.8     & 264.3    &  152.4    & 9.6     &  20.7 & 105 & 1016.27 \\ \hline
\end{tabular}
\end{table}

\subsection{Dark Matter density profiles}\label{sec:profiles}



\begin{figure}[t!]
  \centering
  \begin{subfigure}[b]{0.92\textwidth}
    \includegraphics[width=\textwidth]{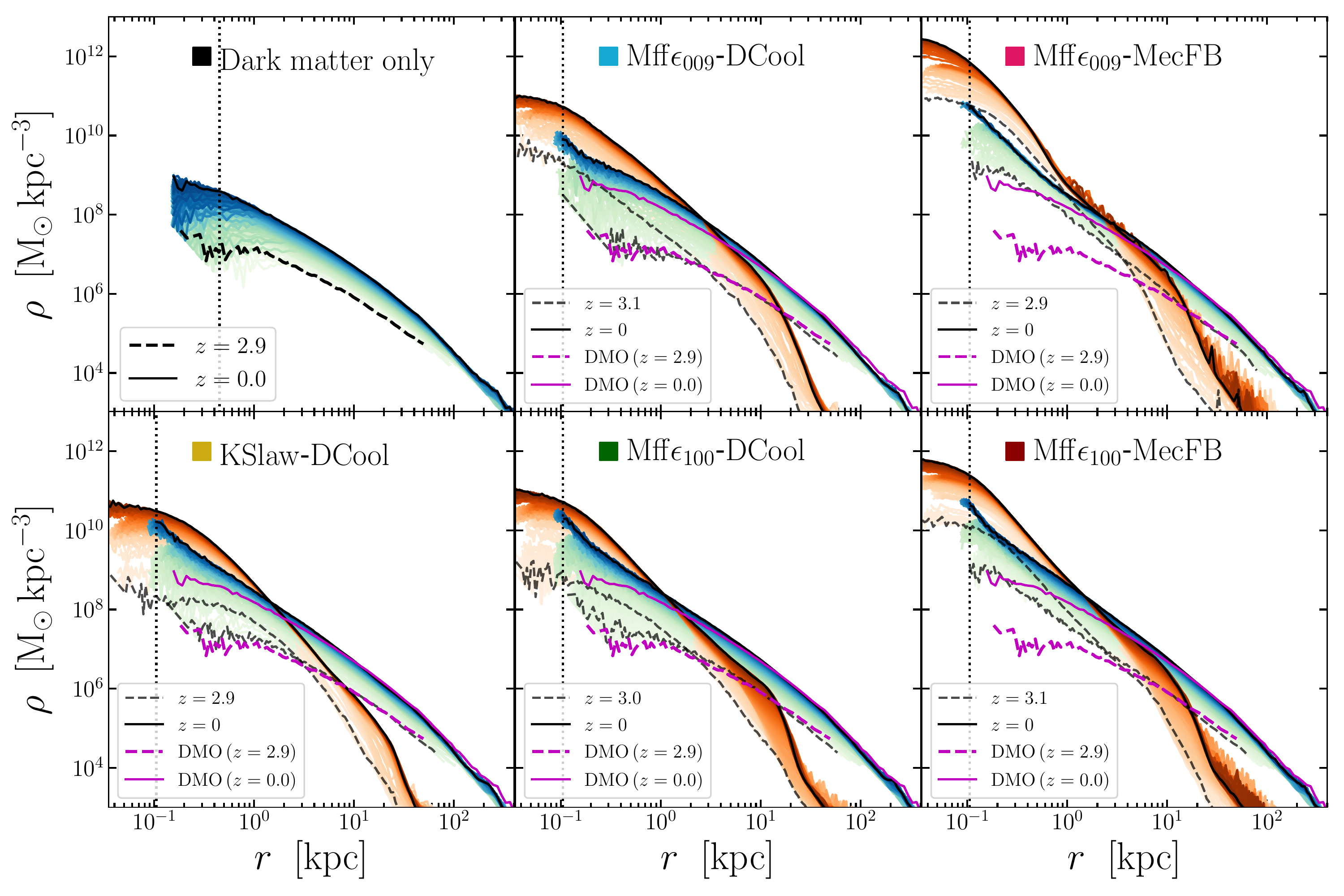}
  \end{subfigure}
  \caption[Dark matter profile History]{Dark matter (blue gradient curves) and stars (red gradient curves) density profiles evolving through time. The $z~3$ and $z=0$ profiles are shown as dashed and solid black lines respectively. The gradient represents the intermediate states between the two extreme lines. The simulation label with its corresponding color patch is given on top of each panel. The violet curves in the hydro panels indicate the high-redshift (dashed line) and final (solid line) DM profile from the DMO run. The vertical dotted line indicates the resolution limit (see discussion in the main text).}
  \label{fig:ProfileHist}
\end{figure}

For both, simulations and observations, the density profiles of DM halos have always been subject to strong debates, in particular, with regard to open questions such as the core-cusp problem \cite{deBlok2001,Salucci2000}, the diversity of observed rotation curves \cite{Oman2015}, the possibly excessive dark matter component in simulated galaxies \cite{Marasco2020}, the contraction by baryons or the DM-profile-flattening caused by baryonic feedback \cite{Pedrosa:2009bt, Governato:2009bg, DiCintio:2013qxa}. Figure \ref{fig:ProfileHist} shows  the DM (blue curve) and the stellar (red curve) spherically averaged density profiles for the DMO run and the five hydrodynamical simulations. The  profile at high redshift ($z\sim3$) is shown with a black dashed curve and the final profile at redshift 0 is shown with a solid line. The profiles at intermediate redshifts are shown with a color gradient. The vertical dotted line represents the (approximate) resolution limit of the simulations. In the case of the DMO simulation, the so-called Power radius $r_{P03} \approx 400\textrm{ pc}$ \cite{2003MNRAS.338...14P} is shown. It evaluates the innermost limit of a self-gravitating structure made out of non-collisional particles, i.e. dark matter particles.


For the hydrodynamical runs, the simulations have a higher resolution in the grid than for the particles.  The limit considered as the minimal reliable scale corresponds to three times the size of the smallest cell in the mesh (labeled $r_{\rm 3hsml} = 3 \cdot \Delta x = 105\textrm{ pc}$). All the values are summarized in table \ref{tab:halo_features}. 

In the region inside the resolution radii, the results have to be taken with caution. Indeed in the DMO run the flattening below $0.4$ kpc is a spurious resolution effect and should not be considered as a core.\\
In order to  facilitate the comparison between the DMO and hydro runs, the DMO DM-profile is added to the hydro simulations panels, again at high redshift (violet dashed line) and at redshift 0 (violet continuous line). It can be seen that for the outer parts of the halo, the DM profiles of the DMO and the hydro runs coincide. Below 30 kpc, as soon as the stellar density becomes comparable to that of DM, the DM profiles in the hydro runs depart from the DM profile in the DMO simulation. 
In the case of \rose\ and \rouge, this effect occurs already at high redshift, due to a strong star formation excess (see fig 3 of paper 1) while in the other simulations, the DM density is close to its DMO counterpart and is altered later during the halo history.\\
The presence of the baryonic gravitational potential steepens the DM density profile. This effect occurs  \emph{differently} in each simulation and gives rise to different concentrations. To define the concentration, one  commonly uses the radius $r_{-2}$ which satisfies the condition $\frac{d(r^2\rho_{DM})}{dr}  \rvert_{r = r_{-2}} =0$. In this way, $r_{-2}$ stands for the radius where the DM density  transits from an inner slope ($\rho \propto -\gamma$) to the outer slope ($\rho \propto -\beta$) and is equal to $-2$\footnote{Conveniently enough, for the NFW profile it holds that $r_{-2}$ is equal to the scale radius.}. The concentration is defined as $c = \frac{r_{\textrm{vir}}}{r_{-2}} $. The values of $r_{-2}$ and $c$ can be found in table \ref{tab:halo_features}. The response of the DM profile to the stellar populations does not occur immediately. Therefore, it is likely that it is the old population of stars that triggers the contraction. Since the stellar bulge is mainly populated by old stars and forms earlier, its mass can be considered as possible driver for DM contraction. In figure \ref{fig:ProfileHist} there are already hints that support such affirmation. Simulations like \rose\ and \rouge\ host very dense central structures at redshift 3 and 0 and show high concentrations and steep central DM profiles. To test this even further, figure \ref{fig:concentrationcorr} of appendix \ref{app:concentration} shows how the concentration correlates with the full stellar mass and with the bulge mass ($r<2.5$ kpc). Here, it is clear that the halo concentration is more sensible to the bulge mass than to the full stellar mass.

\begin{figure}[h!]
\centering
        \includegraphics[width=\linewidth]{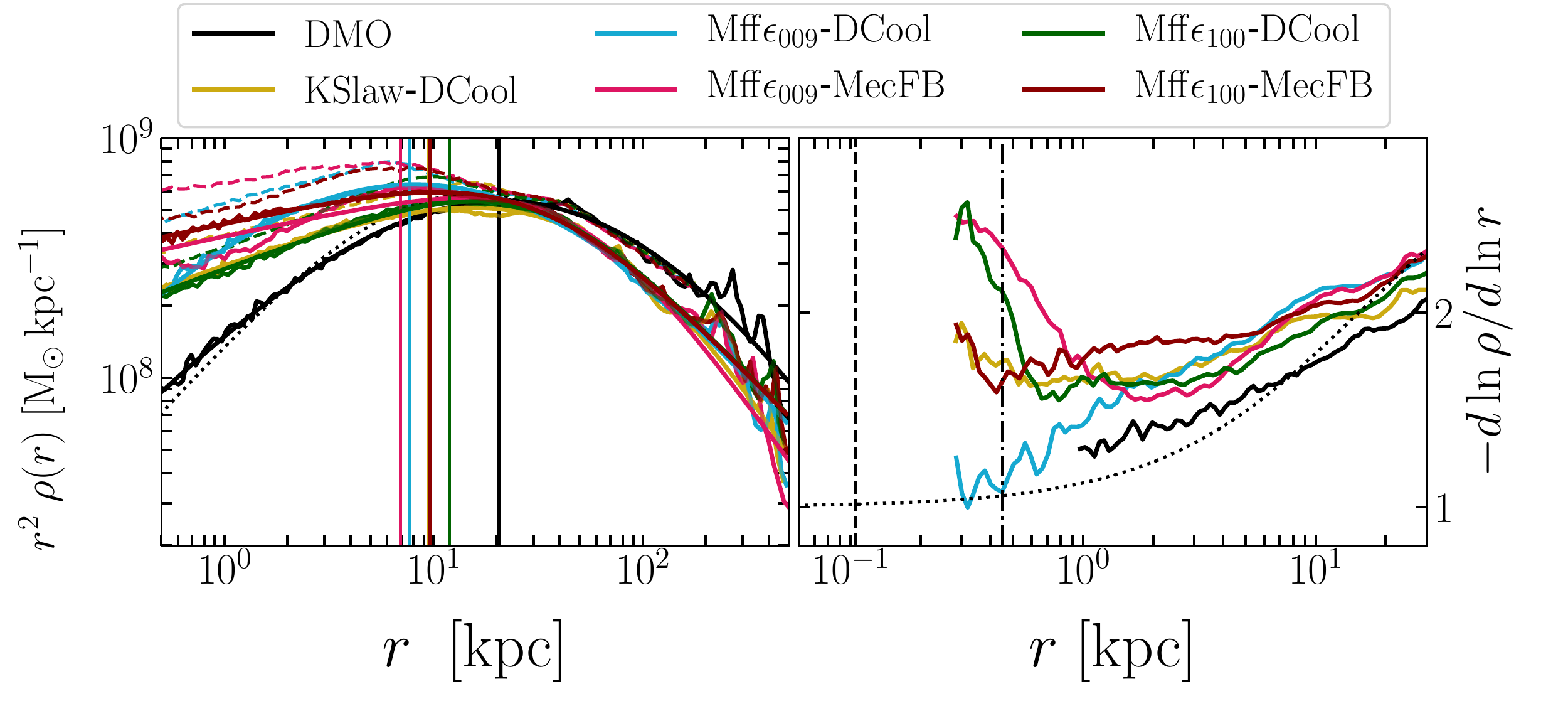}
        \caption{Left: Density profile scaled by $r^2$. The mean density profile is shown in the faint lines and the result of the $\alpha \beta\gamma$-fit on each halo in the dark lines. The vertical lines show the corresponding $r_{-2}$ for each run computed with the mean density profile. Right: The internal slope of the DM profiles. In both panels, the dotted line shows a NFW fit of the DMO profile.
        }
        \label{fig:r2rho_gamma}
\end{figure}

\paragraph{DM profile properties:}\label{sec:DM_profile_properties}
To describe the  DM density profiles (at $z=0$), the generalized $\alpha\beta\gamma$-profile \cite{1996MNRAS.278..488Z} and the Einasto profile \cite{Einasto1965} are considered, using the Bayesian inference tool \texttt{MultiNest} (\cite{Feroz:2007kg,2009MNRAS.398.1601F,Feroz:2013hea}) through the \texttt{PyMultiNest} interface (\cite{2014A&A...564A.125B}) to find the posterior likelihood distribution for the model parameters. The details are explained in appendix \ref{app:FittingProfiles} and the results are given in tables \ref{tab:haloprofiles_NFW} and \ref{tab:haloprofiles_einasto}. One interesting (and rarely explored on DM profiles) benefit of this method is that the posteriors can be used to estimate a confidence band on the resulting fit. This band is shown on top of our fits (see appendix \ref{appe:variance}).\\
It can be argued that the DMO run exhibits a behaviour similar to the NFW profile, while the hydro simulations show clear departures from an NFW behaviour. This result is not surprising since the NFW profile was inferred using DM only halos. 
On the left panel of figure \ref{fig:r2rho_gamma}, the density profile (calculated from the $\alpha\beta\gamma$ fit parameters) is scaled by $r^2$ and the positions of $r_{-2}$ for the simulations are marked by the vertical lines. 
For the sake of comparison, the simulation data, i.e the spherically average density, are shown with the same colors but a stronger transparency.

A highly relevant feature of the DM density profile is its behavior near the halo center. The right panel of figure \ref{fig:r2rho_gamma}, shows 
$\gamma = - \frac{d \ln \rho}{d \ln r}$ as a function of the radius below 10 kpc. The DMO density from the fit is shown as a black line. The vertical dotted and dashed-dotted black lines indicate the resolution limit of the hydrodynamical runs (100 pc) and of the DMO run (400 pc). The DM density profile steepens for all the hydrodynamical simulations in comparison to the DMO simulation. The resulting inner slope $\gamma$ spans from 1.3 to 1.8. The creation of stars in the center of the halo implicates a deepening of the central potential and leads to a contraction of the DM density \cite{Eggen1962}. This aspect is discussed further in appendix \ref{app:contraction}. The implemented feedback mechanisms that are inherently correlated to the stellar formation are not able to counterbalance the contraction dynamics. 
Nevertheless, our results illustrate the fact that even for the same numerical galaxy but with different baryonic physics implementations, the resulting DM profiles are different and do not match the DMO profile in any configuration; naturally, the stronger the baryonic profile, the steeper the DM density profile in our runs.
\section{Phase space distribution}\label{sec:localDMPhaseSpace}


These aspects are of paramount importance to understanding galactic dynamics and dark matter detection strategies.
Moreover, considering the lack of knowledge from observations regarding these particular features it is fully relevant to use simulations, especially in a cosmological context, to infer those assumptions or to calibrate analytical methods. 


\subsection{Mass density and velocity distribution}\label{subsec:massdensdist}

\begin{figure}[t]
\includegraphics[width=\linewidth]{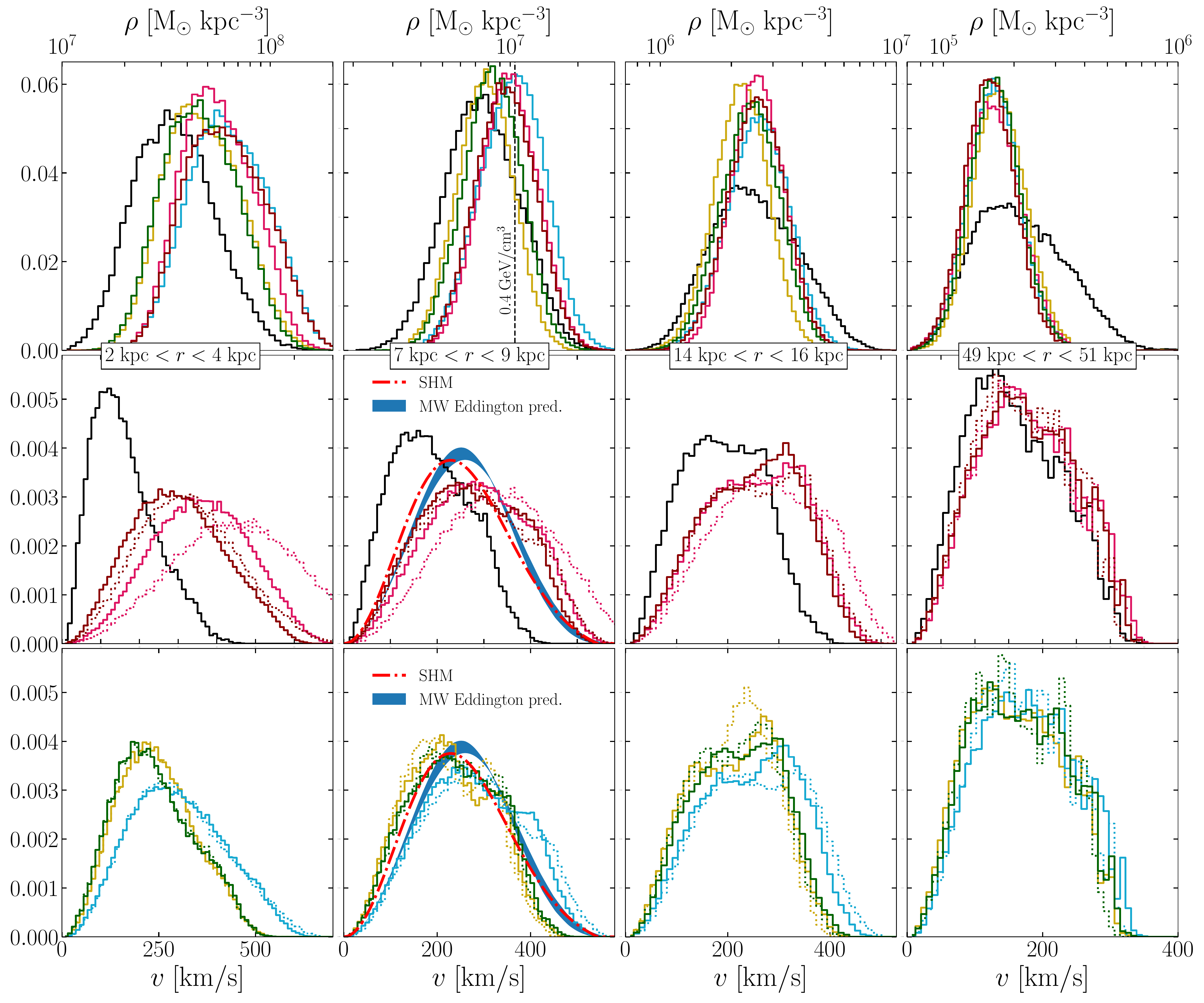}

  \caption[Phase-space distributions]{The normalized distribution of  density (top row) and speed (middle and bottom rows) of dark matter inside $ 2$ kpc thick shells at $r$ = 2,8,15 and 50 kpc. The dashed lines show the distributions at radii defined by $M_{\mathrm{SIMU}}(r)=M_{\mathrm{MW}}(2,8,15 \ \mathrm{and} \ 50 \ \mathrm{kpc})$ that contains the same amount of mass as the MW harbours inside the corresponding radii, following \cite{2017MNRAS.465...76M}. Additionally, for the case of the solar radius (8 kpc, second column from left to right) two references are shown, 
 the standard halo model (red dot-dashed lines) and a band containing several variations of the Eddington inversion (blue, see text for details).}
\label{fig:rhofdvhisto}
\end{figure}


The key features of the DM regarding its detection are its density or mass distribution and its velocity, both are involved in the searches for DM interaction. Whether the aim is to directly identify the interaction of a DM particle inside the detector's volume or to catch an unmistakable signal of DM interaction in a dense and faraway region.




\subsubsection*{Mass distribution}
The upper row of figure \ref{fig:rhofdvhisto} shows the distribution of the dark matter density at $r=$2,8,15 and 50 kpc. The density of the simulation particles are calculated with their local neighbouring.

 Each simulation including baryons presents density distributions that differ significantly from the DMO simulation at all radii. This effect is stronger at large radii where the DMO run shows a slightly higher mean density but a significantly higher dispersion. This situation is due to the weaker tidal effects on DM substructures that result from a shallower central gravitational potential. Therefore, a larger number of particles inside dense sub-halo central regions populate the high density tail of the mass distributions.

 Two effects can be observed. First, in the baryon-dominated region, i.e at radii smaller than the  stellar disc extension (2,8,15 kpc) the DM is contracted (DM mass outside the baryonic-dominated region is brought inside from the outer regions) inducing a shift toward higher mean densities in the hydro runs. This effect is especially noticeable in the bulge region, and is milder far from the center. The shift in the mean density follows the hierarchy of the baryonic component between the hydrodynamics runs, it is stronger for the two runs with mechanical feedback and the Mff$\epsilon_{009}$-DCool run i.e the three most massive stellar components \cite{Nunez-Castineyra:2020ufe}. The density shift is similar at radii beyond 15 kpc for all hydrodynamical runs as the enclosed baryonic mass is comparable. This explains also the common crossing of density already observed(described) on figure \ref{fig:r2rho_gamma}.
 
 
 Then at large radii, the second effect is due to the tidal disruption of subhalos in hydrodynamics runs which is stronger than in the DMO simulation inducing narrower distributions.

The simulations including baryons differ as they get closer to the galactic center while the distributions are similar at large radii (50 kpc). Namely, well inside the regions dominated by the baryons, 2 and 8 kpc, it is then clear that different central baryonic distributions will impact accordingly the central density distribution of dark matter, as seen in the full dark matter density profiles. However, while the mean density is shifted, the width of the distributions stays very similar amongst the hydro runs. 

For all the cases, it is noticeable that the density distributions at 8 kpc have a mean that is in the ballpark of values inferred from observations (even if those values have still large uncertainties, see \cite{2021RPPh...84j4901D}). Namely, considering detection prospects using cosmological simulations, the detection rates, both for direct detection and neutrino telescopes looking for dark matter toward the Sun, are directly proportional to the dark matter density in the "solar neighborhood".




\subsubsection*{Velocity distribution}

Figure \ref{fig:rhofdvhisto} also shows the velocity (speed) distribution at 2, 8, 15 and 50 kpc. Here again the distribution are correlated with the gravitational  potential.

At $r=2$ kpc, the hydrodynamics simulations exhibit a global shift of the mean to higher velocities compared to the DMO run due to the stronger potential of the bulge. The drift follows the hierarchy of the potential of the three cases. The steepest effect corresponds to the strongest bulge of the Mff$\epsilon_{009}$-MecFB simulation.

At $r=8$ kpc, the effect of the discs also matters and the potential of the discs induce also a shift of the mean and the overall distributions. Again the effect is stronger for the Mff$\epsilon_{009}$-DC, Mff$\epsilon_{009}$-MecFB and Mff$\epsilon_{100}$-MecFB simulations accordingly to their strong discs and the impact is weaker for the remaining runs due to the weaker disk, which exhibit a mean that is around the DMO value.


Some simulations have reported the presence of an increase in DM density inside the galactic disc \cite{Read2009, Ling2010b}, the so-called dark disc. This feature tends to be absent in more recent works \cite{Schaller2016,Fattahi2016} with more realistic discs. The dark matter distribution does not show any dark disc features in the five runs. 

%

At $r=15$ kpc, there are still some effects with a shift due to strong discs and wider distributions for Mff$\epsilon_{009}$-DCool, Mff$\epsilon_{009}$-MecFB and Mff$\epsilon_{100}$-MecFB (but almost no effect compared to DMO for the KSlaw-DCool and Mff$\epsilon_{100}$-DCool for which the  distribution is very close to the DMO simulation.

At $r=50$ kpc, away from the disc, where there is mostly DM, the shape of all halos is probably ruled by the presence of sub-halos. To illustrate this, figures \ref{fig:maps-pot} and \ref{fig:mapsfull-YZ}, show two different projections of each halo, the ellipsoid describing its shape in black, and either a white contour where the stellar distribution is or the total virial radius.  

More quantitatively, the position of the most probable value and the mean of the distributions of the hydrodynamical simulations compared to the DMO are inferred by the gravitational potential. Those positions are roughly given by the additional enclosed mass: $\bar{v}_{hydro}(r)-\bar{v}_{DMO}(r)\sim \sqrt{\frac{G}{r}}(\sqrt{M_{hydro}(<r)}-\sqrt{M_{DMO}(<r)})$. Regarding the spread of the distributions, it is connected to $v_{\mathrm{esc}}$ which are higher in hydrodynamical simulations due to the strength of the potential (see figure \ref{fig:vesc-vmax-spherical} and \ref{fig:vesc-vmax-iso}) thus inducing broader distributions compared to the DMO case.
In the appendix \ref{app:fdvfits} figure \ref{fig:fitfdv} shows the results of fitting  each distribution with usual (generalized) Maxwellian and Tsallis functions truncated at the escape velocity (see e.g \cite{Ling2010b,NunezCastineyra2019}). The fit parameters are given in table \ref{tab:vdffits}. while these typical functions show generic problems to account for the top and the tail of the distributions, including the escape velocity naturally improves the matching on the high velocity tail, Particulary for the Tsallis distribution. An overall improvement should come from a careful subtraction of clumps and streams across time as they are likely responsible for the bump-like features in the velocity distributions in the inner halo. 

Finally, at 8 kpc. For the sake of comparison, figure \ref{fig:rhofdvhisto} also shows the popular Maxwellian speed distribution of the Standard Halo Model (SHM) used to derive experimental exclusion curves and a band of the Eddington inversion applied on the MW mass models of \cite{2017MNRAS.465...76M}. To build the band, four cases are considered, $\gamma=0,0.25,0.5,1$ for the central slope.
As expected from the stellar mass of the Mochima simulations \cite{Nunez-Castineyra:2020ufe} compared to the MW, the velocity distributions are close and comparable (mean and spread) with the SHM (though ad-hoc) and even with the Eddington band \cite{Lacroix2018} (derived from observations and mass models \cite{McMillan2017}). But as shown in  \cite{Lacroix2018}  Eddington derivations of $f(v)$ from the gravitational potential of cosmological simulations are only qualitatively able to reproduce the precise shape of real velocity distribution data (but the method performs very well for the moments  \cite{Lacroix2018}), meaning that there is no guarantee that the velocity distribution labeled here {\it Eddington} is close to the real MW one (which is unknown). So even if the velocity distributions of our simulations are in the ballpark, the variability induced by baryonic physics still gives rise to inconclusive predictions for the details of $f(v)$. Fortunately, the detection rates are more sensitive to integrated quantities which are much less sensitive to the exact shape of f(v), reducing uncertainties on detection rates of typically 10-20\%.

Therefore, baryonic physics modifies the phase space distribution of dark matter, especially inside the galaxy. This has consequences on direct detection and neutrino telescopes or gamma indirect detection  for p-wave annihilating dark matter candidates. 

\subsection{Energy distribution - (pseudo) Phase space distribution function}

When halos show some departure from equilibrium, the phase-space densities will not fully depend on integrals of motion. In addition, without isotropy and spherical symmetry, it should depend on quantities other than energy. Nevertheless, as justified by a recent study \cite{Lacroix2020} (see also \cite{Vogelsberger:2008qb}), the mean  pseudo phase space density (PPSD) can be defined by spherically binning $r$ and $v$, with    
$r^{i}_{\mathrm{c}}$ and $v^{j}_{\mathrm{c}}$ being  the central values of the bin and $\Delta r$ and $\Delta v$ their width, leading to the expression; 

\begin{equation}
f(\mathcal{E})^{ij} = \frac{m_{\mathrm{p}} N^{ij}}{4\pi (r_{\mathrm{c}}^i)^2 \Delta r^i 4\pi\, (v_{\mathrm{c}}^j)^2\Delta v^j}, 
\label{eq:PPSD}
\end{equation}

\noindent with $\mathcal{E}^{ij} = \left\langle \Psi \right\rangle_{ij} - \left\langle \frac{v^2}{2}\right\rangle_{ij}$ and where  $\left\langle\Psi\right\rangle_{ij}$ and $\left\langle v^2/2 \right\rangle_{ij}$, $m_{\mathrm{p}}$ and $N^{i j}$ are respectively the mean potential, the mean velocity of the corresponding ($r$,$v$) bin, the DM particle mass and the number of particles inside the bin.

\begin{figure}[t]
\centering
\begin{subfigure}[b]{0.45\linewidth}
\includegraphics[width=\linewidth]{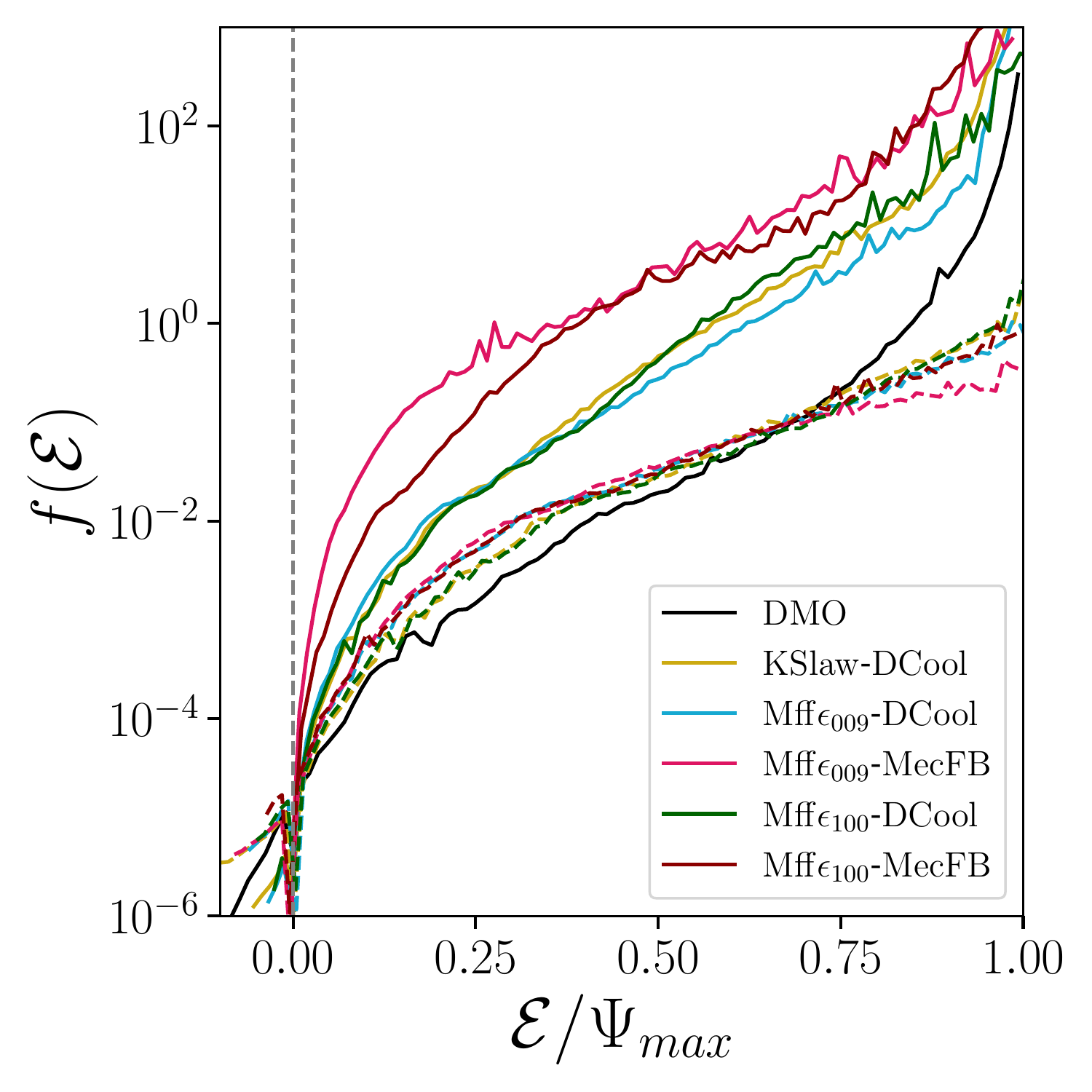}
\end{subfigure}
\begin{subfigure}[b]{0.45\linewidth}
\includegraphics[width=\linewidth]{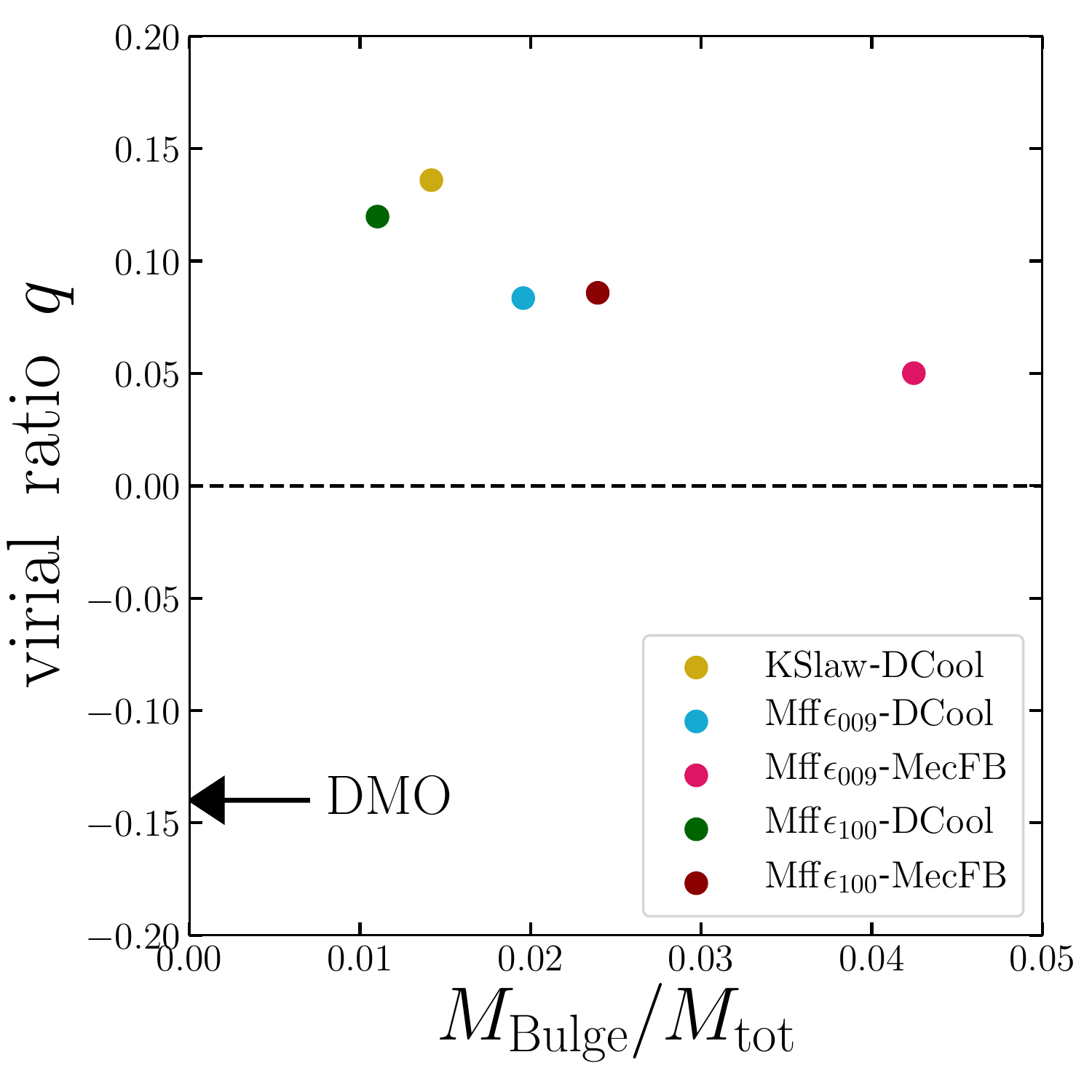}
\end{subfigure}
\caption{Left: Phase space distribution function $f(\mathcal{E})$. The solid lines are built using all particles inside the virial radius and the dashed lines with particles between 3 kpc and the virial radius to exclude the bulge in hydro runs. The vertical dashed line indicates the limit of unbound particles. Right: Equilibrium parameter $q$ versus bulge-to-total mass ratio.}
\label{fig:f_epsilon}%
\end{figure}

The results are shown on the left of figure \ref{fig:f_epsilon}. The shapes of $f(\mathcal{E})$ are different for each simulation.
Indeed, each hydrodynamical run shows departure from the DMO case. 
The star-formation/SN-feedback balance and the resulting baryonic distributions are determining the energy distribution of DM particles. Namely, the contraction of the dark matter profile increases the potential energy of the particles in the central part (i.e below the crossing between DM density and baryon density). Consequently, $f(\mathcal{E})$ reaches higher values at high energy for the hydrodynamical simulations than in the DMO run. To further highlight this, the figure shows as dashed curves the energy distribution of the particles that are beyond 3 kpc from the centre, therefore excluding the most energetic particles. Two effects can be noticed: this curve is dominated by the central population of particles and the feedback scheme induces different energy distributions. A discrimination is observed between the three simulations with delayed cooling and the two with mechanical feedback for which the dark matter cusp is steeper resulting in particles with higher energy. This discrimination was also pointed in the baryonic density of the stellar bulge in these simulations \cite{Nunez-Castineyra:2020ufe}. In addition, one should notice that this behaviour of $f(\mathcal{E})$ could be the opposite for strong enough SN feedback disrupting the DM cusp \cite{Mollitor:2014ara}. The resulting $f(\mathcal{E})$ is then flatter than the DMO one as can be seen in \cite{Lacroix2020} (see also \cite{Lacroix2018,Widrow_2000}).
Figure \ref{fig:f_epsilon} also shows the virial ratio $q=\frac{2E_{Kin}}{E_{pot}}+1$ (see e.g  \cite{Zjupa2017}) as a function of the bulge-to-total mass ratio. Indeed, those two quantities are indicators of the $f(\mathcal{E})$ shape discrimination. More violent feedback (less adiabatic) gives rise to higher $q$ parameters (less equilibrium) and lower bulge mass. The similar shape of the DMO curve with violent feedback cases might be due to the more triaxial distribution. 

\begin{figure}[t]
\centering
\begin{subfigure}[b]{0.92\linewidth}
\includegraphics[width=\linewidth]{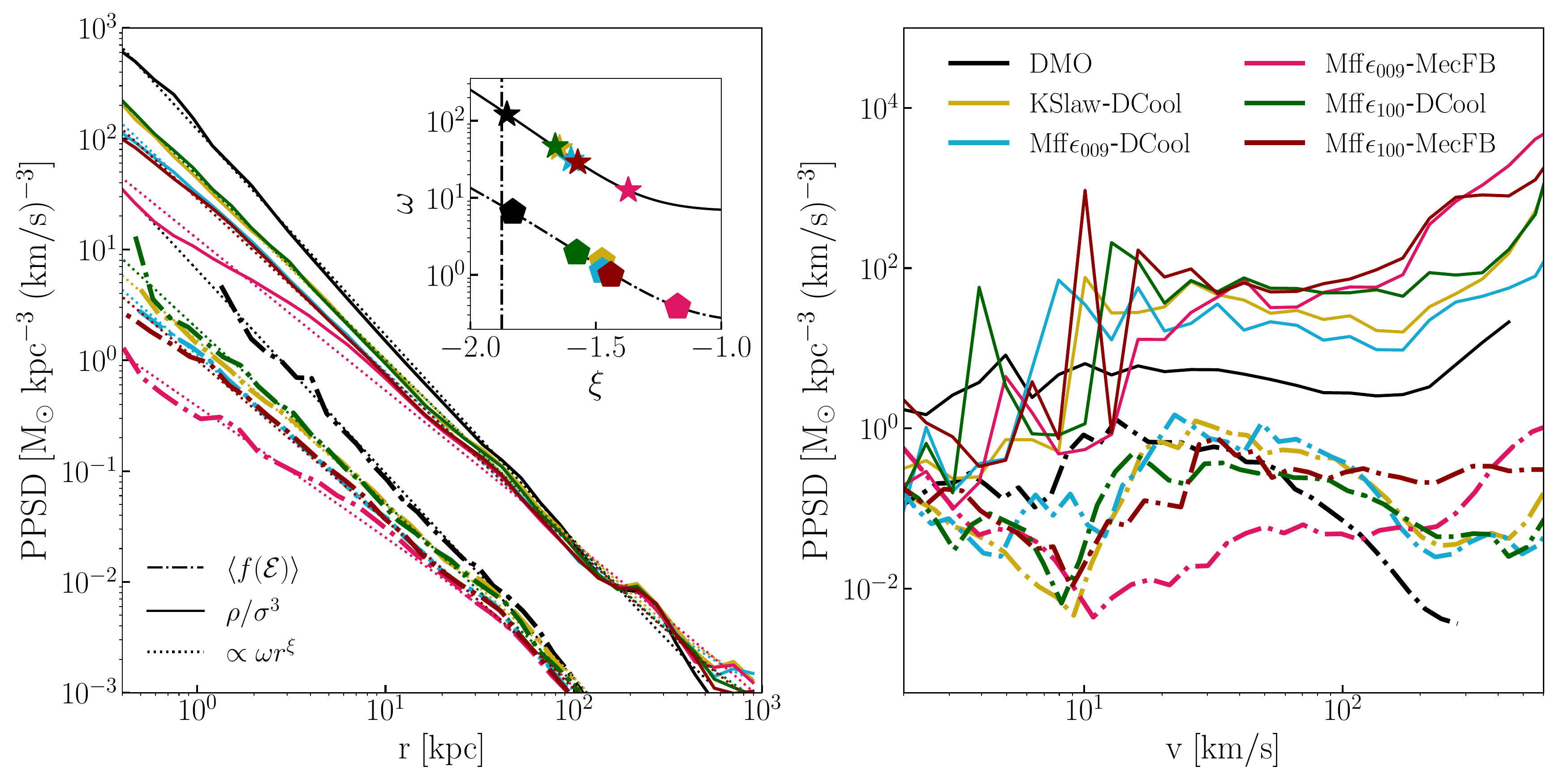}
\end{subfigure}

\caption{
Radial (left) and velocity (right) ) pseudo phase space density profile comparison. $f(\mathcal{E})$ (dot-dashed line), $\rho(r)/\sigma(r)^3$ (solid line) and a power law fit $\propto \omega r^{\xi}$ (dotted line). The results of the fit are shown in the inner panel.
}

\label{fig:f_epsilon_profiles}
\end{figure}

For the sake of comparison, we also consider the PPSD estimator, calculated as the ratio of the density and the cube of velocity dispersion, $\rho(r)/\sigma(r)^3$ \cite{Bertschinger1985} where the velocity dispersion is computed as the square sum of the tangential and radial velocity dispersion. These PPSD profiles were found to behave as a radius power law, $\propto r^{\xi}$, with an index $\xi=-1.875$ for dark matter only simulations \cite{Taylor2001,Ludlow2010}. Figure \ref{fig:f_epsilon_profiles} shows the PPSD profiles in solid lines with a power law fit in dotted lines. Similarly, the dashed-dot lines show the PPSD profile stacked from equation \ref{eq:PPSD}. While the two quantities are parallel with a similar spread between the different simulations, there is a shift in their normalization as the $\rho(r)/\sigma(r)^3$ approximation does not take into account the volume element in velocity space. Nevertheless, the indexes of the power law are in agreement between both approaches, and the DMO run fit is in complete agreement with the results found by \cite{Taylor2001} (shown as a vertical black line in the sub-panel). The right panel of figure \ref{fig:f_epsilon_profiles} shows the velocity profile of the pseudo phase space distribution function. As seen in  section \ref{subsec:massdensdist}, the stronger central potentials  induce a boost on the dark matter velocity distribution towards higher values.


\begin{figure}[t]
\centering
\begin{subfigure}[b]{0.45\linewidth}
\includegraphics[width=\linewidth]{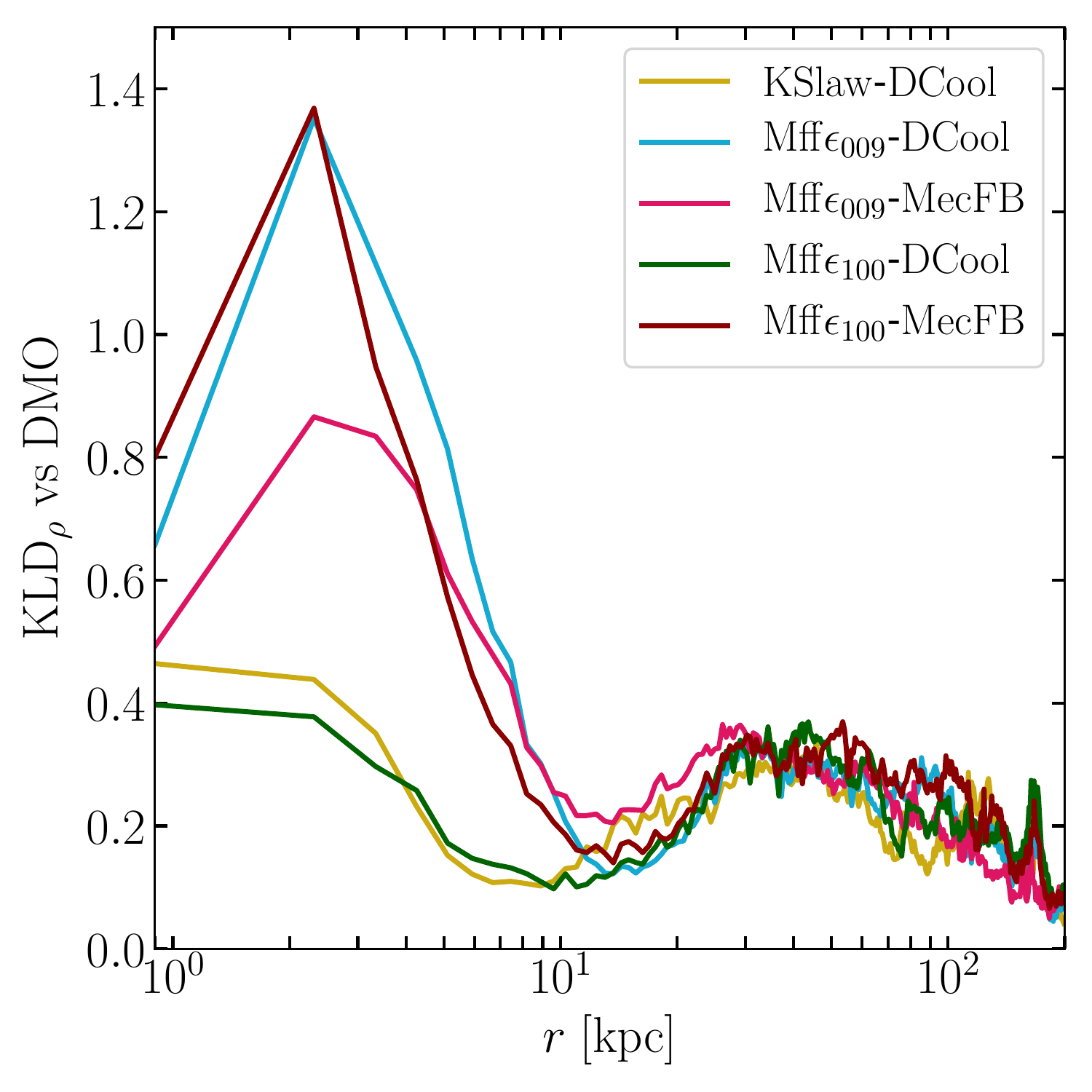}
\end{subfigure}
\begin{subfigure}[b]{0.45\linewidth}
\includegraphics[width=\linewidth]{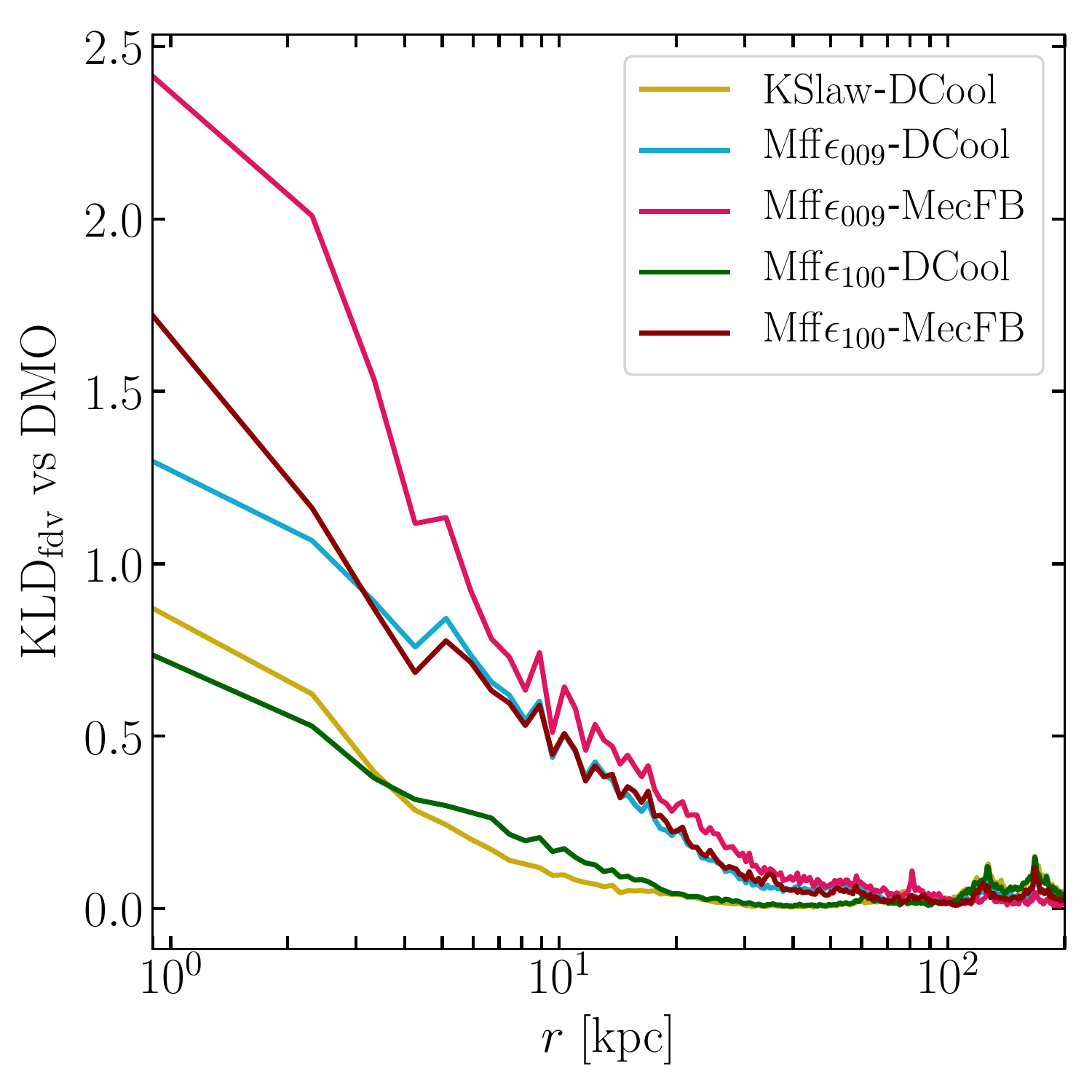}

\end{subfigure}

\caption{Kullback-Leibler divergence of the mass density distribution (left) and  the speed distribution (right) of the five hydrodynamics simulations with respect to the DMO run. All shells are spherical and isopopulated by $10^4$ dark matter particles.}
\label{fig:kullback-leibler}

\end{figure}

\section{Summary - Conclusions}\label{sec:summary}

This work presents a global analysis of the dark matter halo resulting in the re-simulations of the same Milky-Way size galaxy. All simulations have the same initial conditions, one run includes only dark matter and the other five include baryons. Those hydro runs differ in the models used to describe stellar physics. The results for all five cases is a  spiral disc galaxy and the detailed analysis of the baryonic component was presented in \cite{Nunez-Castineyra:2020ufe}. 

Baryonic effects on the DM halo are observed in two manners, common to all halos and singular to the specific baryonic distributions.
 
The mere presence of baryons, regardless of the baryonic physics implementation, pulls DM into the central regions and induces a common impact on the following features of the DM halo in all halo realizations:
\begin{itemize}
    \item \textbf{Sphericity in the central halo}: The inner halo is more spherical in the runs including baryons than in the DMO case due to the presence of the baryonic component and the deepening of the central potential. For r<30 kpc the DMO halo has $S\sim 0.55$ and in the hydro runs the halo have $S\sim 0.8$ (figure \ref{fig:SandT}).
    
    \item \textbf{Halo concentration}: Halos that include baryons have higher concentration that the DMO halo by factors between 1.5 and 3 (see table \ref{tab:halo_features}).
    
    \item \textbf{Central density profile}: The in-falling dark matter induces a cusp in the central density profile with different values for $\gamma$ that are all above 1 (figures \ref{fig:ProfileHist}, \ref{fig:r2rho_gamma} and table \ref{tab:haloprofiles_NFW}).
    
    \item \textbf{Density distribution}: DM particles with $r\gtrsim 14$~kpc, bounded to sub-halos in the DMO run, will no longer be inside sub-halos as most of them are destroyed in the hydro runs, therefore, the high-density tail of the density distributions are shifted to lower densities (figure \ref{fig:rhofdvhisto}) 
    
    \item \textbf{Velocity distribution}: The presence of baryons boosts the velocity of DM particles in the inner halo, the mean velocity is shifted towards higher values as well as the high-velocity (higher escape velocities). As a result, in the hydro runs most of the velocity distributions around the solar neighborhood are interestingly not so far from predictions inferred from a MW mass model (figure \ref{fig:rhofdvhisto}).
    \item \textbf{Energy distribution}: Similarly to the velocity, the energy distribution in all halos with baryons is boosted to higher energies, nevertheless this effect is dominant in the very central halo, r<3 kpc (figure \ref{fig:f_epsilon}). 
\end{itemize}

Additional analyses have been performed and resulted in small or not significant differences among the DM halos:

\begin{itemize}
    \item \textbf{Disc alignment}: This is an observation that concerns mainly the baryons and how they compare to the alignment of the DM halo in the DMO simulation. The plane where the galactic disc is formed is roughly the same in all cases. It is likely that the variation observed (around 30 degrees, see figures \ref{fig:anglesVS} and \ref{fig:maps-pot}) is related to small changes in the merger history, particularly, to the time of impact for the same merger in the different simulations. It is hard to relate this to the impact of baryonic physics as it can also be related to the inner stochasticity of the simulations. 
    \item \textbf{Halo edge}: Regardless of the different baryonic physics, the ``end of the halos'' coincides within a 5\% difference (section \ref{sec:Edge}). Additionally, using the escape velocity as a binding criterion different definitions for the halo edge are confronted, pointing toward the definition given in \cite{Bryan:1997dn} even if the notion of halo edge is less localized when relaxing sphericity. 
\end{itemize}

In the present analysis, however, it is observed that some punctual effects due to the different baryonic physics and the resulting gravitational influence can result in discernible differences in the DM distribution, such as:

\begin{itemize}
    \item \textbf{Triaxiality in the outer halo}: The DMO run does not show a stable difference with the baryonic runs in the shape of the outer halo as it does in the inner halo. The two cases with extreme SFRs induce less sub-halo destruction than in the other runs (see figure \ref{fig:SandT}). The presence of the surviving sub-halos results in halos with similar triaxilities to that of the DMO halo. Interestingly, these two cases do not share any baryonic implementation between them. 
    \item \textbf{The phase space distribution}: two main trends are observed in the phase space distribution linked to the different feedback implementations. The energy boost caused by the baryon-induced deepening of the central potential is concentrated inside 3 kpc, so much so that the energy distribution is dominated by DM particles inside this limit. Nevertheless, there is a significant difference related to the SN feedback. Indeed the two SN feedbacks tested here affect very differently the early stages of the formation of the galaxy with mechanical feedback allowing a higher early SFR that results in more massive stellar bulges. These early bulges have time to impact the energy distribution of the halo. 
    \item \textbf{The Stellar bulge relation to the central cusp}: The early bulge drives the formation of a cusp and its steepness. While cuspy density profiles are observed in all hydro simulations a correlation between the concentration of the halo and the mass of the stellar bulge is also observed. The bulge is the result of early star formation and its mass is related to the strength of the SN feedback.
\end{itemize}

Our main results on dark matter distribution are compared and synthesized over the entire radial range on figure \ref{fig:kullback-leibler}. We evaluate for the five hydrodynamics simulations, the departure from the DMO distribution for the density profile and the speed distribution with the Kullback-Leibler divergence \cite{10.1214/aoms/1177729694} \footnote{This metric is defined by $D_{\textsc{KL}} (P|Q) \equiv \sum_{i} P(i) \log\left(\frac{P(i)}{Q(i)}\right) $ and is used to evaluate the difference between two probability distributions ($D_{\textsc{KL}}(P|Q=P) = 0$, the higher the $D_{\textsc{KL}}$ value, the more differences between the two distributions).}. Naturally, it confirms that differences are more pronounced in the baryonic matter extension area, typically inside 20 kpc. A hierarchy of the different baryonic potentials is also seen. Indeed, if all distributions show net deviations from DMO, the denser objects have larger $D_{\textsc{KL}}$ values. 
The tested combinations of baryonic physics strategies exert modifications on the distribution of galactic dark matter mediated by the evolving gravitational potential. The delayed cooling scheme is a very effective description while mechanical feedback modeling is based on SN explosion phases. Both approaches give rise to excesses in early star formation, around $z\sim 6-2$ which is more dramatic for the mechanical feedback. This tension could be appeased by the introduction of early feedback processes such as AGN or stellar winds, which absence might be hidden by the overly efficient delayed cooling implementation. Regarding star formation, models with low efficiency, $\epsilon$, counter-intuitively lead to more prominent bulges (see figure 7 of paper 1) and more contracted dark matter profiles.

This paper emphasizes the importance of baryonic physics on dark matter halo properties and illustrates the related variability and uncertainties with up-to-date hydrodynamical cosmological simulations of Milky-Way analogs. This suggests taking with caution strong predictions related to dark matter detection often derived or inspired from numerical simulations. Nevertheless, even if such numerical objects are not \emph{the} Milky Way, they represent very consistent frameworks for dark matter studies and show that understanding galaxy formation is also of prime importance to control dark matter distribution features and  related phenomenology and detection aspects. Namely, even if predictions using blindly DM selections in cosmological simulations are relevant and interesting to have estimations, some caveats remain.

Moreover, a detailed comparison between the simulations and the real Milky Way data is mandatory to highlight such approaches and weigh the messages.\\
At the Milky Way halo  scale,  one set of baryonic physics induces specific features on the dark matter distribution. This makes  simple popular assumptions like Maxwellian velocity assumptions or  NFW/Einasto DMO-inspired profiles low probable and not realistic. Indeed, additional physics will modify further the resulting dark matter distribution and complexify the equilibrium between contraction effect and cusp destruction by feedback. Among those processes, AGN, MHD, or cosmic rays are expected to have non-negligible effects on galaxy formation and the resulting dark matter distribution in the halo. Such improvements in galaxy formation understanding are likely in the perspective of GAIA, Ton size direct detection experiments as well as the next generation of gamma, neutrino, and cosmic ray indirect detection experiments. Baryonic physics improvements are also expected to address some remaining debated questions in cosmological simulations like the hot orbit problem, the plane of satellites, and the formation of bars.

\acknowledgments

We thanks Jean-Charles Lambert for his support in numerical computing aspects.
Centre de Calcul Intensif d’Aix-Marseille is acknowledged for granting access to its high-performance computing resources. This work benefited of the scientific environment from the ANR project ANR-18-CE31-0006 (GaDaMa).

 \bibliography{mainhalo} 

\bibliographystyle{JHEP}
\appendix
\section{The shape tensor} \label{appe:shape}

\begin{figure}[t]
\includegraphics[width=0.87\linewidth]{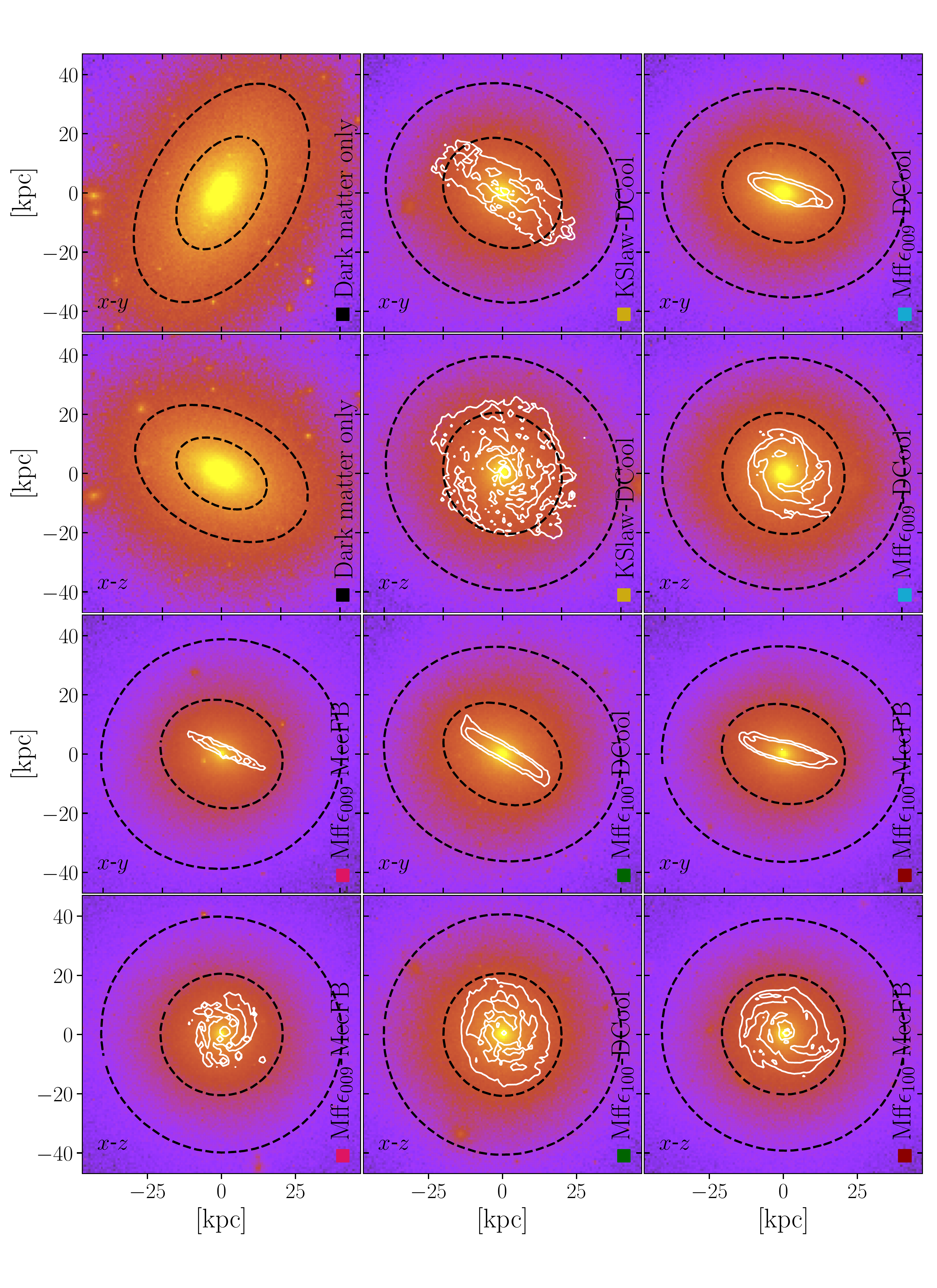}

  \caption{Two projections of the DM mass of the inner halo for the six Mochima runs. The plane of the projections (indicated in the lower left corner of each panel) is in the original system of the simulation box common to all runs. The resulting ellipsis that describes the shape of the inner halo (black dashed lines), and the distribution of young stars (white contours) are shown on top of the mass map.}
\label{fig:maps-pot}

\end{figure}

\begin{figure}
  \begin{minipage}[c]{0.70\textwidth}
    \includegraphics[width=\textwidth]{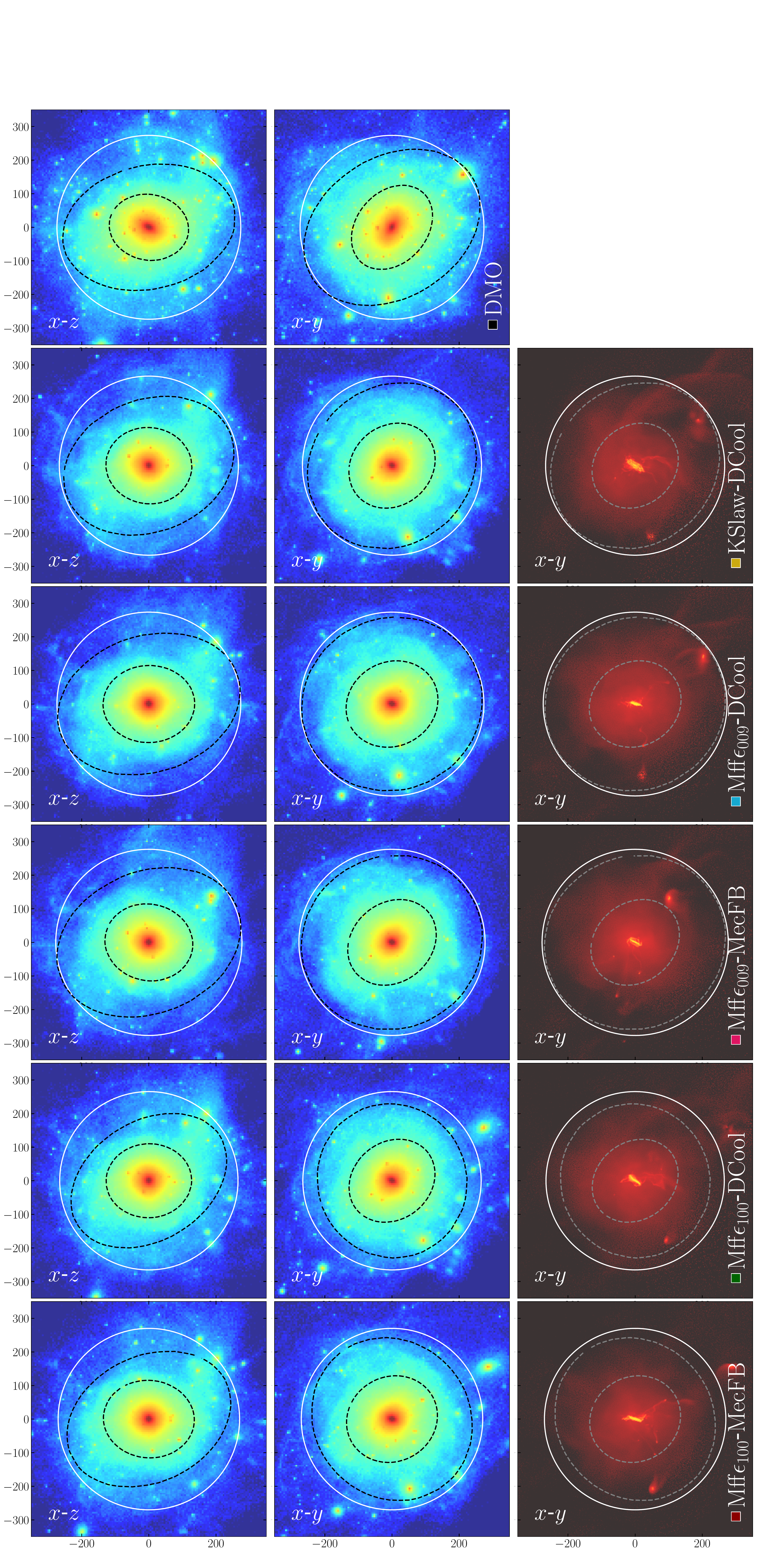}
  \end{minipage}\hfill
  \begin{minipage}[c]{0.3\textwidth}
    \caption{Two projections of the DM mass distribution in the left and center column, and the projection of the gas mass in the right column for the hydro runs. The planes of the projections are in the coordinate system of the simulations box and are labeled in the lower left corner of each image. The virial radius $r_{97}$ is shown in white and an ellipsoid computed with the dark matter mass is shown in dashed lines. The ellipses are computed by forcing the major semi-axis equal to the virial radius and half of the viral radius.
    } \label{fig:mapsfull-YZ}
  \end{minipage}
\end{figure}

\begin{figure}[!ht]
  \centering
  \begin{subfigure}[b]{0.85\linewidth}
    \includegraphics[width=\linewidth]{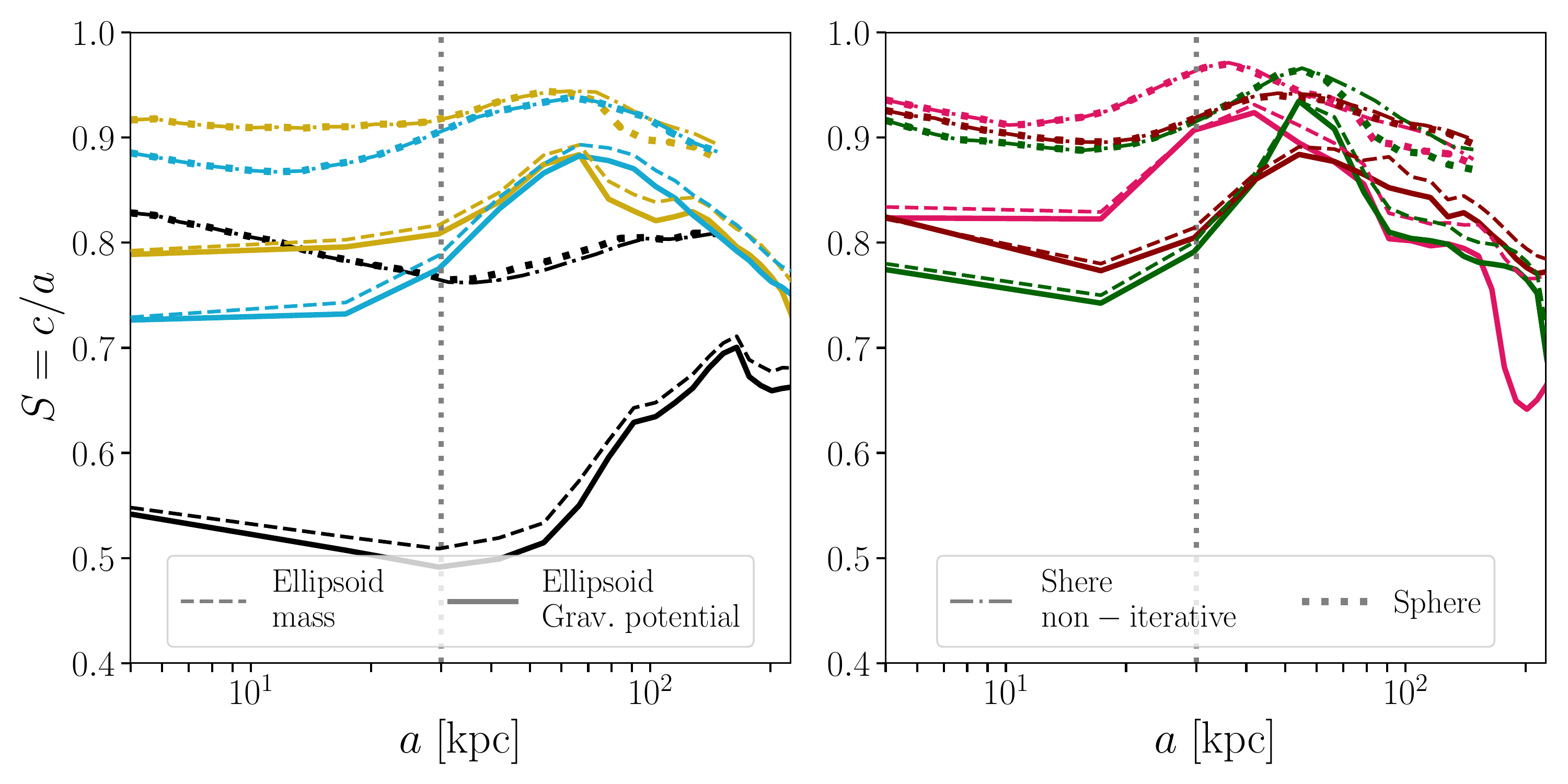}
  
  \end{subfigure}
    \begin{subfigure}[b]{0.9\linewidth}
       \includegraphics[width=\linewidth]{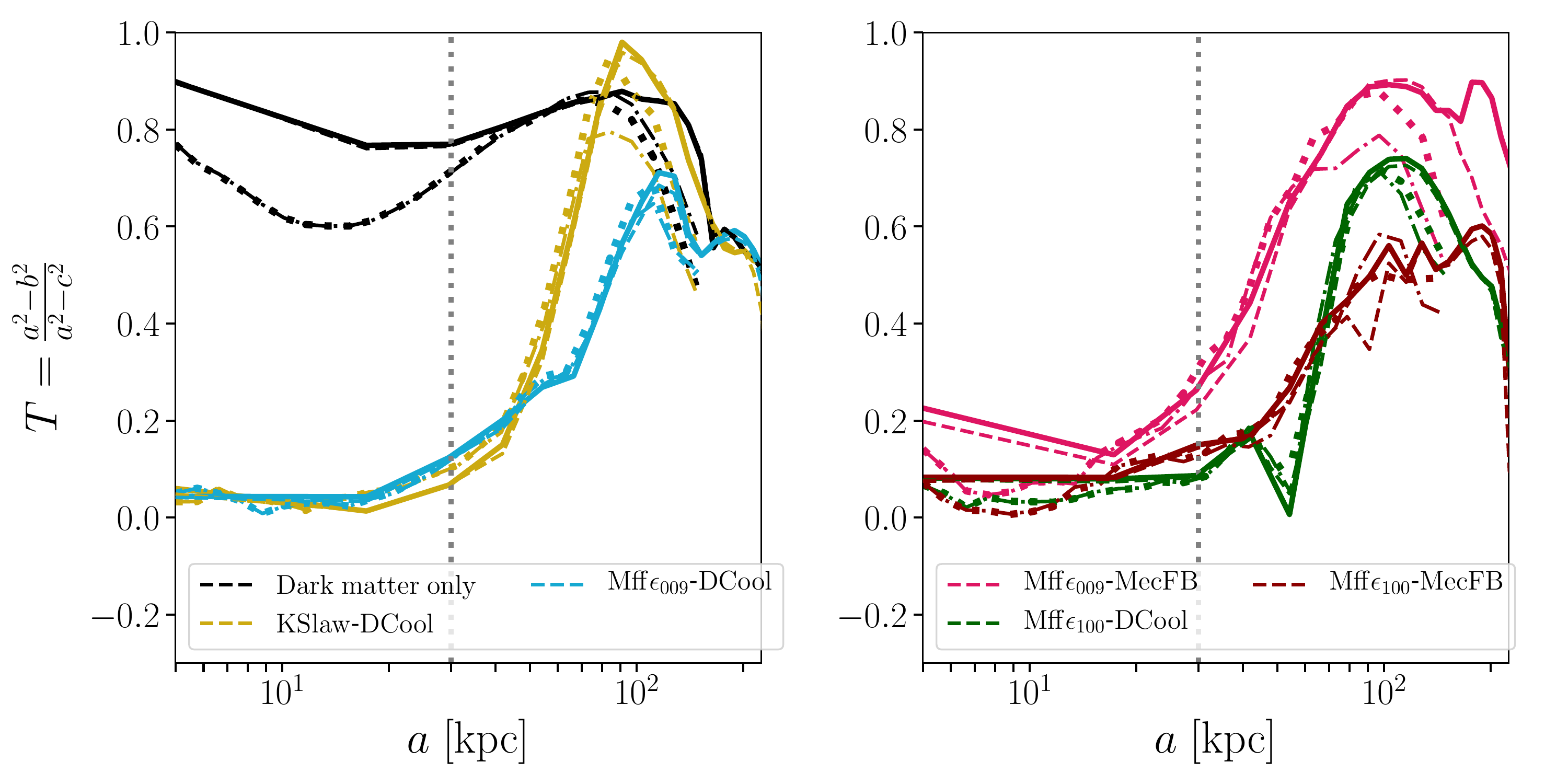}

    \end{subfigure}

  \caption[ Sphericity and Triaxiallity all techniques]{Same as in figure \ref{fig:SandT} but using different observables and volumetric selections, Ellipsoids in mass (dashed) and gravitational potential (solid), iterative sphere(dotted) and non-iterative sphere (dash-dotted) using mass.The vertical dotted line separates the two regions
of interest, the inner halo region (a<30 kpc) and the outer halo region (a>30 kpc) as described
in the text.}\label{fig:SandTallTech}

\end{figure}


The triaxiality of a distribution of points like a halo, based on a typical weight or observable, $o$ e.g the mass of each point, can be characterized by diagonalizing the so-called shape tensor:


\begin{equation}
\mathcal{S}_{i,j} = \frac{\sum_k o_k (x_k)_i (x_k)_j}{\sum_k o_k} \,\,.
\end{equation}

\noindent The subindex $k$ iterates over all the points (DM particles in our case) inside a given volume. $o_k$ corresponds to the value of the observable related to the point $k$. Finally, $(x_k)_i$ is the  $i$-th component of the position vector $\vec{x}$ of the point, in a coordinate system with the origin set in the centre of the halo. 
The tensor eigenvectors point in the direction of the semi-principal axis of the ellipse with norm $a>b>c$ ($a=b=c$ is obtained for a spherical distribution) and its eigenvalues can be written as $a^2/3$, $b^2/3$ and $c^2/3$.

This computation can be done with different volumetric selections. When the distribution is not fully spherical and smooth the specific sub-selection of points used to compute the radial evolution of the semi-principal axes could change the overall result. The first approach to obtain the semi-principal axes is to consider all points inside a certain radius $r$. This spherical selection biases the resulting axes, therefore, a new ellipsoidal selection need to be done following the inferred axes. The re-selection and re-calculation of the axes can be done iteratively until the resulting axes converge. In our work, convergence is assumed when the mean difference between the resulting and previous set of semi-principal axes is less than $0.1\%$. To avoid drastic divergence due to local asymmetries (a massive sub halo inside a DM halo for example) all the axes are rescaled ensuring always that $a=r$, where $r$ is the initial radius of the iteration. This procedure can also be done using shells and homeoid instead of spherical an ellipsoidal selection to focus on the local shape. For a smooth distribution of points, the most effective volume selection is a converging homeoid but in the case of a distribution with important sub-structures, an ellipsoidal selection is more adequate\cite{2011ApJS..197...30Z}. The reason for this is that a sub-halo that falls inside a homeoid can drastically change the results, the sub-halo effect is softened by the central mass in the full ellipsoid.

Figure \ref{fig:maps-pot} shows in black dashed lines the resulting ellipsoids coming out from this procedure applied on the six simulated halos. Additionally, figure \ref{fig:SandTallTech} shows the results for $T$ and $S$ using four different approaches;- a spherical volumetric selection where the eigenvalues are computed only once, i.e. the non-iterative method (dash-dotted lines). - A spherical volumetric selection but the procedure is repeated until convergence is reached (dotted lines).- An ellipsoidal volume selection with the mass of the dark matter inside the selection (dashed lines) or - the total gravitational potential measured in the position of the dark matter particles contained in the selection (solid lines). In both cases with ellipsoidal selections, the computation is iterative.

For the spherical selection, both methods yield  basically the same result and force a high value of $S$, this is not surprising since the selection is always spherical. 
In the outskirt of the halo, the presence of massive sub-halos tends to accentuate $T$. This effect is slightly more apparent if the triaxiality is calculated using the gravitational potential, particularly, in the case with higher baryonic content in subhalos. Indeed this component of the simulations was not directly used so far. On the other hand, if $T$ is calculated using the particles mass these effects are washed out. An example of such situation is shown in the triaxiality of the Mff$_{\epsilon099}$-MecFB run in the outer halo (figure \ref{fig:SandTallTech}).

\section{The edge of halos beyond spherical selections.} \label{appe:vesc-isopot}
The shape of the outer halo is not drastically impacted by the central collapse induced by the presence of the baryons. This can be seen on the convergence observed on the $S$ and $T$ parameters for r>100 kpc in figure \ref{fig:SandT}. Nevertheless, the position of the orbiting sub-halos is not exactly the same and their presence will impact the Triaxility calculations depending on the method that is used (see appendix \ref{appe:shape}).  

\begin{figure}[t]
\begin{center}

\includegraphics[width=\linewidth]{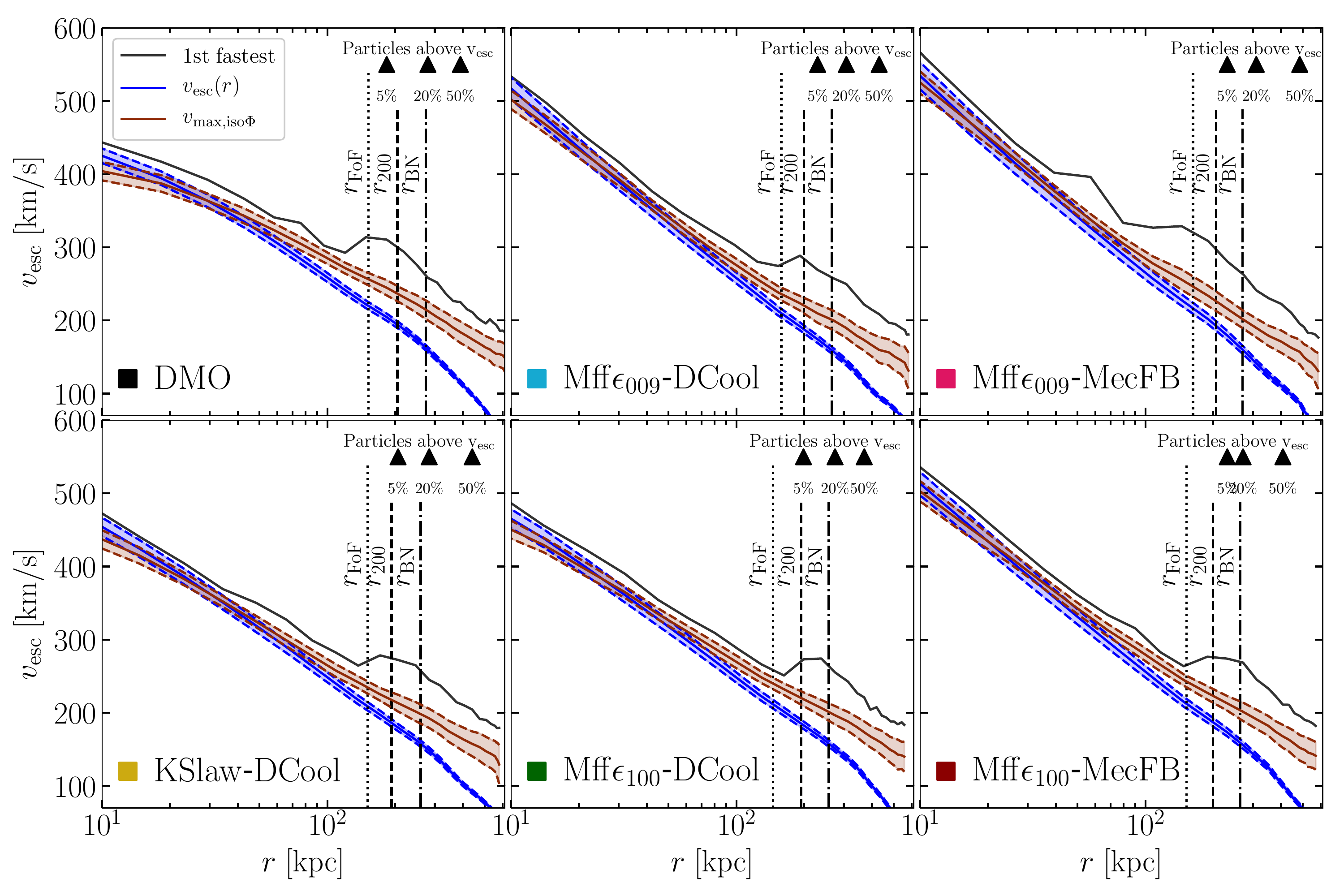}
\caption{Same as figure \ref{fig:vesc-vmax-spherical} with isopotential selections. The overlapping of $v_{max}$ and $v_{esc}$ suggests a less localized and determined halo extension.}
\label{fig:vesc-vmax-iso}
\end{center}
\end{figure}

This can be observed in figure \ref{fig:mapsfull-YZ} which shows two projections of each DM halo and one projection of the gas content. Here the virial radius is shown in white and in a dashed black ellipsis the computed shape of the halo for a fixed major semi-axis $a= r_{\rm BN}, r_{\rm BN}/2$. Additionally, as an illustration, the gas distribution is shown in one of the projections for the full halo. The outer shape shows some variability and it seems to be related to the position of the sub-halos. This claim needs to be confirmed in a careful study of the sub-halo population which is out of the scope of this paper. Nevertheless, it is clear that the distribution of the sub-halos changes from simulation to simulation, this suggests that the differences in the baryonic content impact the merger history of the central halo. Furthermore, we evaluate the outer boundary of the halo beyond sphericity. To this end, isopotential selections are considered instead of radial shells and look at the relationship between the escape speed, the mean maximal velocity, and the velocity of the fastest particle in the selection. The results are shown in figure \ref{fig:vesc-vmax-iso}. In most cases, the mean escape velocity and the mean maximal velocity follow each other more closely in the isopotential selection than in the spherical case. All the usual characteristic radii stand now after the  overlapping region of $v_{max}$ and $v_{esc}$, suggesting a slightly reduced halo extension. The two velocities disentangle in the outer halo likely due to the presence of sub-halos, or more importantly, the departure from the smooth section of the halo. This explains why the disentangling happens at a smaller radius for the DMO run where there are more surviving sub-halos.

\begin{figure}[t]
\begin{center}

\includegraphics[width=0.5\linewidth]{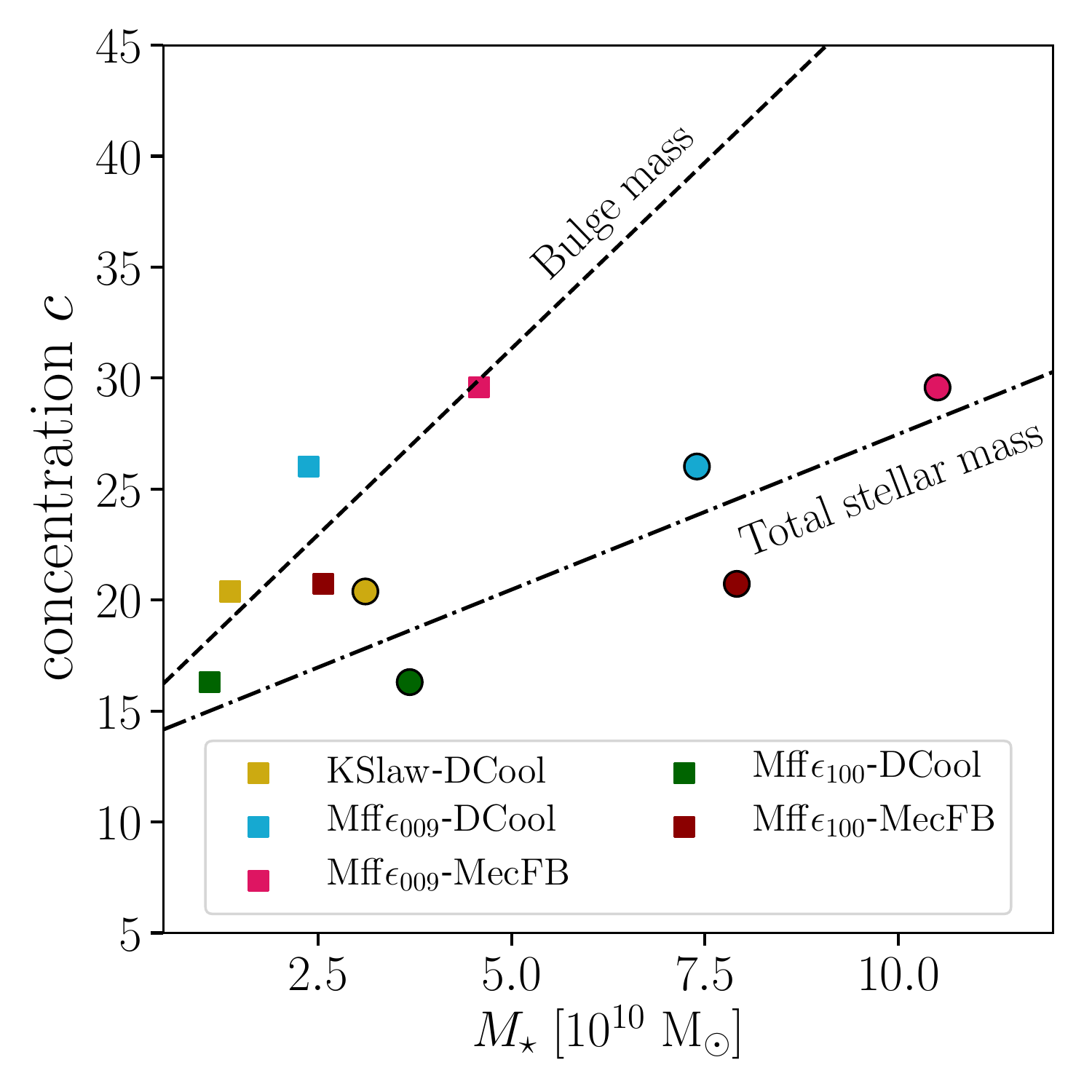}
\caption{Halo concentration with respect to the total stellar mass (circles) or to the central bulge mass (squares). A linear fit is shown in dot-dashed lines for the total stellar masses and in dashed ot the bulge masses. The resulting coefficients sate a stronger correlation of the concentration with the bulge than with the total stellar mass.}
\label{fig:concentrationcorr}
\end{center}
\end{figure}


\section{Halo concentration correlation with bulge mass}\label{app:concentration}

 The early formation of a stellar bulge ($z=3$), drives the contraction of the halo into a  steeper cusp. 
 For the sake of clarity, figure \ref{fig:concentrationcorr} highlights the correlation between the total and the bulge-only stellar masses with the concentration parameters. 
 The concentration values are much more sensitive to the bulge mass than to the total stellar mass. Actually, this is not a surprising result, our galaxies have similar bulge and disc masses, while the bulges are concentrated in a smaller, centrally located volume (see figure \ref{fig:SandT}).
 Two couples of our simulations serve as a good example to support this bulge-cusp relation, two sets of stellar mass twins i.e. galaxies with similar stellar mass (within 15\%); 1) \blue\ and \rouge (the former is 1.4\% smaller than the latter), and 2) \jaune\ and \verte  (the former is 16\% smaller than the latter). The first couple differs on the bulge: already at $z=3$ (dash line of the baryonic profile in figure \ref{fig:ProfileHist}) in \rouge\ the central baryonic component extends up to 300 pc at a density  $10^{10}$~M$_{\odot}$, while in \blue\ the baryonic density  profile does not exceed this value anywhere, this difference on the central bulge is mirrored in the DM profile as \rouge\ ends up having a cuspier profile ($\gamma= 1.82$) than \blue ($\gamma = 1.37 $) at $z=0$. Alternatively, the second couple has almost the same exact inner baryonic profile at z=3, but a very different stellar disc at $z=0$. Since the stellar disc is formed at late times in galactic history, this difference does not have time to affect the DM halo by z=0. As a result, the two DM profiles are fully consistent with each other above 300 pc (with $\gamma = 1.70$ for \jaune\ and $\gamma = 1.69$ for \verte).  Below 300 pc however, the baryonic profile of \verte\ is steeper in the center but enough to induce extra pinching of the very central DM profile. Nevertheless, this region is close to the resolution limit and the central difference 



\section{Fitting the density profile}\label{app:FittingProfiles}

\begin{figure}[ht!]
  \centering
  \begin{subfigure}[b]{0.4\linewidth}
    \includegraphics[width=\linewidth]{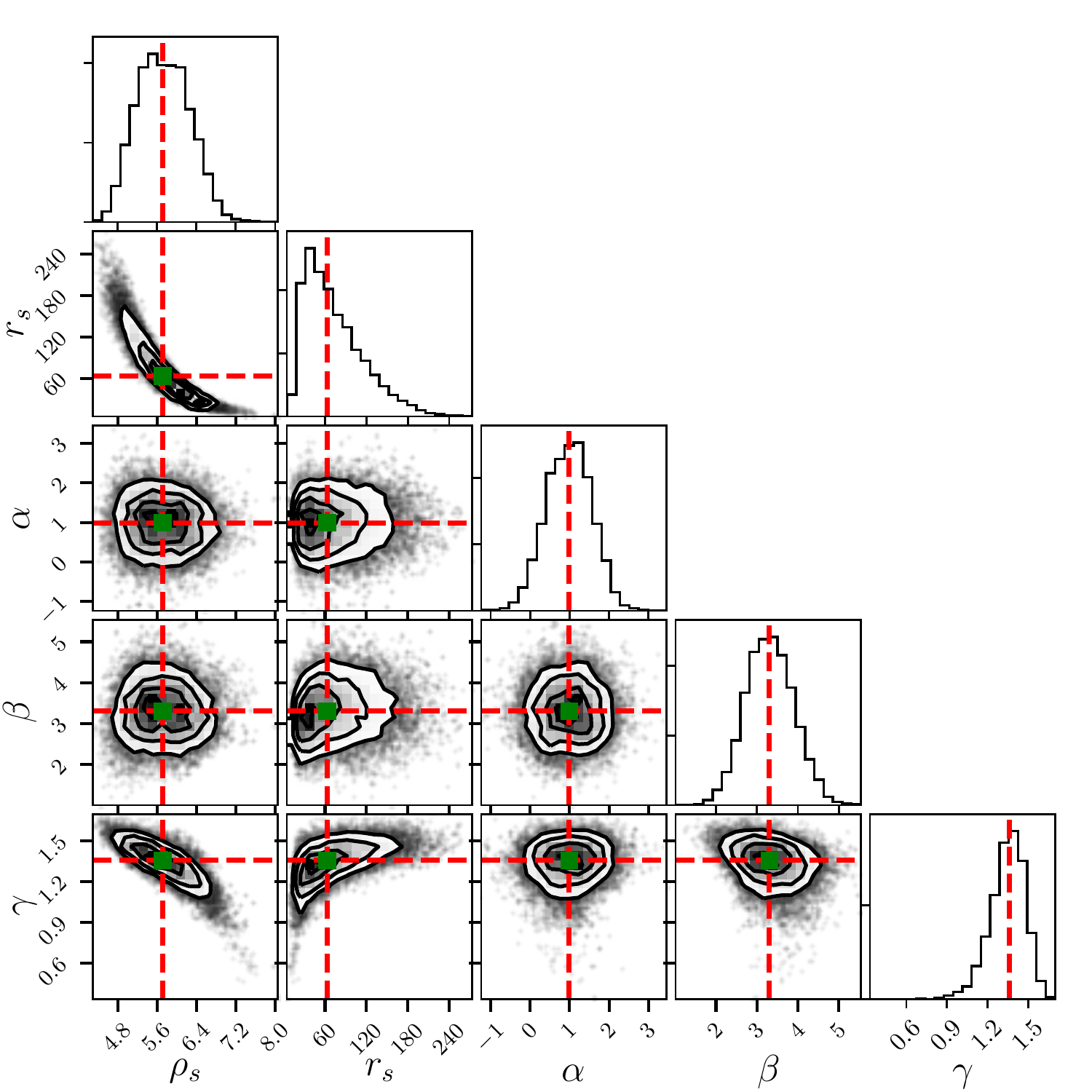}
    \caption{DMO}
    \label{fig:parameter_DMO}
  \end{subfigure}
   \begin{subfigure}[b]{0.4\linewidth}
    \includegraphics[width=\linewidth]{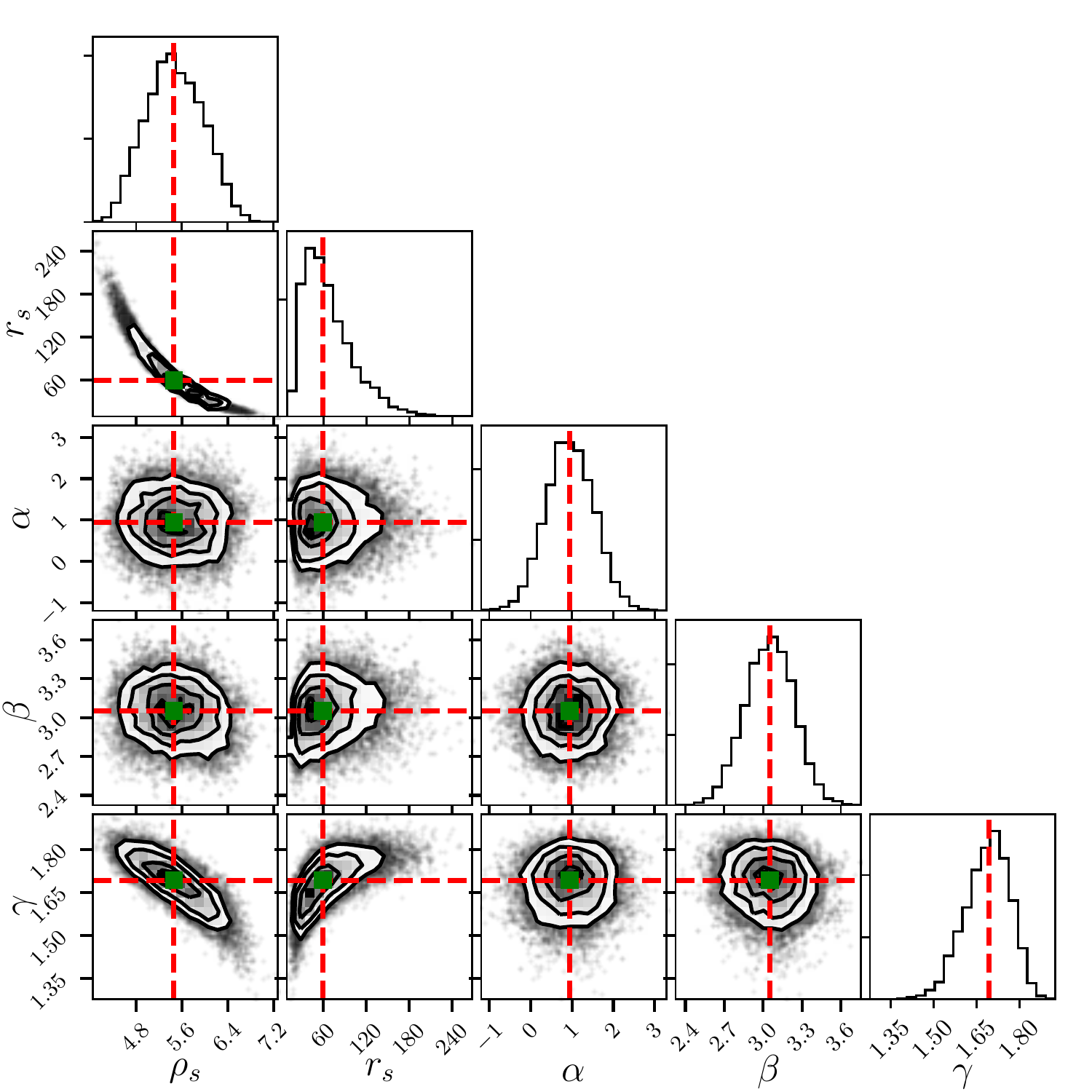}
    \caption{\textcolor{colorYELLOW}{KSlaw-DCool}}
    \label{fig:parameter_SF0DC}
  \end{subfigure}
   \begin{subfigure}[b]{0.4\linewidth}
    \includegraphics[width=\linewidth]{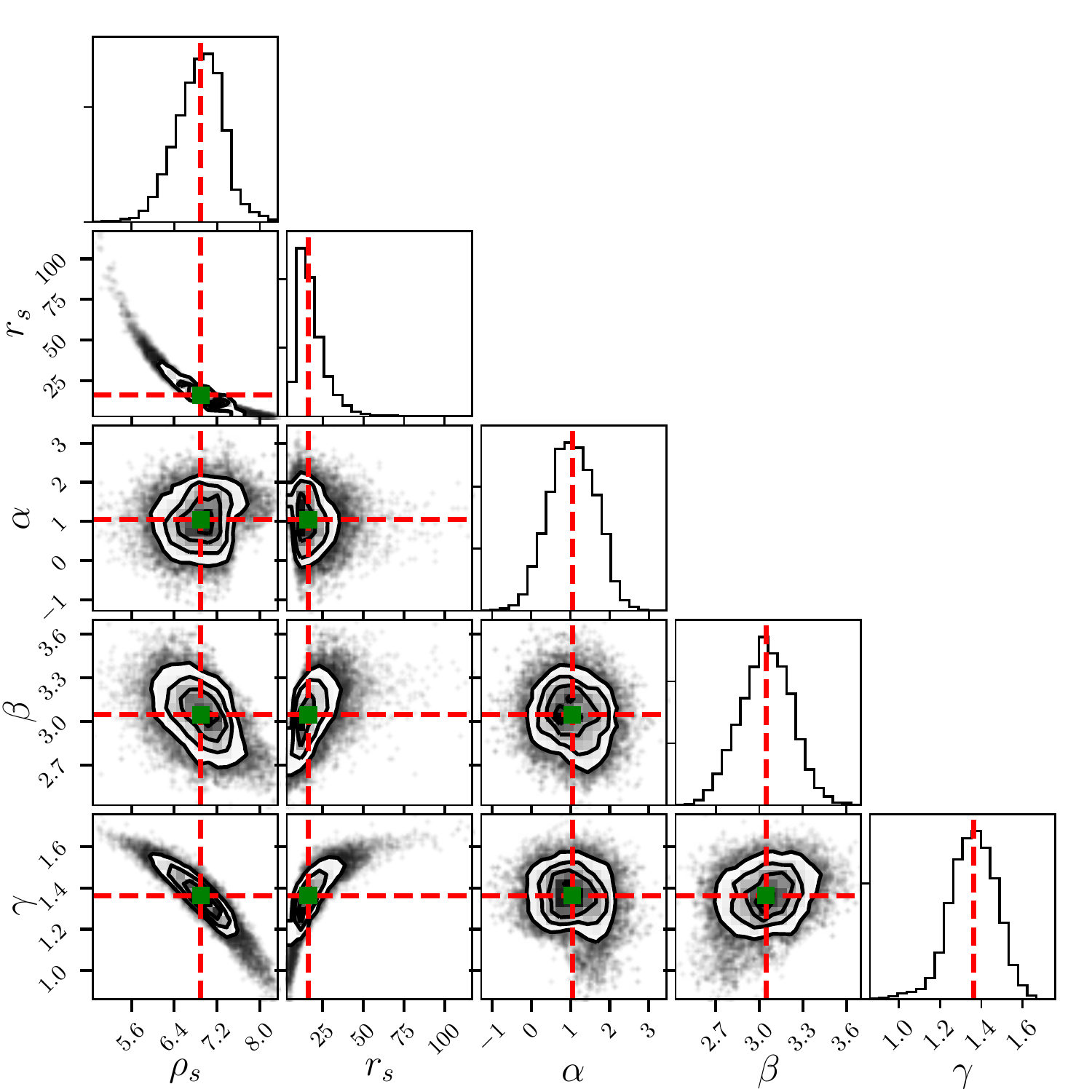}
    \caption{\textcolor{colorBLUE}{Mff$\epsilon_{009}$-DCool}}
    \label{fig:parameter_SF1DCe1}
  \end{subfigure}
  \begin{subfigure}[b]{0.4\linewidth}
    \includegraphics[width=\linewidth]{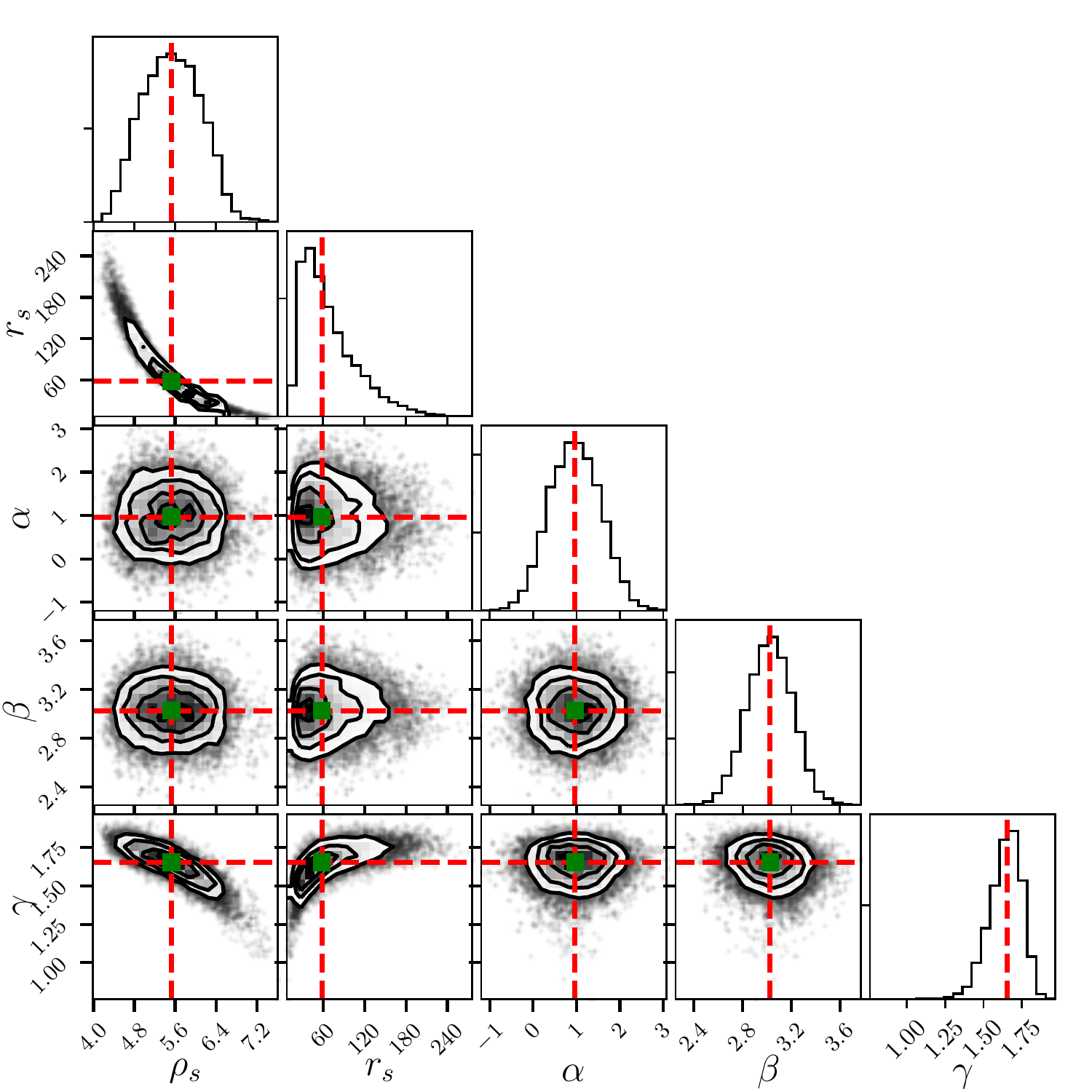}
    \caption{\textcolor{colorGREEN}{Mff$\epsilon_{100}$-DCool}}
    \label{fig:parameter_SF1DCe1}
  \end{subfigure}
  \begin{subfigure}[b]{0.4\linewidth}
    \includegraphics[width=\linewidth]{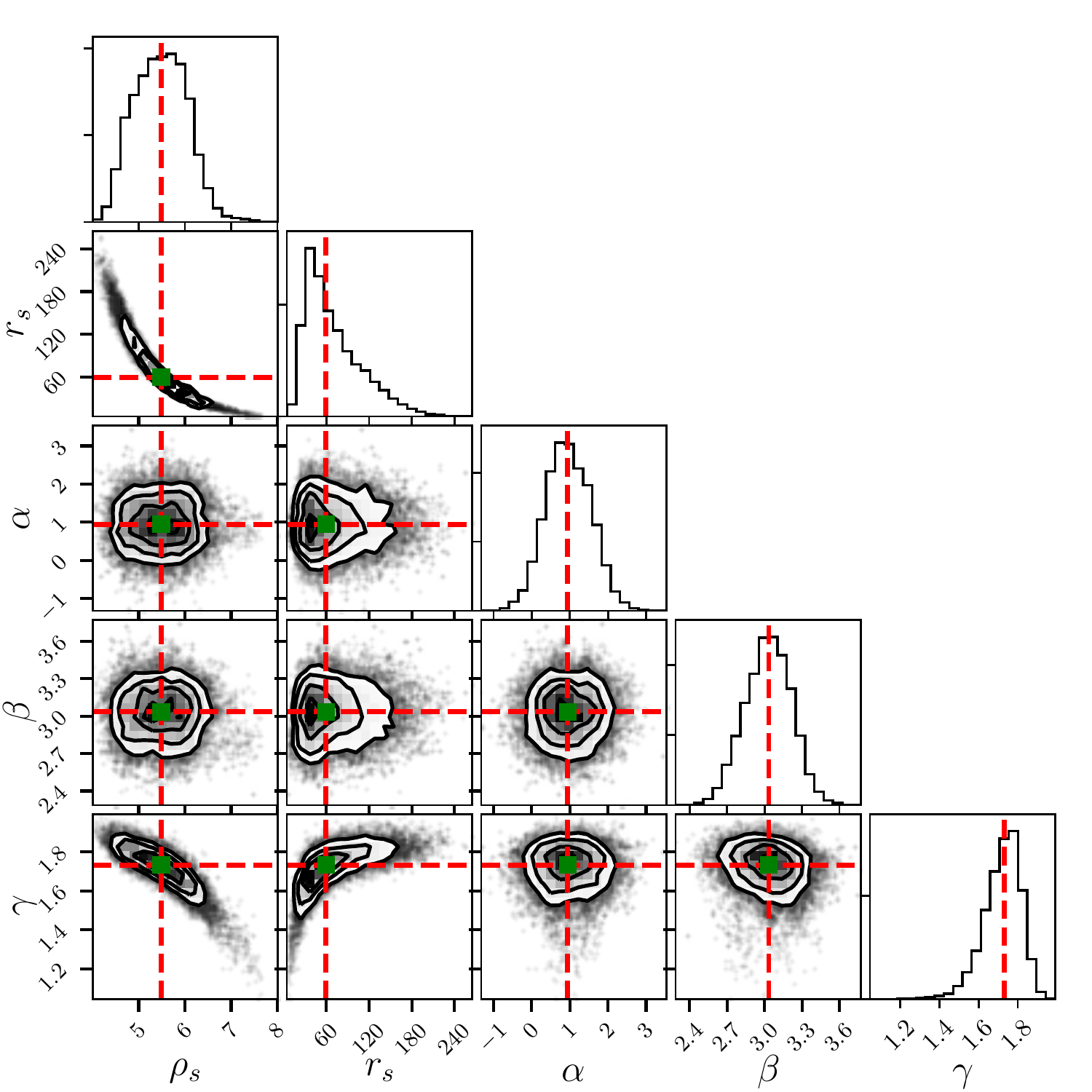}
    \caption{\textcolor{colorROSE}{Mff$\epsilon_{009}$-MecFB}}
    \label{fig:parameter_SF1ME}
  \end{subfigure}
  \begin{subfigure}[b]{0.4\linewidth}
    \includegraphics[width=\linewidth]{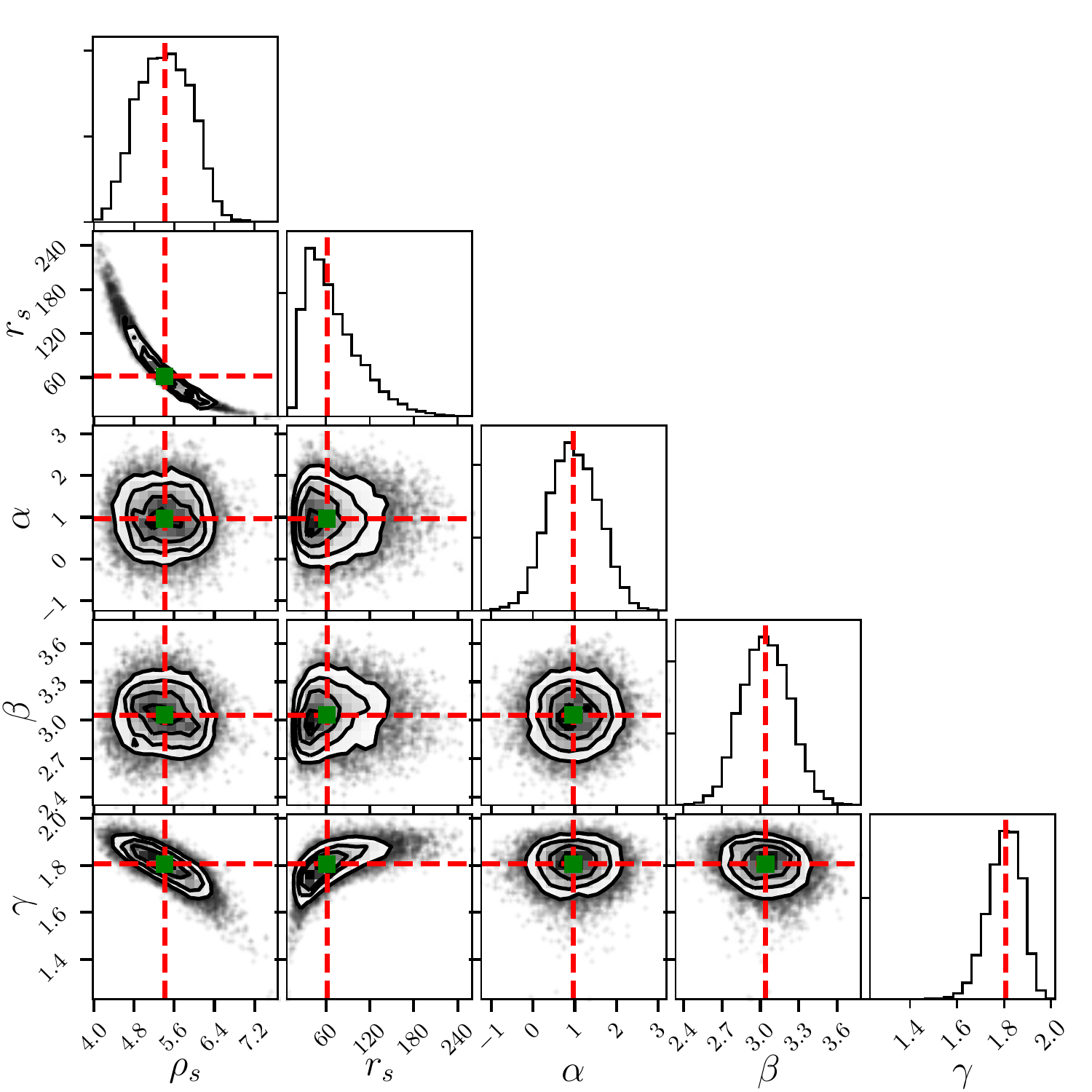}
    \caption{\textcolor{colorRED}{Mff$\epsilon_{100}$-MecFB}}
    
    \label{fig:parameter_SF1MEe1}
  \end{subfigure}
     
    \caption{The posterior distributions of the parameters for the $\alpha\beta\gamma$-profile fitted over the different simulations.}
    \label{fig:MCMCfit_posteriors}
\end{figure}

\begin{figure}[ht!]
  \centering
  \begin{subfigure}[b]{0.4\linewidth}
    \includegraphics[width=\linewidth]{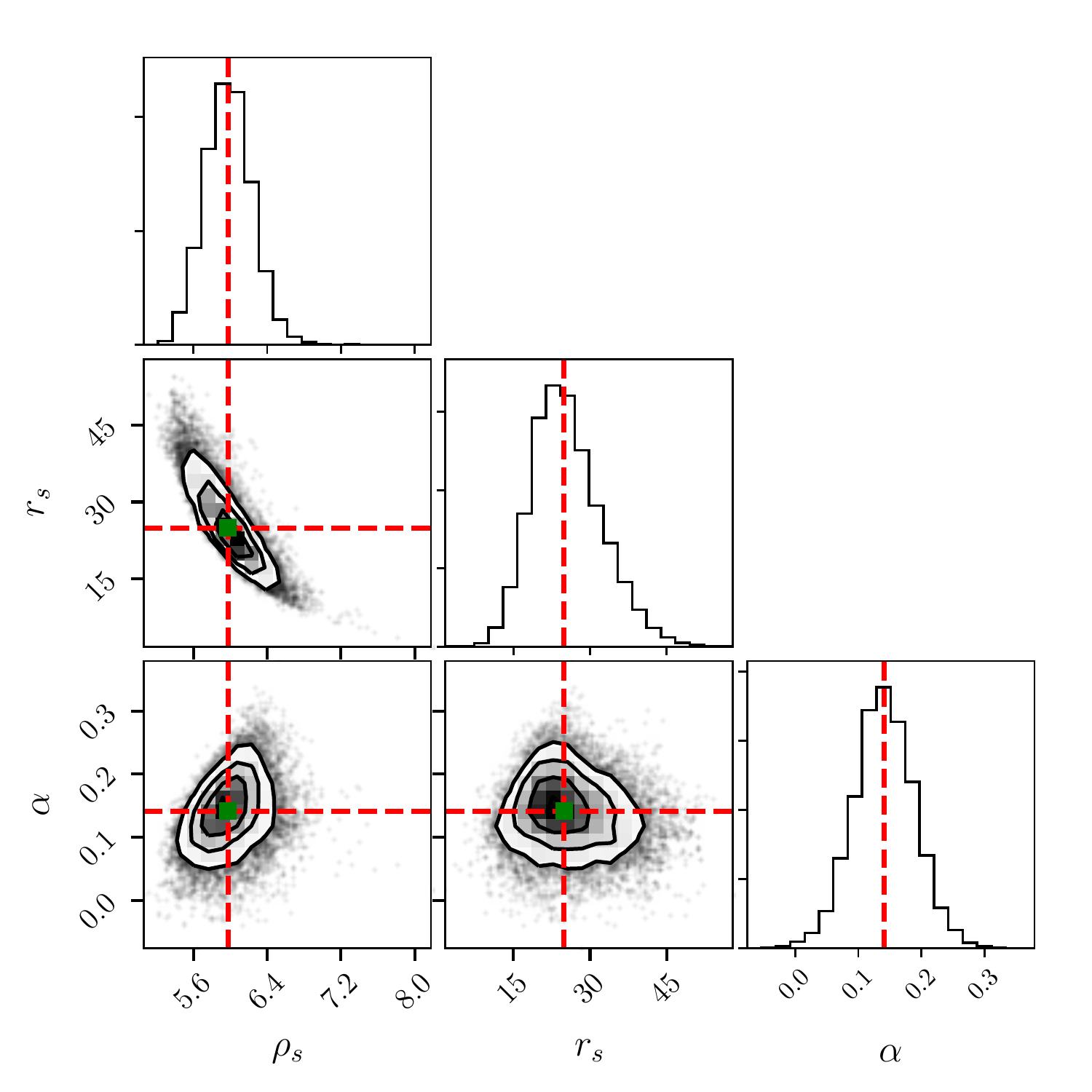}
    \caption{DMO}
    \label{fig:parameter_DMO_Einasto}
  \end{subfigure}
   \begin{subfigure}[b]{0.4\linewidth}
    \includegraphics[width=\linewidth]{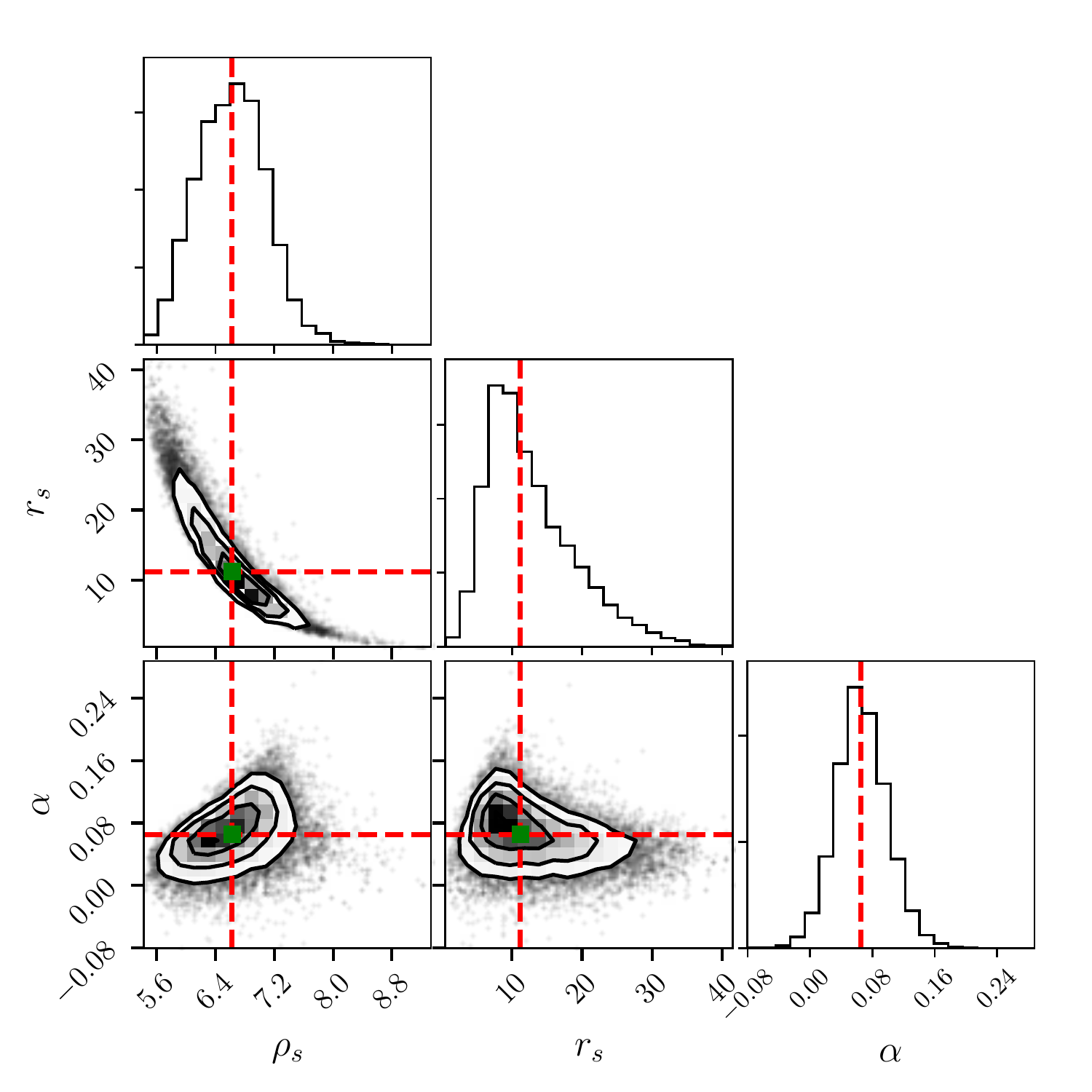}
    \caption{\textcolor{colorYELLOW}{KSlaw-DCool}}
    \label{fig:parameter_SF0DC_Einasto}
  \end{subfigure}
  \begin{subfigure}[b]{0.4\linewidth}
    \includegraphics[width=\linewidth]{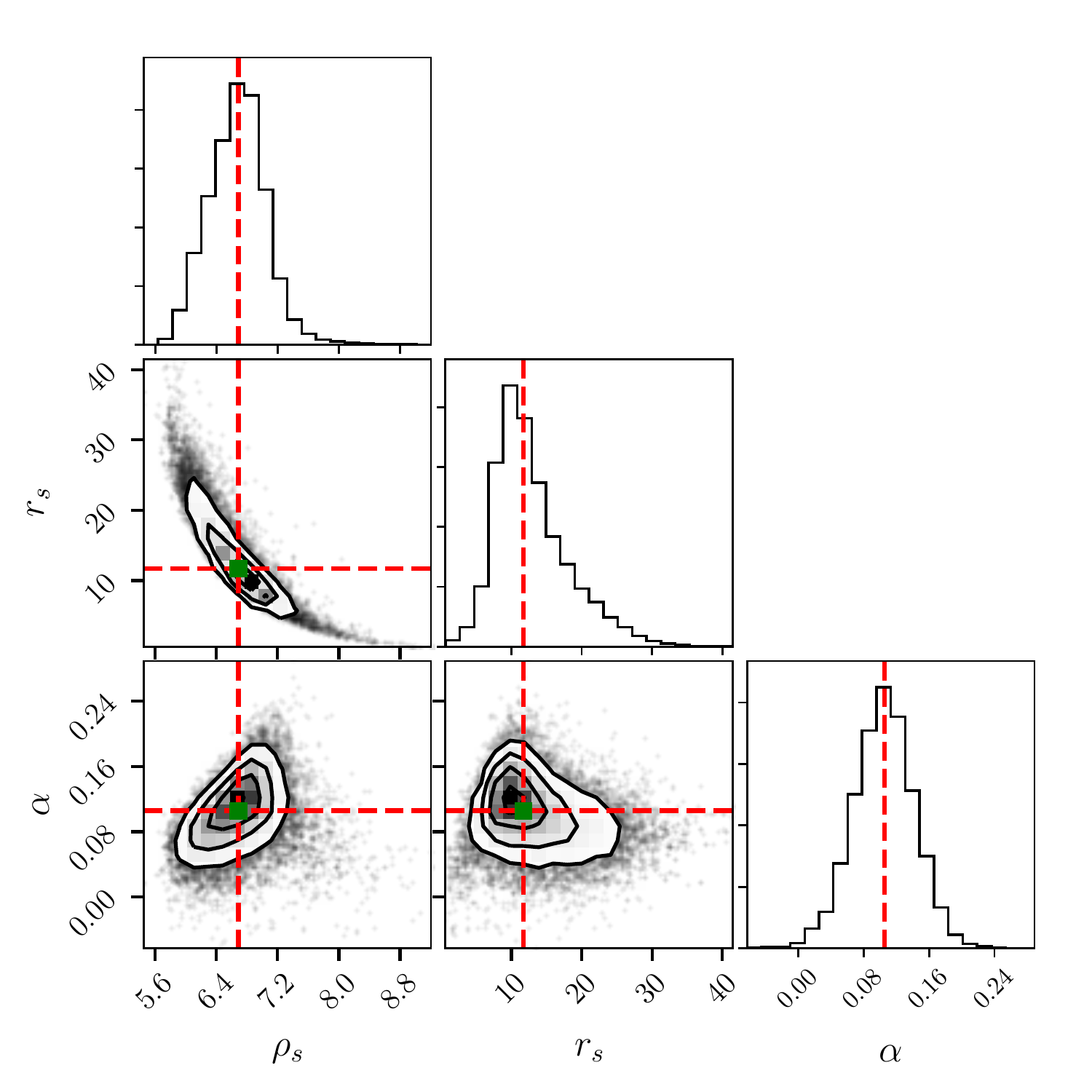}
    \caption{\textcolor{colorBLUE}{Mff$\epsilon_{009}$-DCool}}
    \label{fig:parameter_SF1DC_Einasto}
  \end{subfigure}
  \begin{subfigure}[b]{0.4\linewidth}
    \includegraphics[width=\linewidth]{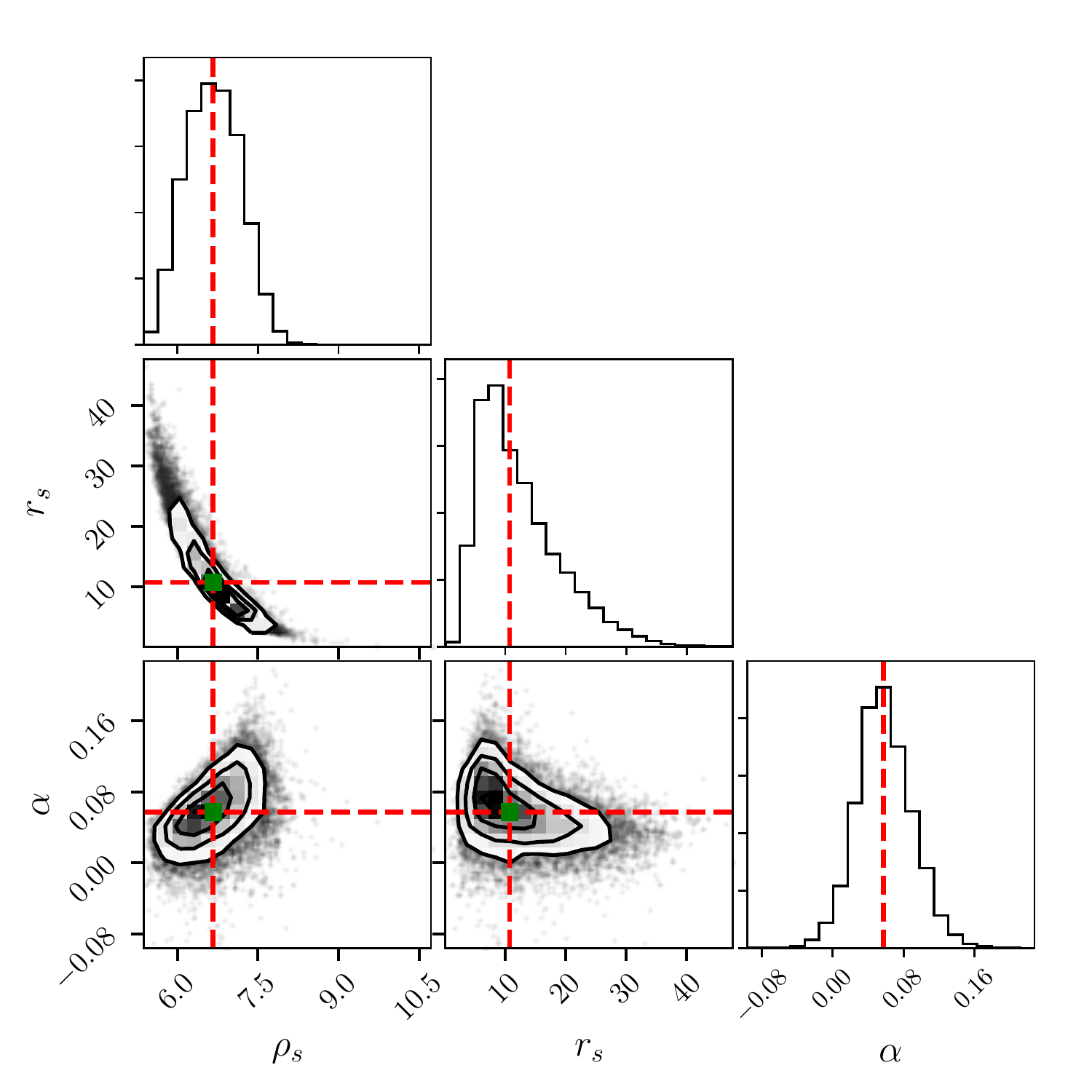}
    \caption{\textcolor{colorGREEN}{Mff$\epsilon_{100}$-DCool}}
    \label{fig:parameter_SF1DCe1_Einasto}
  \end{subfigure}
  \begin{subfigure}[b]{0.4\linewidth}
    \includegraphics[width=\linewidth]{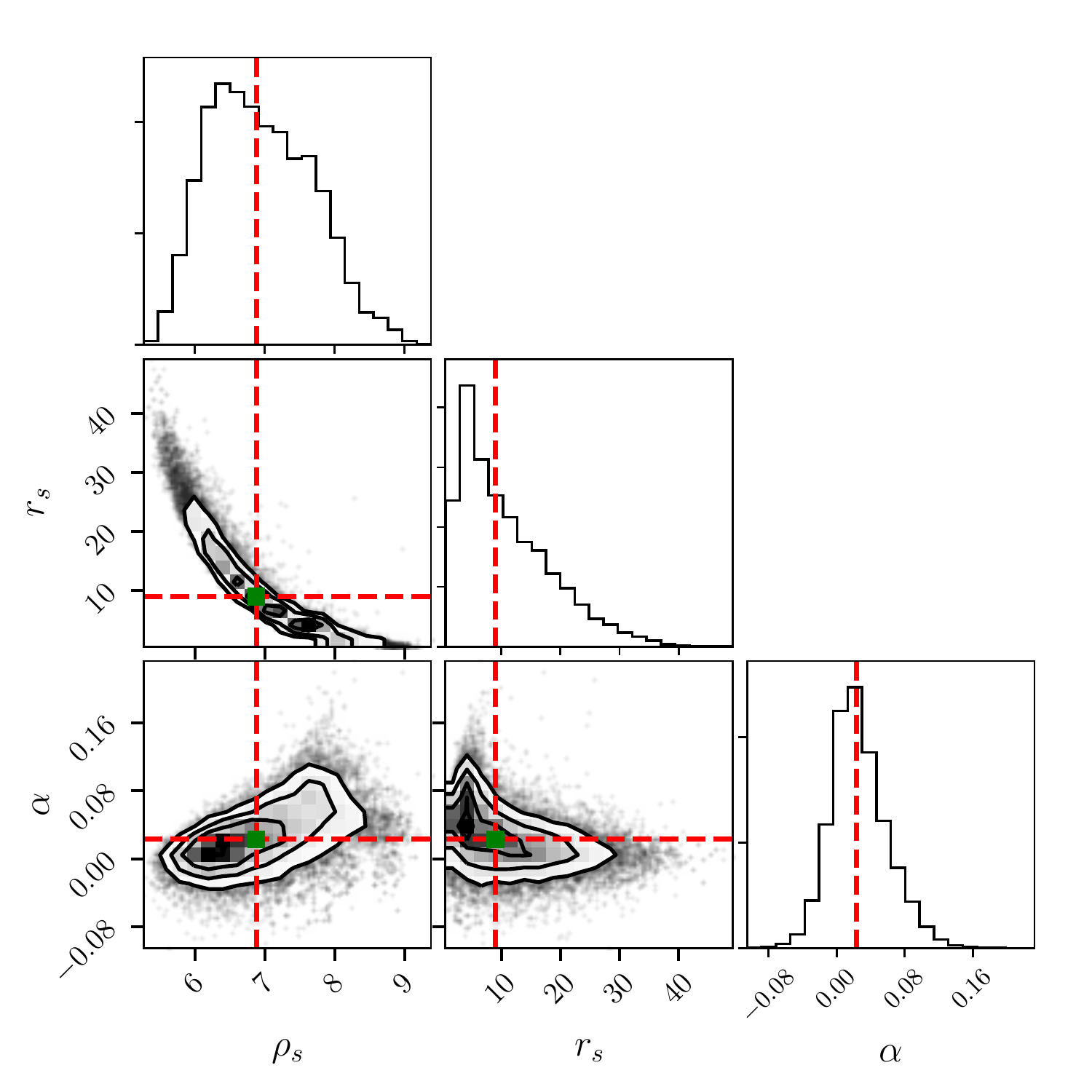}
    \caption{\textcolor{colorROSE}{Mff$\epsilon_{009}$-MecFB}}
    \label{fig:parameter_SF1ME_Einasto}
  \end{subfigure}
  \begin{subfigure}[b]{0.4\linewidth}
    \includegraphics[width=\linewidth]{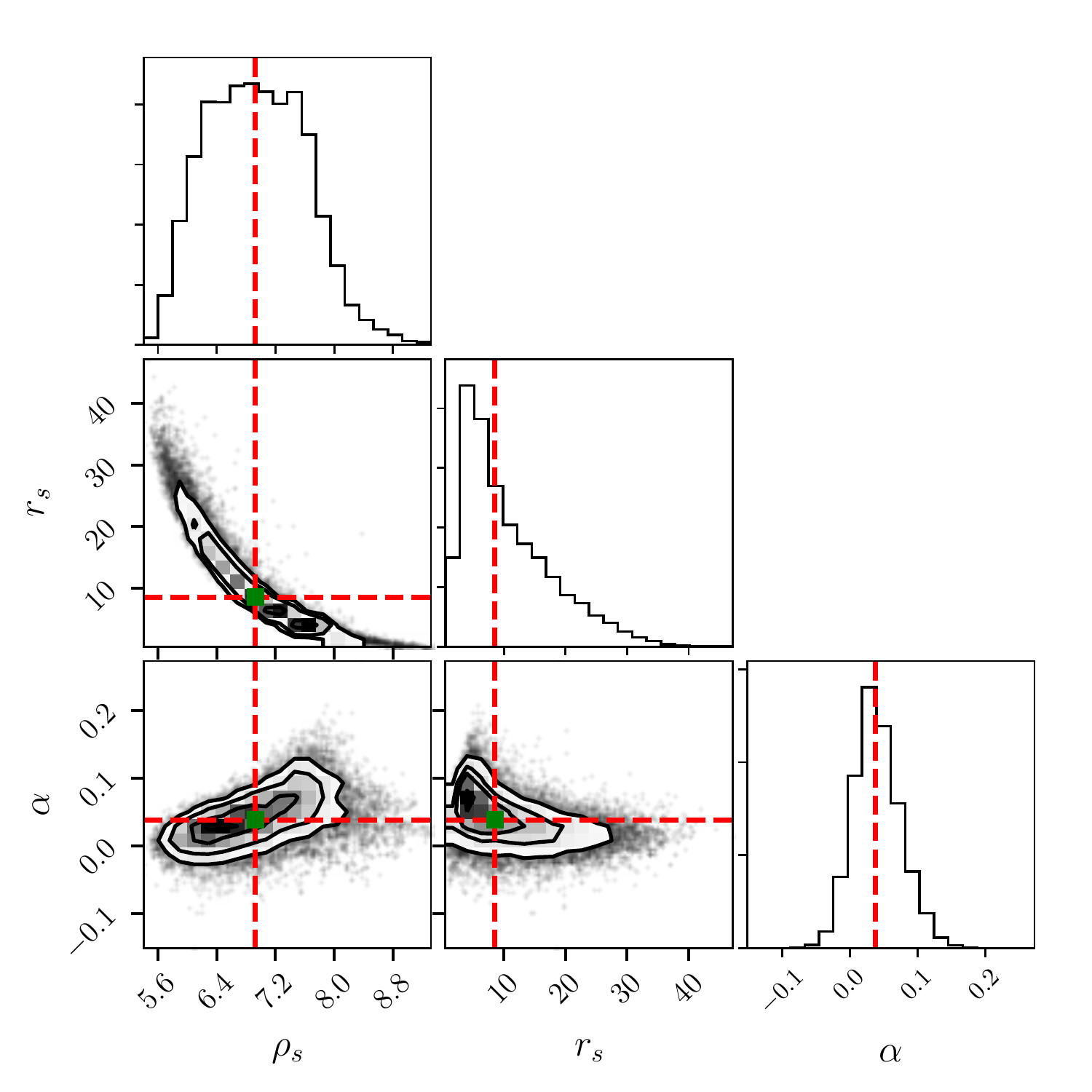}
    \caption{\textcolor{colorRED}{Mff$\epsilon_{100}$-MecFB}}
    \label{fig:parameter_SF1MEe1_Einasto}
  \end{subfigure}
    
    \caption{The posterior distributions of the parameters for the Einasto profile fitted over the different simulations.} 
  
  \label{fig:MCMCfit_posteriors_Einasto}
\end{figure}

The DM density profile can be described with the generalized $\alpha\beta\gamma$-profile \cite{1996MNRAS.278..488Z} :
\begin{equation}\label{eq:nfw_gen}
\rho(r,\rho_s,r_s,\alpha,\beta,\gamma) = \frac{\rho_s}{\left(\frac{r}{r_s}\right)^{\gamma} \left(1+\left(\frac{r}{r_s}\right)^{\alpha}\right)^{(\beta-\gamma)/\alpha}}
\end{equation}
\noindent where $\rho_s$ and $r_s$ are the scale density and the scale radius, respectively. Here, the density in the inner region ($r\ll r_s$) scales as $r^{-\gamma}$ and in the outskirt ($r\gg r_s$) of the halo it scales as $r^{-\beta}$. Note that the NFW profile is recovered when $(\alpha, \beta,\gamma) = (1,3,1)$. \\
Another widely-used profile is the Einasto fit \cite{Einasto1965}
\begin{equation}\label{eq:einasto}
 \rho(r,\rho_{-2},r_{-2},\alpha) = ,\rho_{-2} \exp\left( -\frac{2}{\alpha} \left[ \left( \frac{r}{r_{-2}} \right)^{\alpha} -1\right] \right)
\end{equation}
where $\rho_{-2}$ and $r_{-2}$ are the density and the radius at the point where the local slope is -2.\\ 
Recently, it has been shown that alternative functions (e.g. Dekel profile \cite{2020MNRAS.499.2912F}) are also able to fit the DM density profile that is subject to baryonic effects with high accuracy.\\
The Bayesian inference tool \texttt{MultiNest} (\cite{Feroz:2007kg,2009MNRAS.398.1601F,Feroz:2013hea})is used through the \texttt{PyMultiNest} interface (\cite{2014A&A...564A.125B}) to find the posterior likelihood distribution for the model parameters. The data to be analyzed is then the mean DM density in 39 radial, in logspace equidistant, bins that range from the resolution limit to the virial radius. The dispersion of the data including the standard deviation of the DM densities in each bin is taken into account assuming gaussian noise. 
Gaussian priors are used for the parameters ($\alpha$, $\beta$, $\gamma$, $\rho_s$ and $r_s$ for the $\alpha\beta\gamma$-profile, and $\rho_{-2}$, $r_{-2}$ and $\alpha$ for the Einasto profile) that are centered at reasonable initial guesses and have <<large>> standard deviations to make sure that the priors are not too restrictive.
Several tests have been performed to be sure that changing the priors does not alter the results.\\
The posterior distributions of the model parameters for all six simulations are shown on figure \ref{fig:MCMCfit_posteriors} for the $\alpha\beta\gamma$-profile and on figure \ref{fig:MCMCfit_posteriors_Einasto} for Einasto profile. The probability density distributions of each parameter is shown where the red dashed line marks the median of the distribution. 
Then, the joint probability distributions of the parameters is also shown, and mark the position of the combined medians by the green square. The  contours correspond to $1.0 \sigma$, $1.5\sigma$, and $2\sigma$ confidence levels, where $\sigma$ is the standard deviation of a two-dimensional normal distribution.

The profiles are shown in figures \ref{fig:MCMCfit_profile} and \ref{fig:MCMCfit_profile_Einasto}. And tables \ref{tab:haloprofiles_NFW} and \ref{tab:haloprofiles_einasto} show the median values and the value of the $68\%$ confidence intervals of the posterior distributions for the parameters of /the $\alpha\beta\gamma$-profile and the Einasto profile respectively . Note that for the DMO simulation $\alpha=1$. This is not a drastic measure due to the degeneracy of the parameters of this fit and allows us to recover the NFW profile which is related to DMO simulations. \\

\begin{table}[ht!]
\centering
\caption{ Median values and $68\%$ confidence interval of the posterior distributions of the $\alpha\beta\gamma$ parameters fitting the DM halo profiles.}
\begin{tabular}{|c|c|c|c|c|c|}
\hline
                          & $r_s$            & $\log{(\rho_s)}$ & $\alpha$  & $\beta$                        & $\gamma$                     \\ \hline
DMO                       & 63.3$\pm$42.78 & 5.71$\pm$6.39        & 1.0$\pm$0.57 &      3.3$\pm$0.58                      & 1.35$\pm$0.13                      \\ \hline
\textcolor{colorYELLOW}{KSlaw-DCool}               & 59.53$\pm$35.06 & 5.46$\pm$0.47        & 0.94$\pm$0.59 & 3.04$\pm$0.19                      & 1.70$\pm$0.06                      \\ \hline
\textcolor{colorBLUE}{Mff$\epsilon_{009}$-DCool} & 17.03$\pm$8.40     & 6.84$\pm$0.38        & 0.95$\pm$0.56 & 3.06$\pm$0.17                      & 1.37$\pm$0.10                      \\ \hline
\textcolor{colorGREEN}{Mff$\epsilon_{100}$-DCool} & 73.24$\pm$40.42     & 5.30$\pm$0.45        & 1.02$\pm$0.59 & \multicolumn{1}{l|}{3.02$\pm$0.18} & \multicolumn{1}{l|}{1.69$\pm$0.06} \\ \hline
\textcolor{colorROSE}{Mff$\epsilon_{009}$-MecFB} & 68.09$\pm$41.48     & 5.34$\pm$0.53        & 1.02$\pm$0.60 & 3.03$\pm$0.19                      & \multicolumn{1}{l|}{1.71$\pm$0.10} \\ \hline
\textcolor{colorRED}{Mff$\epsilon_{100}$-MecFB }& 63.95$\pm$37.16     & 5.37$\pm$0.49        & 1.01$\pm$0.59 & 3.04$\pm$0.19                      & 1.82$\pm$0.06                      \\ \hline
\end{tabular}\label{tab:haloprofiles_NFW}
\end{table}

\begin{table}[ht!]
\centering
\caption{ Median values and $68\%$ confidence interval of the posterior distributions of the Einasto parameters fitting the DM halo profiles.}
\begin{tabular}{|c|c|c|c|}
\hline
                          & $r_s$            & $\log{(\rho_s)}$ & $\alpha$  \\ 
\hline
DMO & 24.91$\pm$6.83 & 5.98$\pm$0.26  &  0.14$\pm$0.05                       \\ 
\hline
\textcolor{colorYELLOW}{KSlaw-DCool} & 11.20$\pm$6.37& 6.62$\pm$0.38 & 0.07$\pm$0.03               \\ 
\hline
\textcolor{colorBLUE}{Mff$\epsilon_{009}$-DCool} &11.71$\pm$5.38 &6.69$\pm$0.38 & 0.11$\pm$0.04                     \\ 
\hline
\textcolor{colorGREEN}{Mff$\epsilon_{100}$-DCool} & 10.73 $\pm$6.70 & 6.66$\pm$0.50 & 0.11$\pm$0.03\\ 
\hline
\textcolor{colorROSE}{Mff$\epsilon_{009}$-MecFB} &9.00$\pm$7.75 & 6.88$\pm$0.73 & 0.02$\pm$0.03 \\ 
\hline
\textcolor{colorRED}{Mff$\epsilon_{100}$-MecFB} & 8.53$\pm$7.32 & 6.93$\pm$0.67 & 0.04$\pm$0.03        \\ 
\hline
\end{tabular}\label{tab:haloprofiles_einasto}
\end{table}

Figure \ref{fig:MCMCfit_profile} (Figure \ref{fig:MCMCfit_profile_Einasto}), show the $\alpha\beta\gamma$ profile (Einasto profile) where  the medians of the posterior distributions of the parameters is used. Other possible choices (i.e. the average of the distribution or the most probable value) were tested but these recurrently yielded higher $\chi^2$ values. The error bars indicate  1 standard deviation of the posterior samples at the corresponding radii calculated as described in section \ref{appe:variance}. The results of our model is confronted with the data from the simulations, the black line is the DM density profile and the shaded area shows 1 standard deviation of the DM density in the radial bins. For orientation, the two vertical dotted lines indicate the boundaries of the bayesian inference method, i.e. $r_{\rm res}$ and $r_{\rm Vir}$. The residuals are shown in the lower section of each panel.

\section{Variance of a random variable function} \label{appe:variance}



\begin{figure}[ht!]
  \centering
  \begin{subfigure}[b]{0.32\linewidth}
    \includegraphics[width=\linewidth]{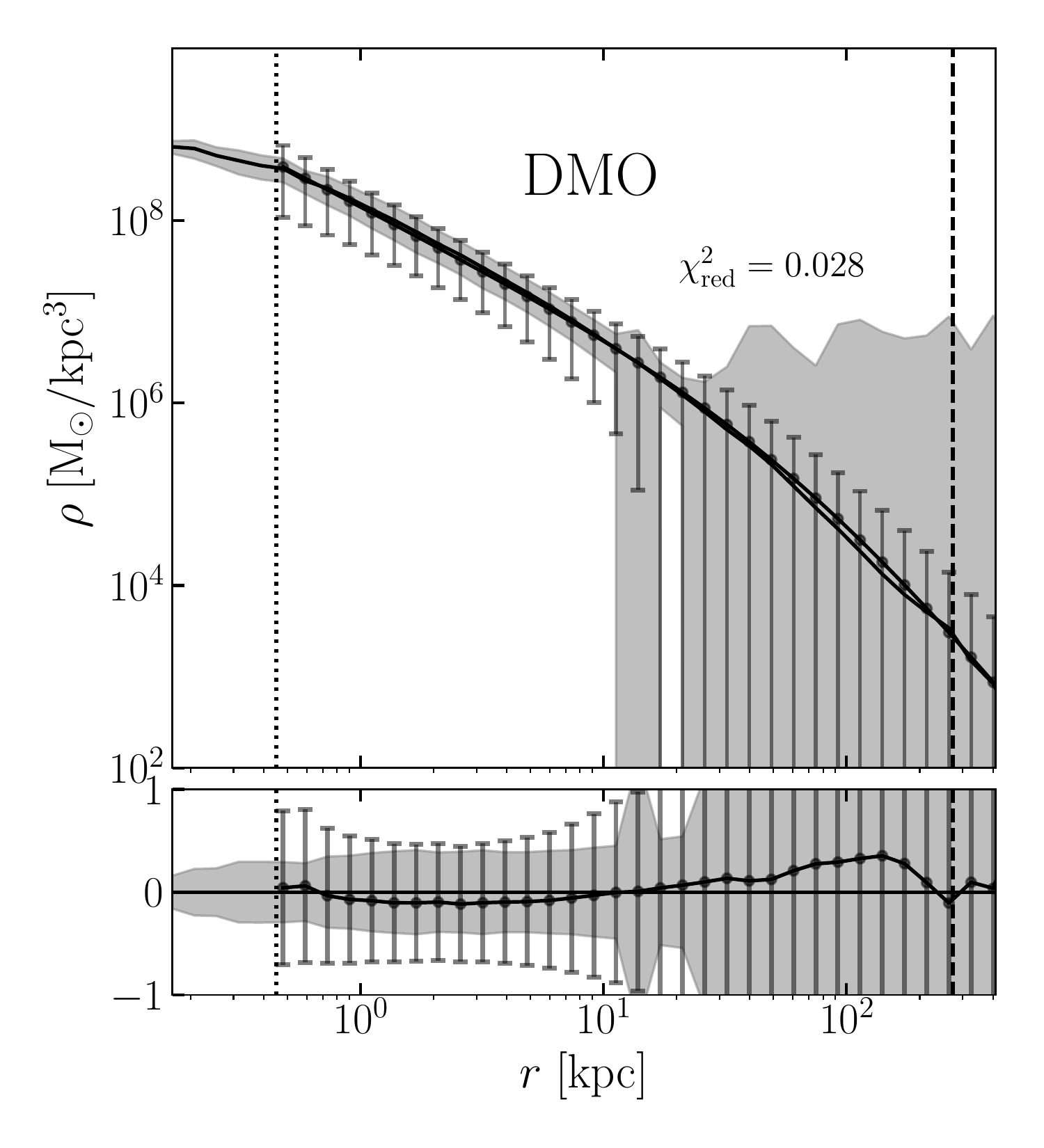}
    \caption{}
    \label{fig:MCMCfit_profile_DMO}
  \end{subfigure}
  \begin{subfigure}[b]{0.32\linewidth}
    \includegraphics[width=\linewidth]{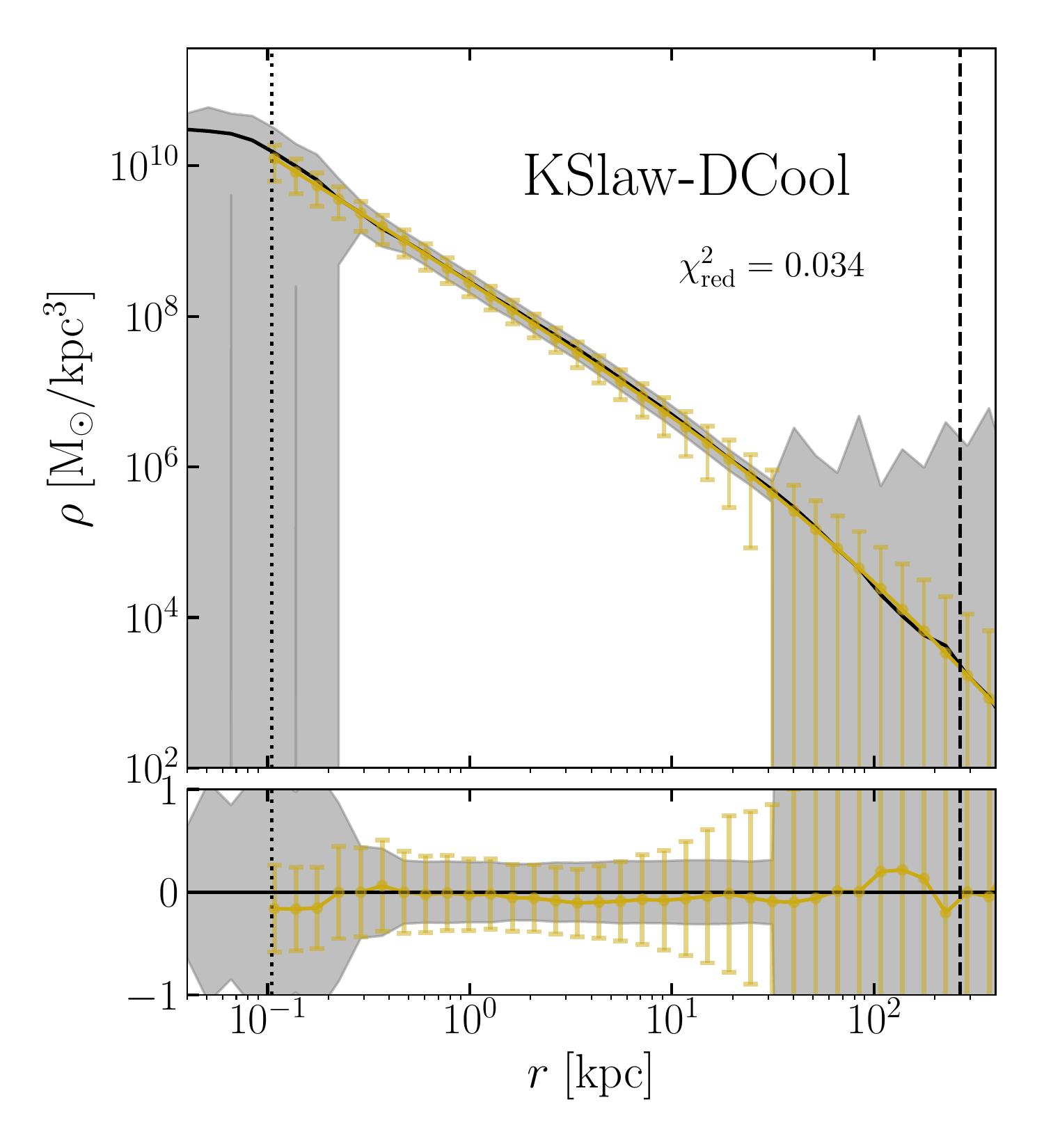}
    \caption{}
    \label{fig:MCMCfit_profile_SF0DC}
  \end{subfigure}
  \begin{subfigure}[b]{0.32\linewidth}
    \includegraphics[width=\linewidth]{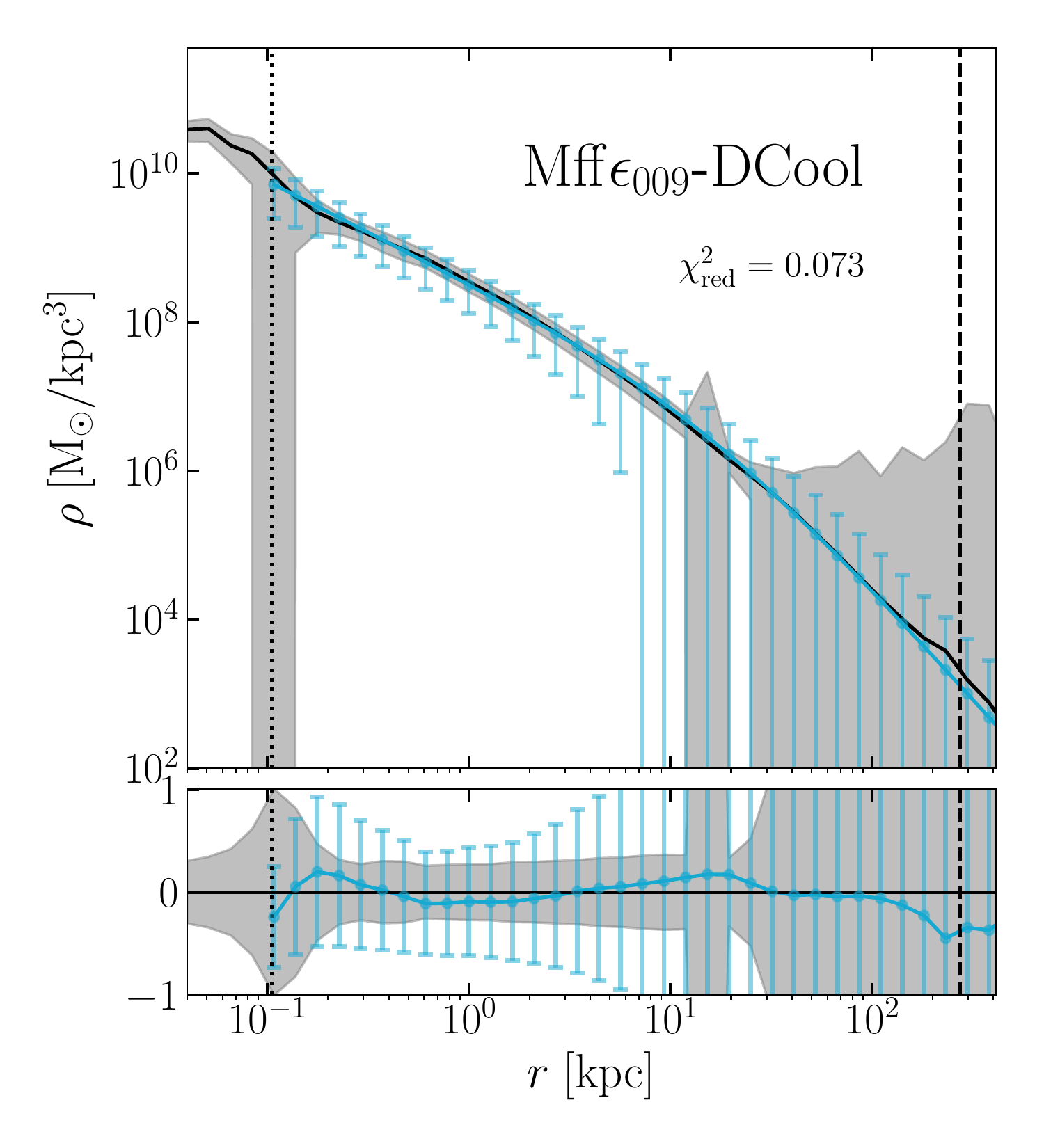}
    \caption{}
    \label{fig:MCMCfit_profile_SF1DC}
  \end{subfigure}
  \begin{subfigure}[b]{0.32\linewidth}
    \includegraphics[width=\linewidth]{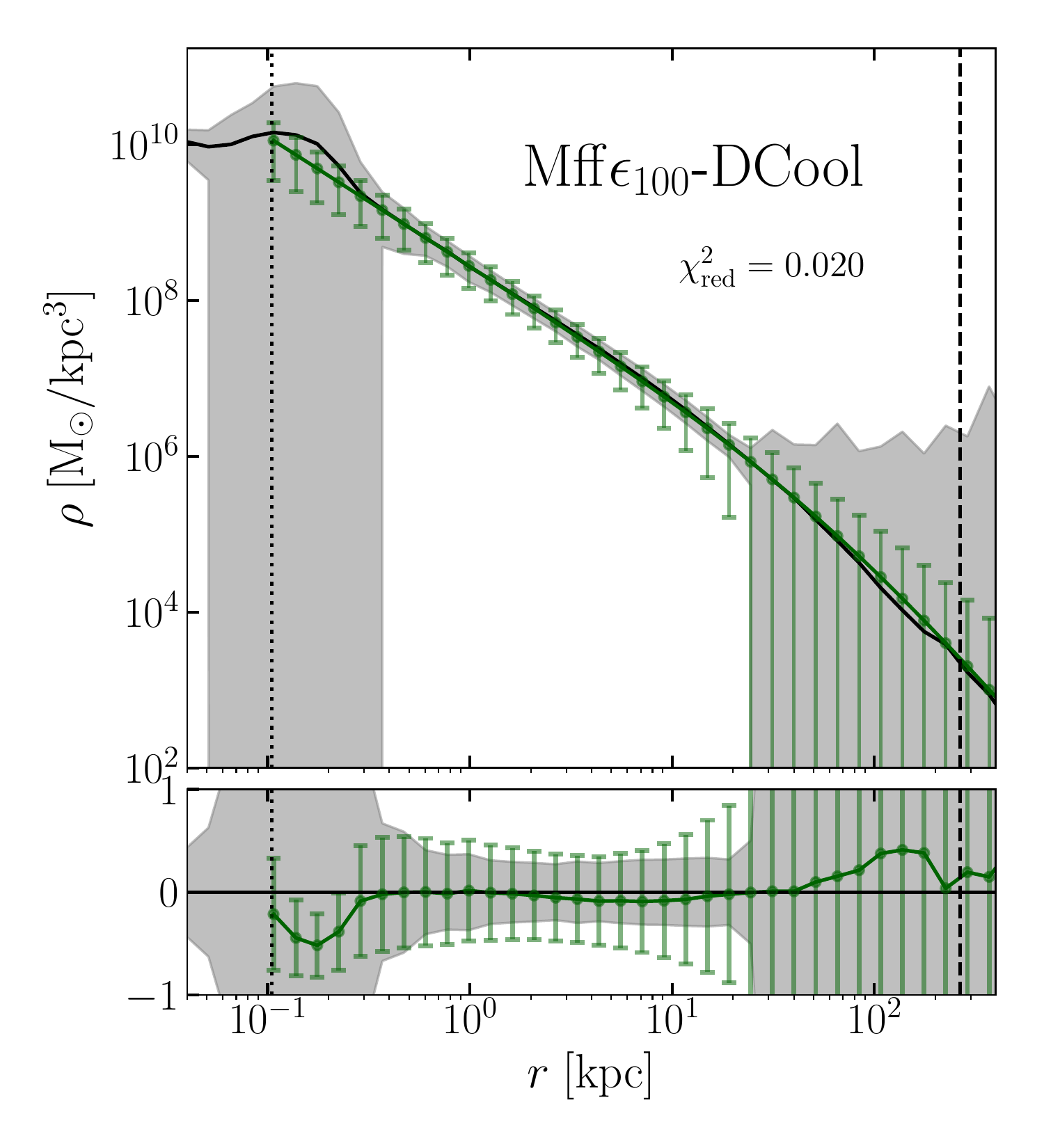}
    \caption{}
    \label{fig:MCMCfit_profile_SF1DCe1}
  \end{subfigure}
  \begin{subfigure}[b]{0.32\linewidth}
    \includegraphics[width=\linewidth]{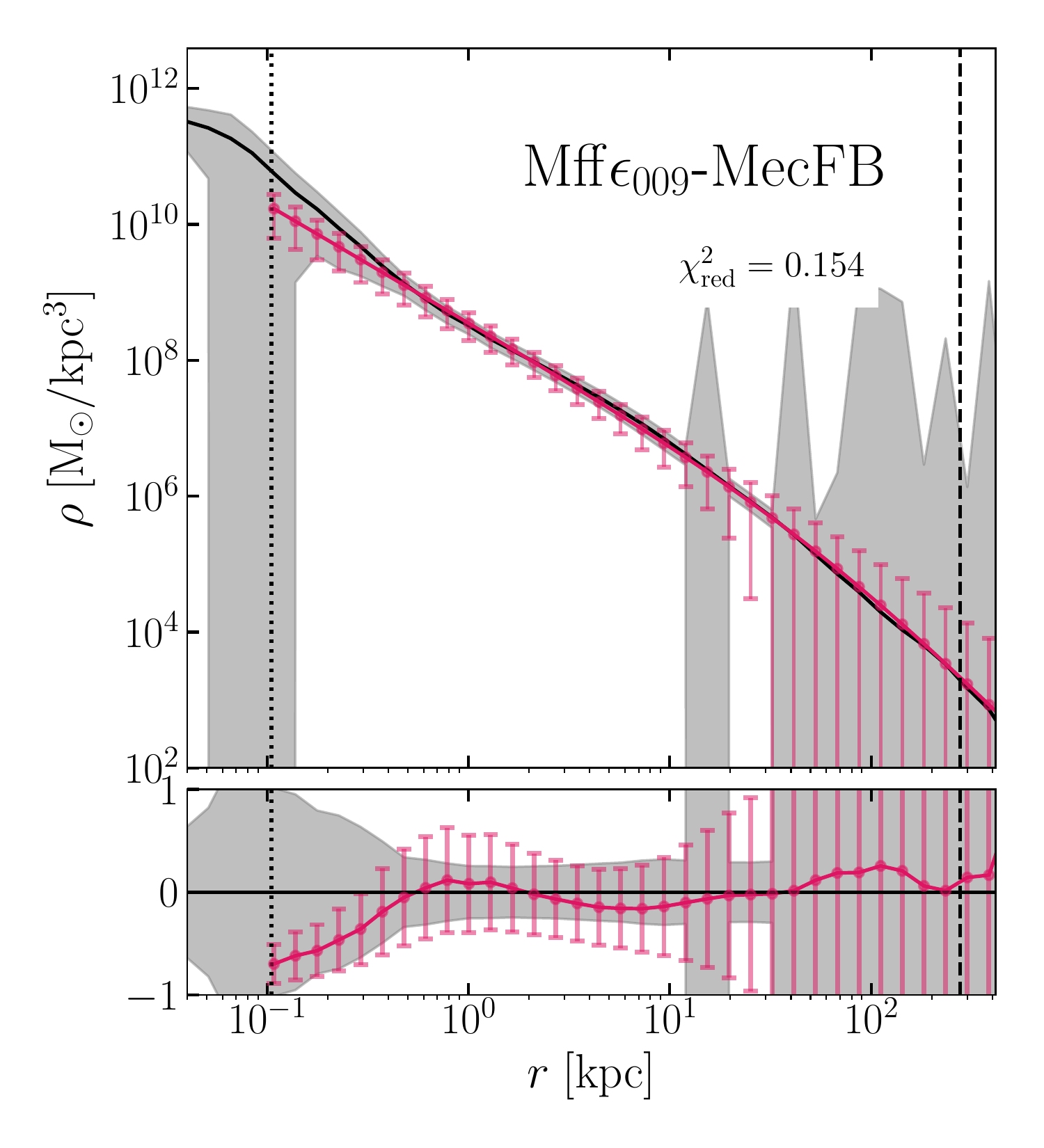}
    \caption{}
    \label{fig:MCMCfit_profile_SF1ME}
  \end{subfigure}
  \begin{subfigure}[b]{0.32\linewidth}
    \includegraphics[width=\linewidth]{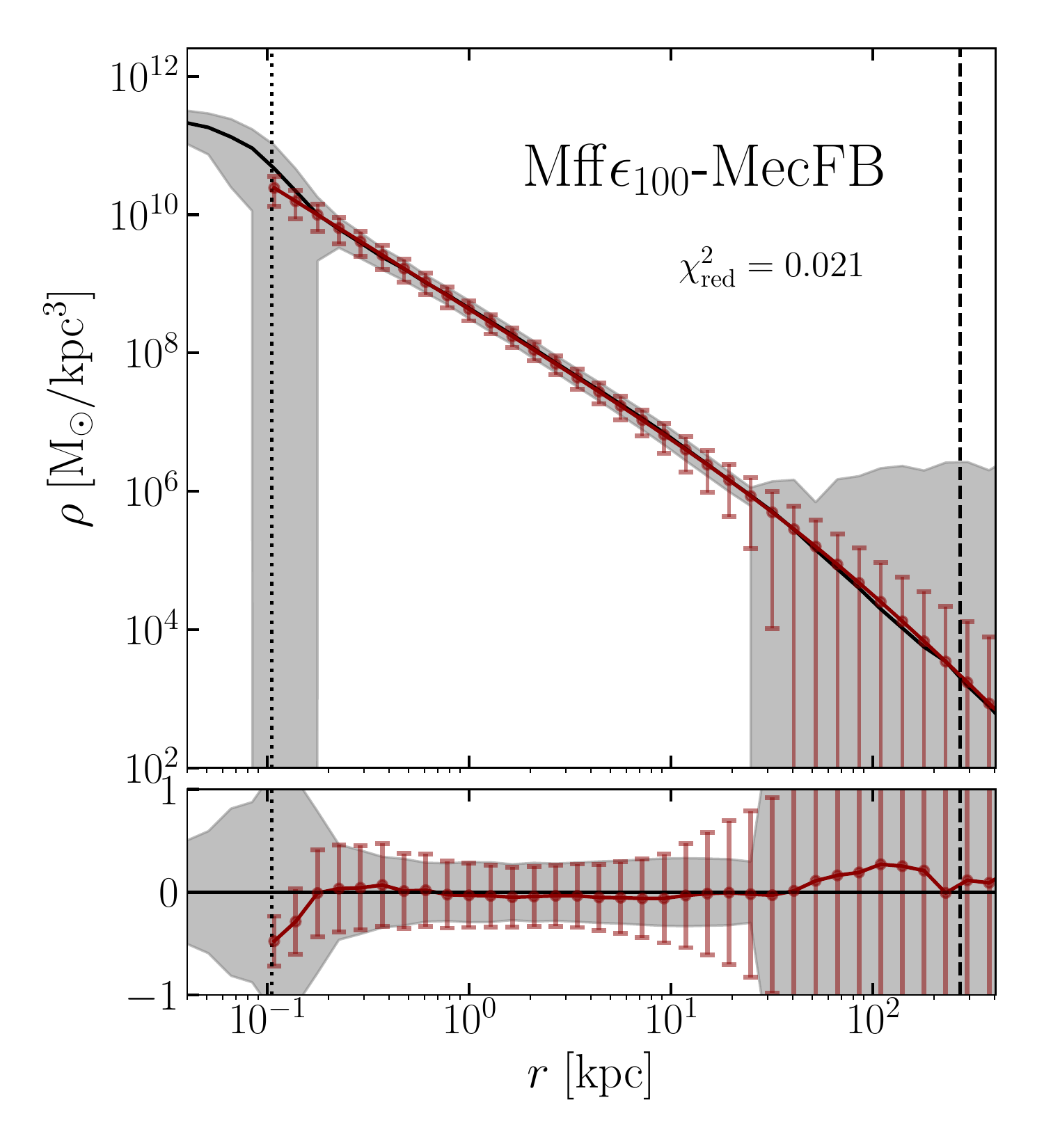}
    \caption{}
    \label{fig:MCMCfit_profile_SF1MEe1}
  \end{subfigure}
  
  \caption{The black line shows the mean DM density as a function of the radius in the 6 simulations. The shaded area gives 1 standard deviation of the DM density for the corresponding radii. The colored points are the best fit of the $\alpha\beta\gamma$-profile obtained from the posteriors from the Bayesian inference. The colored error bars represent the 68\% confidence level of the posterior. The vertical lines indicate the limits where the fit has been performed. The residuals are shown in the bottom panels.}
  \label{fig:MCMCfit_profile}
\end{figure}

\begin{figure}[ht!]
  \centering
  \begin{subfigure}[b]{0.32\linewidth}
    \includegraphics[width=\linewidth]{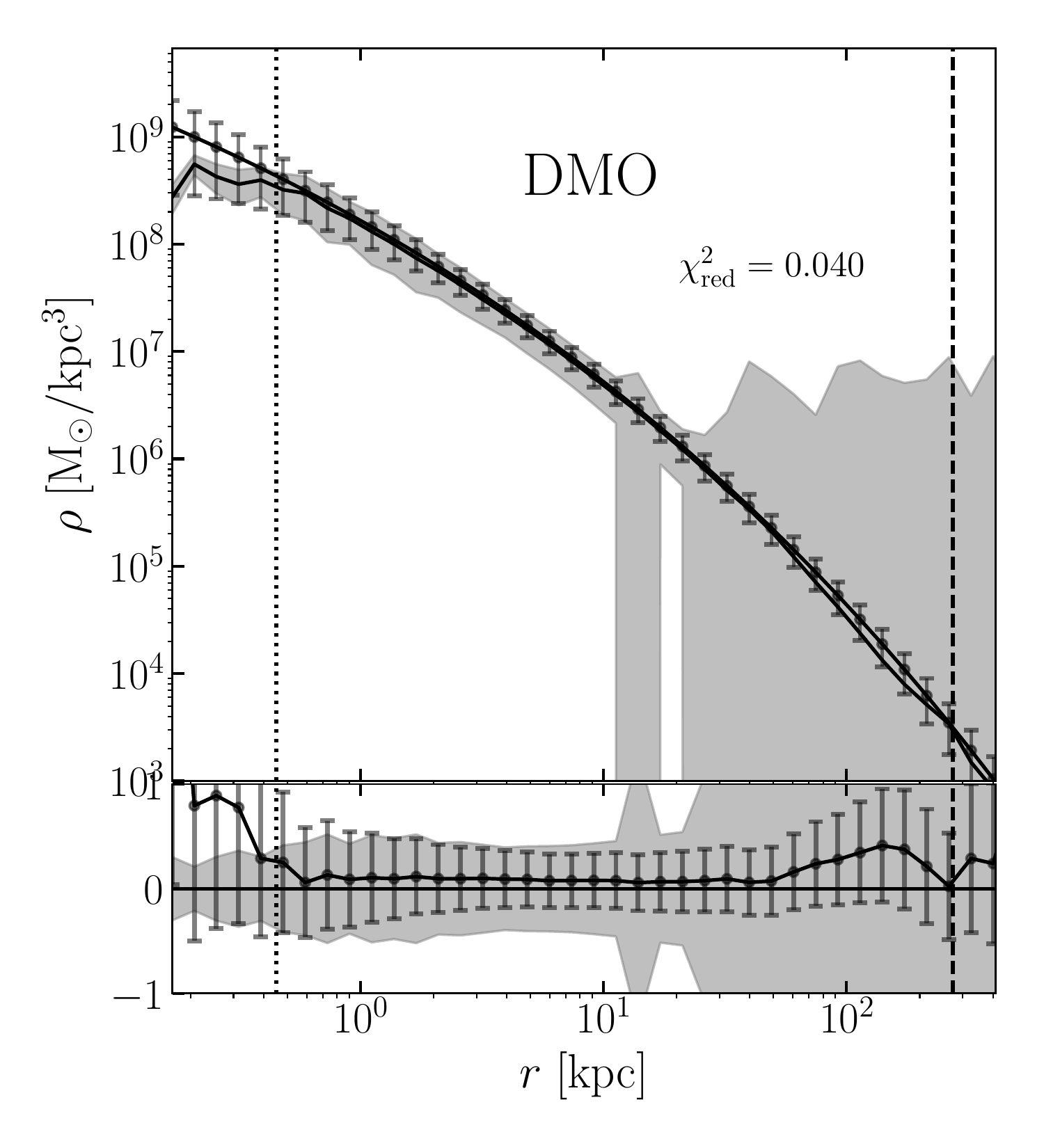}
    \caption{}
    \label{fig:MCMCfit_profile_DMO_Einasto}
  \end{subfigure}
  \begin{subfigure}[b]{0.32\linewidth}
    \includegraphics[width=\linewidth]{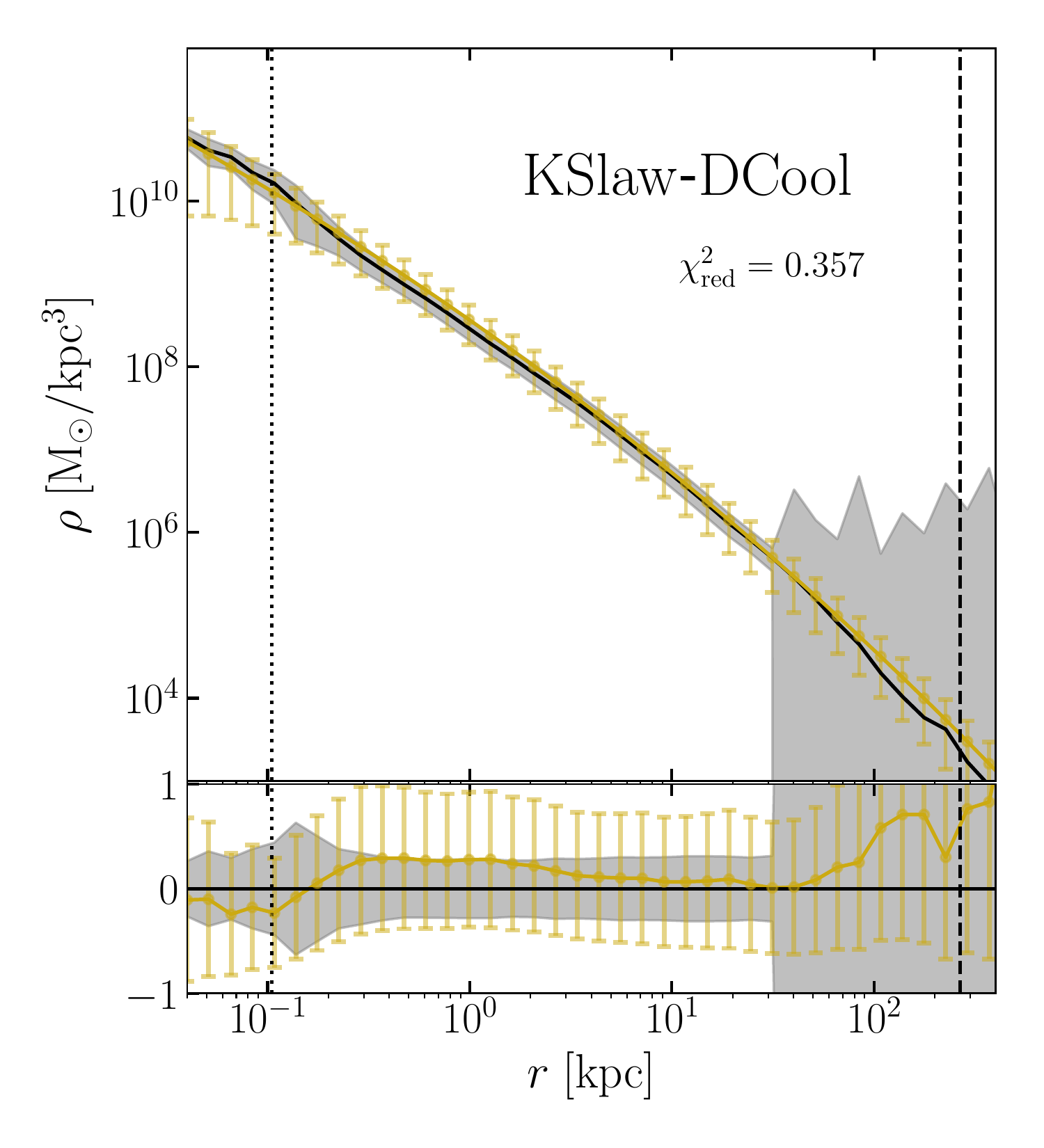}
    \caption{}
    \label{fig:MCMCfit_profile_SF0DC_Einasto}
  \end{subfigure}
  \begin{subfigure}[b]{0.32\linewidth}
    \includegraphics[width=\linewidth]{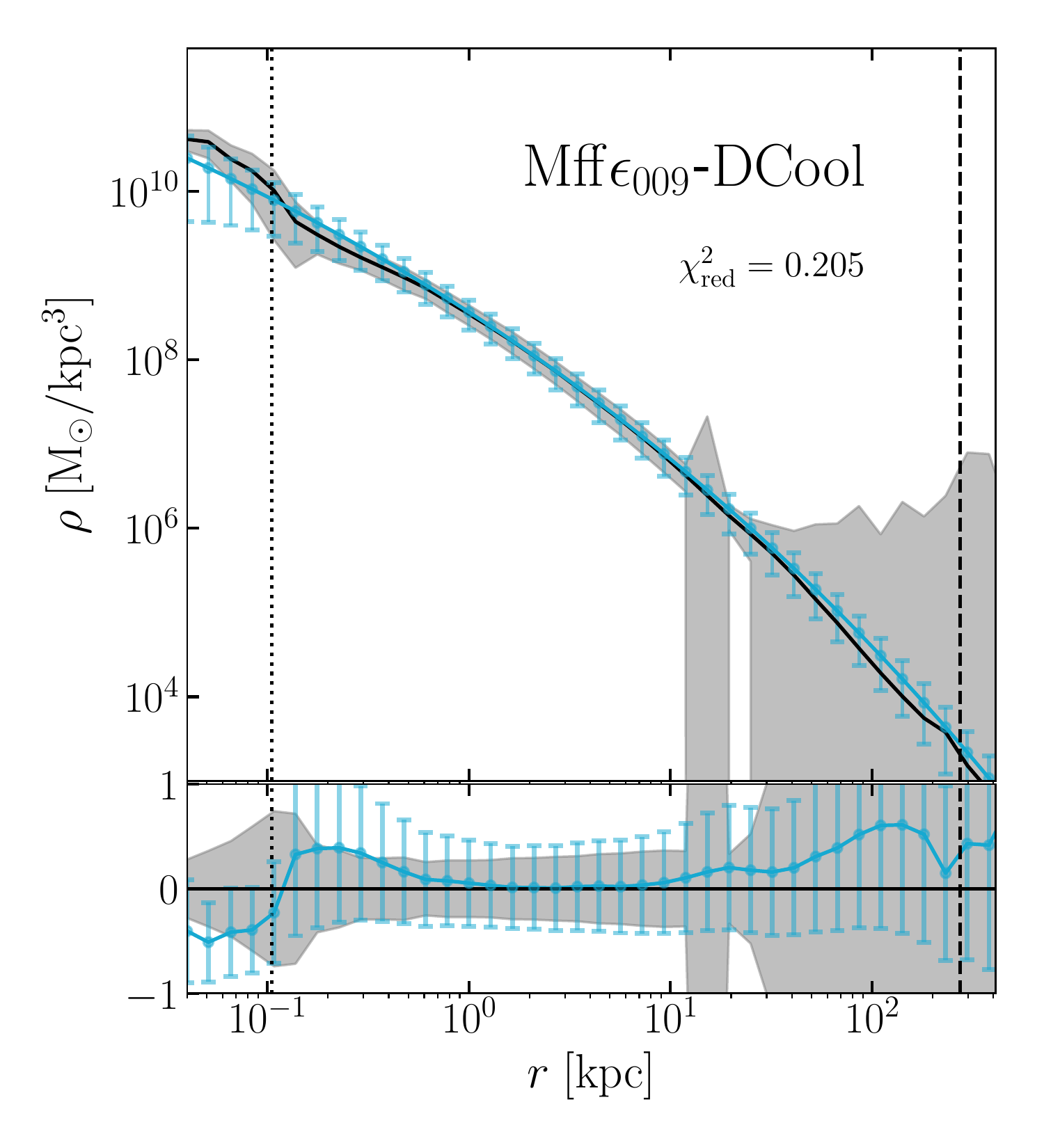}
    \caption{}
    \label{fig:MCMCfit_profile_SF1DC_Einasto}
  \end{subfigure}
  \begin{subfigure}[b]{0.32\linewidth}
    \includegraphics[width=\linewidth]{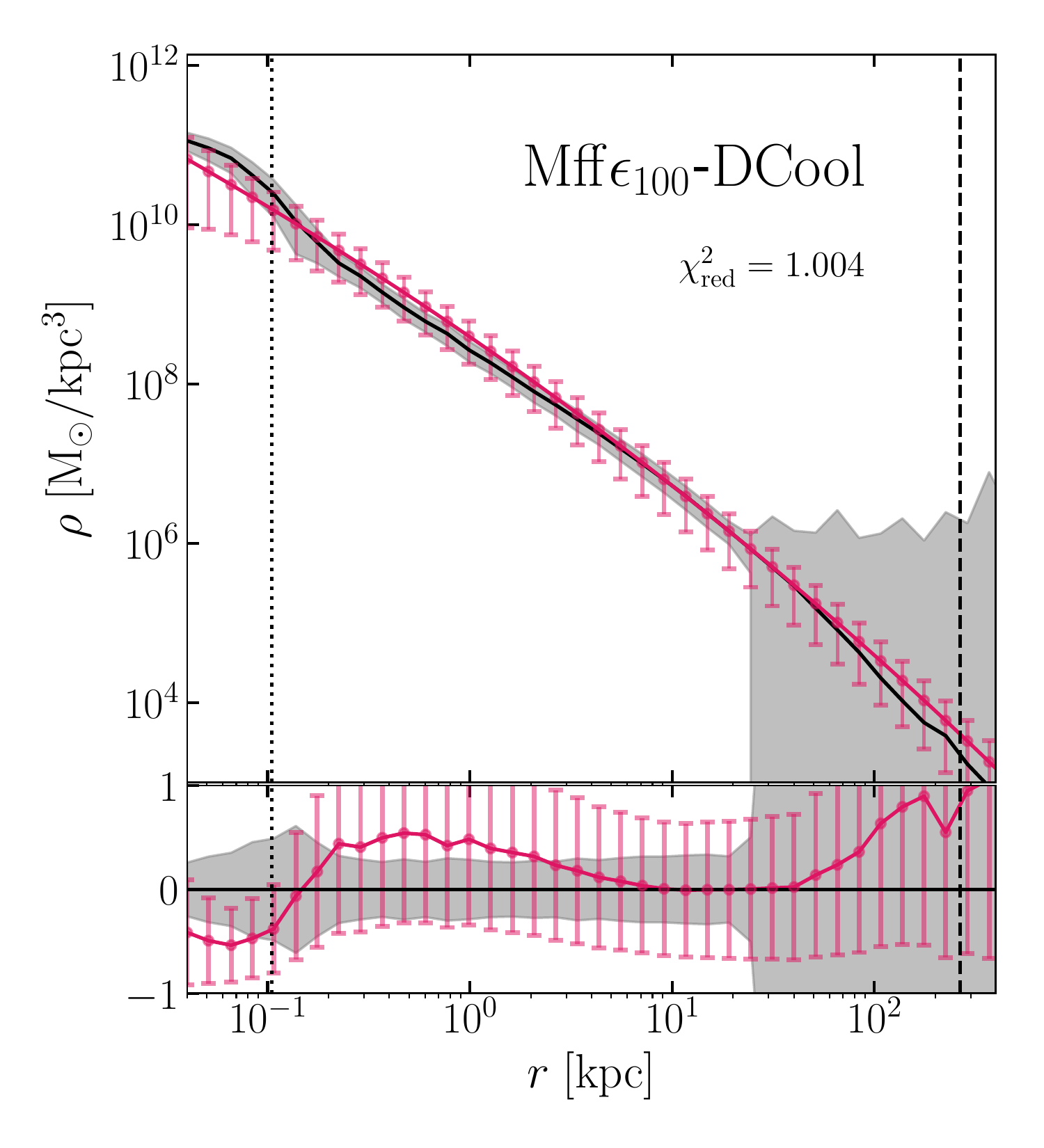}
    \caption{}
    \label{fig:MCMCfit_profile_SF1DCe1_Einasto}
  \end{subfigure}
  \begin{subfigure}[b]{0.32\linewidth}
    \includegraphics[width=\linewidth]{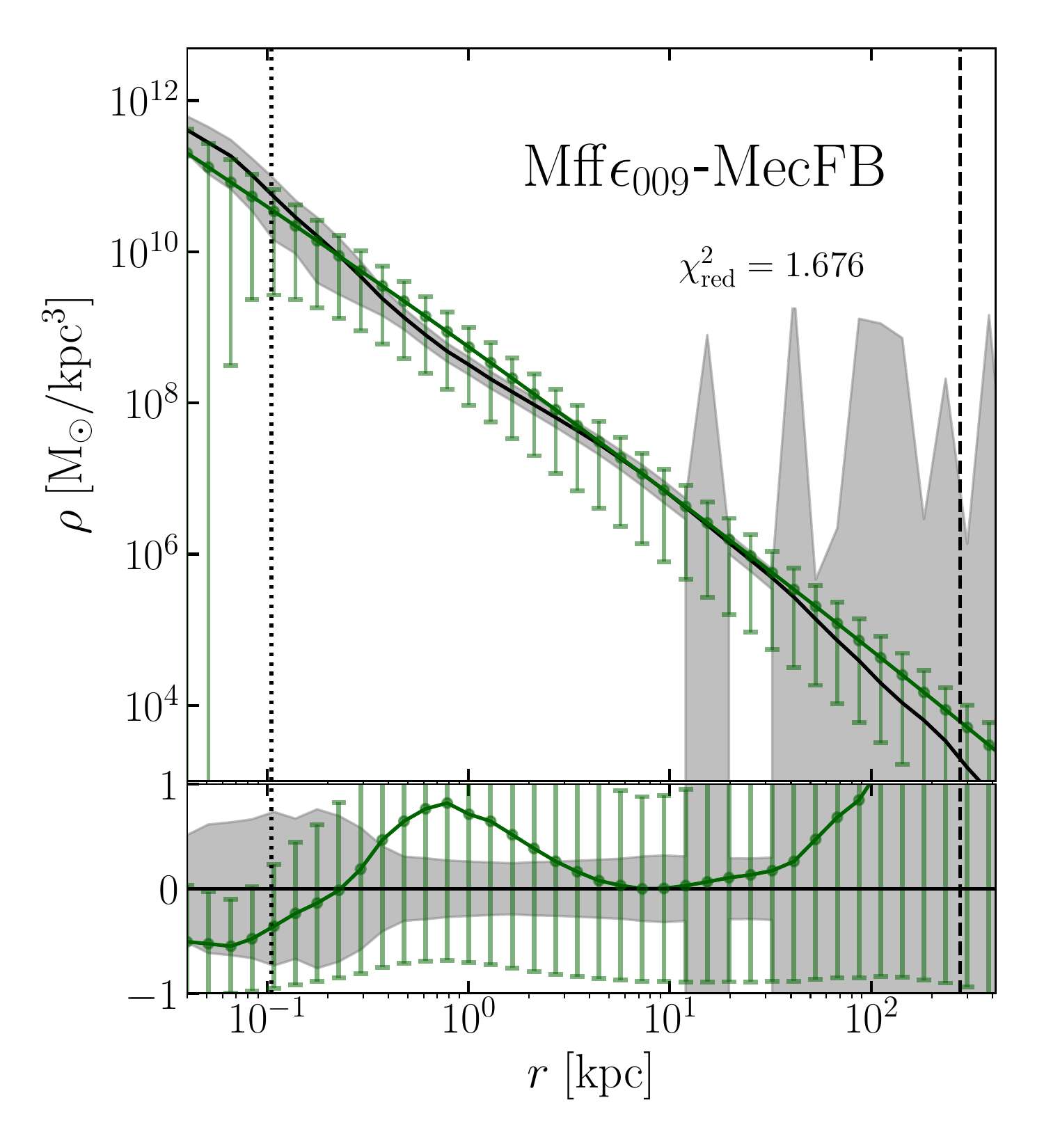}
    \caption{}
    \label{fig:MCMCfit_profile_SF1ME_Einasto}
  \end{subfigure}
  \begin{subfigure}[b]{0.32\linewidth}
    \includegraphics[width=\linewidth]{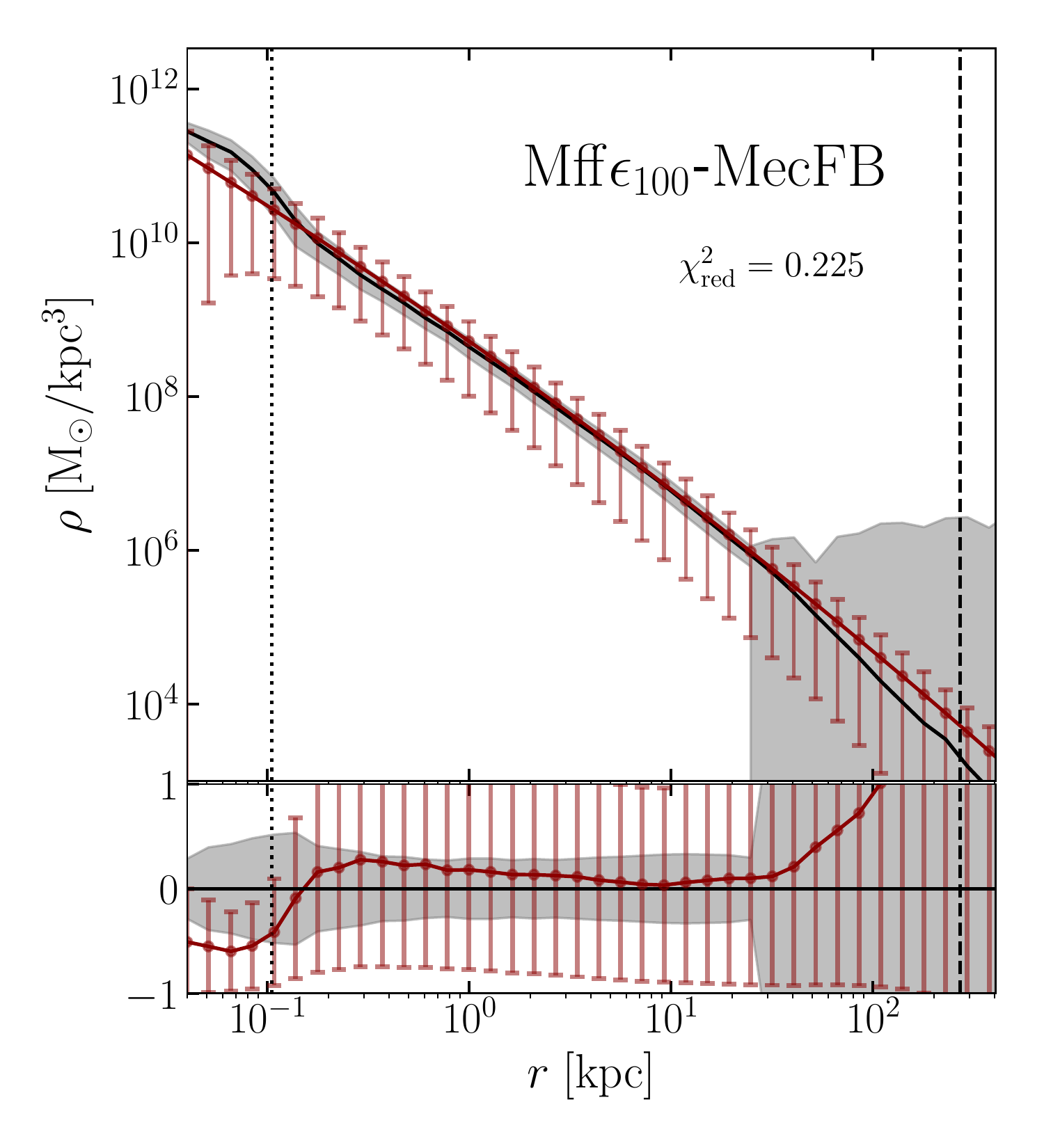}
    \caption{}
    \label{fig:MCMCfit_profile_SF1MEe1_Einasto}
  \end{subfigure}
  
  \caption{Fit of the density profiles using the Einasto parameterization. See figure \ref{fig:MCMCfit_profile} for explanations.}
  \label{fig:MCMCfit_profile_Einasto}

  \label{fig:MCMCfit_parameters_SF1DC}
\end{figure} 

While a data fit serves as a description of the measurements it is important to consider the degeneracy of the fit itself and its inner errors. This can be done in the case of an MCMC sampling for example by calculating the error band associated with the posteriors. Such calculation can be done as follows, let us consider a profile function $f(r,\vec{x})$ that depends on the coordinate r and on a vector made out of sets of random variables for each of the parameters of the profile $\vec{x}:{\rho_s,r_s,\alpha,\beta,\gamma}$. Each distribution of random variables is taken from the posterior of the MCMC sampling done minimizing a $\chi^2$ function on the simulation data. Take for example the generalized Zhao profile:
   
\begin{equation}
f(r,\vec{x}) = \frac{\rho_s}{\left(\frac{r}{r_s}\right)^{\gamma} \left( 1+ \left(\frac{r}{r_s}\right)^{\alpha}\right)^{\frac{\beta - \gamma}{\alpha}}}
\end{equation}

Since $\vec{x}$ is made of random variables, then it is acceptable to assume that the expectation value $\mathrm{E}[f(r,\vec{x})]$ is the $f(r,\vec{x})$ evaluated at the mean of the set of random variable i.e $\mathrm{E}[f(r,\vec{x})]\simeq f(r,\vec{\mu}) $. It is important to remark that this is an approximation that assumes gaussian distributions of random variables and that the obtained posterior distributions are not always gaussian. Nevertheless, is an interesting exercise that shows the degeneracy of the typical fits done on DM density profiles.  
Now, let us consider the elements of the covariance matrix defined as:
\begin{equation}
C_{ab} = \frac{1}{N-1}\sum_{i=1}^N (x_{a,i}- \mu_a)(x_{b,i} - \mu_b)\;\;,
\end{equation}
where the $a,b$ indices run over the profile parameters and $i$ over the each element of the posterior distribution of said parameter. Additionally, we can approximate the expectation value of the square of $f(r,\vec{x})$ as 
\begin{equation}
\mathrm{E}[f^2(r,\vec{x})](r) \simeq f^2 (r,\vec{\mu}) + \sum_{a,b}^n \frac{df(r)}{fx_a} \frac{df(r)}{fx_b}\bigg|_{\vec{x} = \vec{\mu}} C_{a,b} \;\;.
\end{equation}

\noindent Then from the definition of the standard deviation, we can build the standard deviation for our final MCMC set of posteriors as
\begin{equation}
\sigma^2(r) \simeq \sum_{a,b}^n \frac{df(r)}{d fx_a} \frac{df(r)}{d fx_b}\bigg|_{\vec{x} = \vec{\mu}} C_{a,b}.
\end{equation}
This calculation has been done on the fits carried over the density profiles shown in our simulations and the resulting uncertainty band is shown in figures \ref{fig:MCMCfit_profile} and \ref{fig:MCMCfit_profile_Einasto}. This procedure is important to understand the high degree of degeneracy of the typical fitting approaches.



\section{Contraction of the DM profile}
\label{app:contraction}

\begin{figure}[h!]
\centering
        \includegraphics[width=\linewidth]{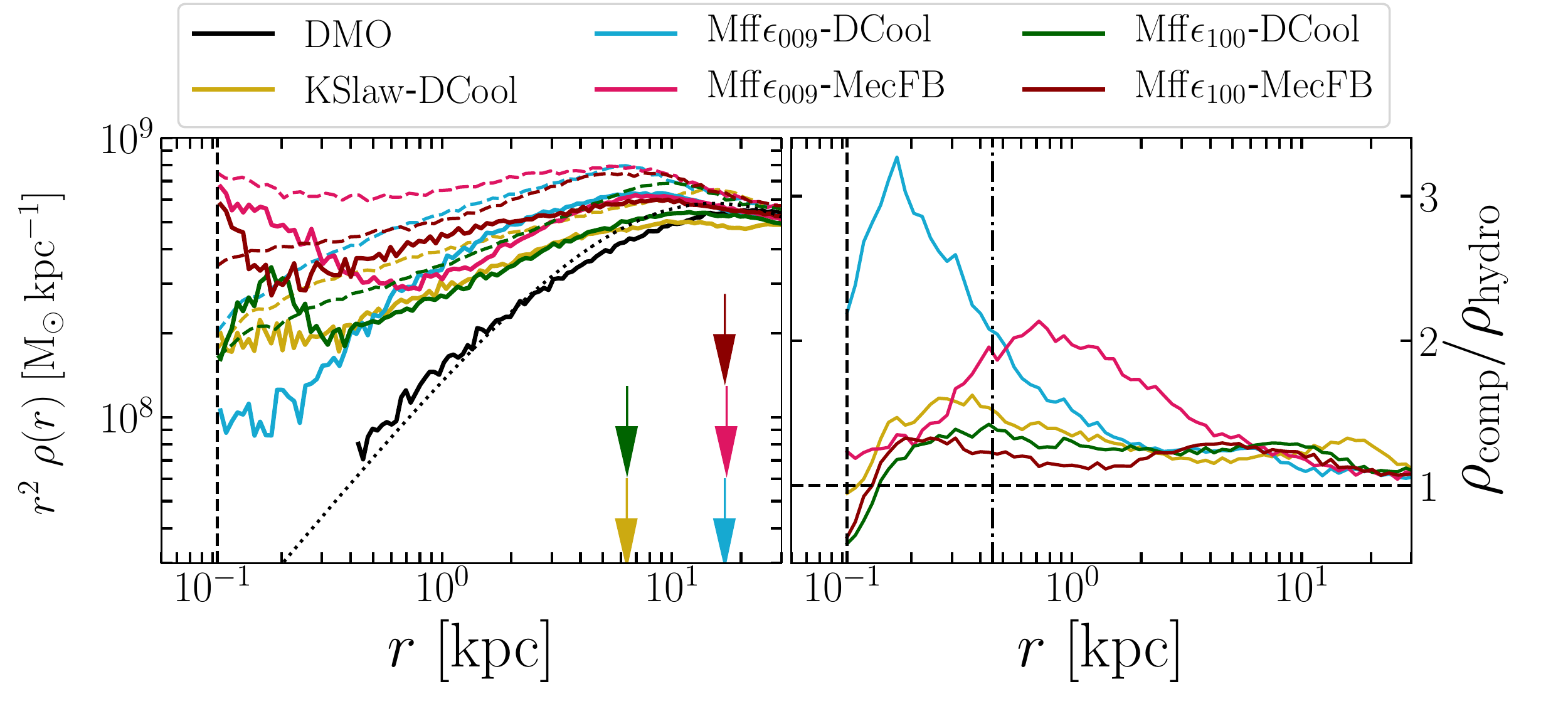}
        \caption{Left: Zoom of the inner dark matter profiles (scaled by $r^2$) below the dark matter-baryon crossing $\lesssim 30$~kpc. The compressed profiles of DM in the five hydrodynamics runs are added as dashed lines. The black dotted line is the DM density from the DMO fit, used as extrapolation below 400 pc. The arrows indicate the radius at which the stellar density dominates the DM density in the hydrodynamical simulations. Right: The ratio between the (calculated) compressed DM density and the (measured) density in the hydrodynamical runs. The vertical black dashed lines correspond to $r_{\rm 3hsml}$ and the vertical black dash-dotted line to $r_{\rm P03}$. }
        \label{fig:r2rho_compression}
\end{figure}

Studies of the impact of different baryonic physics implementation on the DM profile have demonstrated that, if SN feedback is strong enough, it is capable of changing a cusp ($\gamma >1$) into a core ($\gamma  \sim 0$) \cite{Maccio2012,Teyssier2013,Mollitor:2014ara}. Such effects could reconcile observations for the MW that favour a flat DM profile \cite{Nesti2013}. Particularly Pontzen and Governato (2012)\cite{Pontzen2012} have provided an analytical model for how gas motions are responsible for cusp-core transformations. However, this requires stronger SN feedback than the implementations used in this suite of simulations. As an illustration of this difference, one could compare the simulations done by Mollitor et al. (2014)\cite{Mollitor:2014ara} which have the same numerical environment as our KSlaw-DCool run except for one main parameter, the dissipation time scale of the SN  
in the Delayed Cooling implementation. They take a fixed characteristic time of 20 Myr, while the approach followed here is the resolution-dependent formulation from Dubois et al.(2015)\cite{Dubois2015} (see equation A8)  that yields a value of $\sim 1$ Myr which constrains the SN feedback to be weaker.

To illustrate quantitatively the effect of baryons on the DM density profile, the dark matter profiles of the hydrodynamical simulations with the  adiabatic contraction \cite{Blumenthal1986} of the  DMO dark matter profile computed with the respective baryonic distributions are compared.Assuming angular momentum conservation $M(r)r=cst$, ($M(r)$  being the  total  mass  enclosed within  radius $r$) new positions of DM particles $r_f$ are attributed by $[M_{dm}(r_i)]r_i = [M_{dm}(r_f)+M_{baryons}(r_f)]r_f$ which can be solved iteratively to obtain the contracted dark matter profile. Below the resolution limit, the DMO profile is extrapolated with the $\alpha\beta\gamma$ fit for the compression calculations to avoid the spurious numerical flattening  already mentioned. Figure \ref{fig:r2rho_compression}) shows the results where, again, the DM density scaled by $r^2$ is shown by the continuous color curves. The DMO density is shown as black continous curve and the fitting profile (used for the extrapolation below the DMO resolution limit) is the dotted black curve. Notice that the resolution limit for the hydro runs is below the radial range of the graph. 
The contracted profiles then are given by the dashed colored curves.
 From the crossing between $\rho_{DM}$ and   $\rho_{baryons}$ (indicated by the colored arrow), the contraction starts and is fairly well reproduced by the classical algorithm down to the resolution limit except for the Mff$\epsilon_{009}$-DCool and  Mff$\epsilon_{009}$-MecFB simulations. There, the densest and more localised star formation sites (see \cite{Nunez-Castineyra:2020ufe}) generate a more efficient feedback effect giving rise to a less cuspy profile in the hydro run compared to the adiabatic contraction calculation. This can be seen more clearly on figure   \ref{fig:r2rho_compression}d)
 where the ratio between the (calculated) contracted density profile and the (measured) DM density profile is given. All the compressed profiles, and by far the simulations Mff$\epsilon_{009}$-DCool and  Mff$\epsilon_{009}$-MecFB, overestimate the DM densities. This means that the non-trivial interplay bewteen baryonic processes shapes the differences in the DM profiles between the simulations.\\
The differences could find their origin in the fact that one of the hypotheses of the contraction model (e.g. spherical symmetry, adiabaticity) is not fulfilled. The improved contraction model \cite{Gnedin:2004cx} which accounts for orbital eccentricities of particles is tested but it could not improve the predictions of the contraction models. Further improvements e.g. \cite{Callingham:2020ips,Freundlich2020} could also be tested in the future. Alternatively, one could try to quantify the energy injections. These efforts though are beyond the scopes of this paper.


\section{Fits of the speed distribution}
\label{app:fdvfits}

\begin{figure}
\begin{center}
\includegraphics[width=0.8\linewidth]{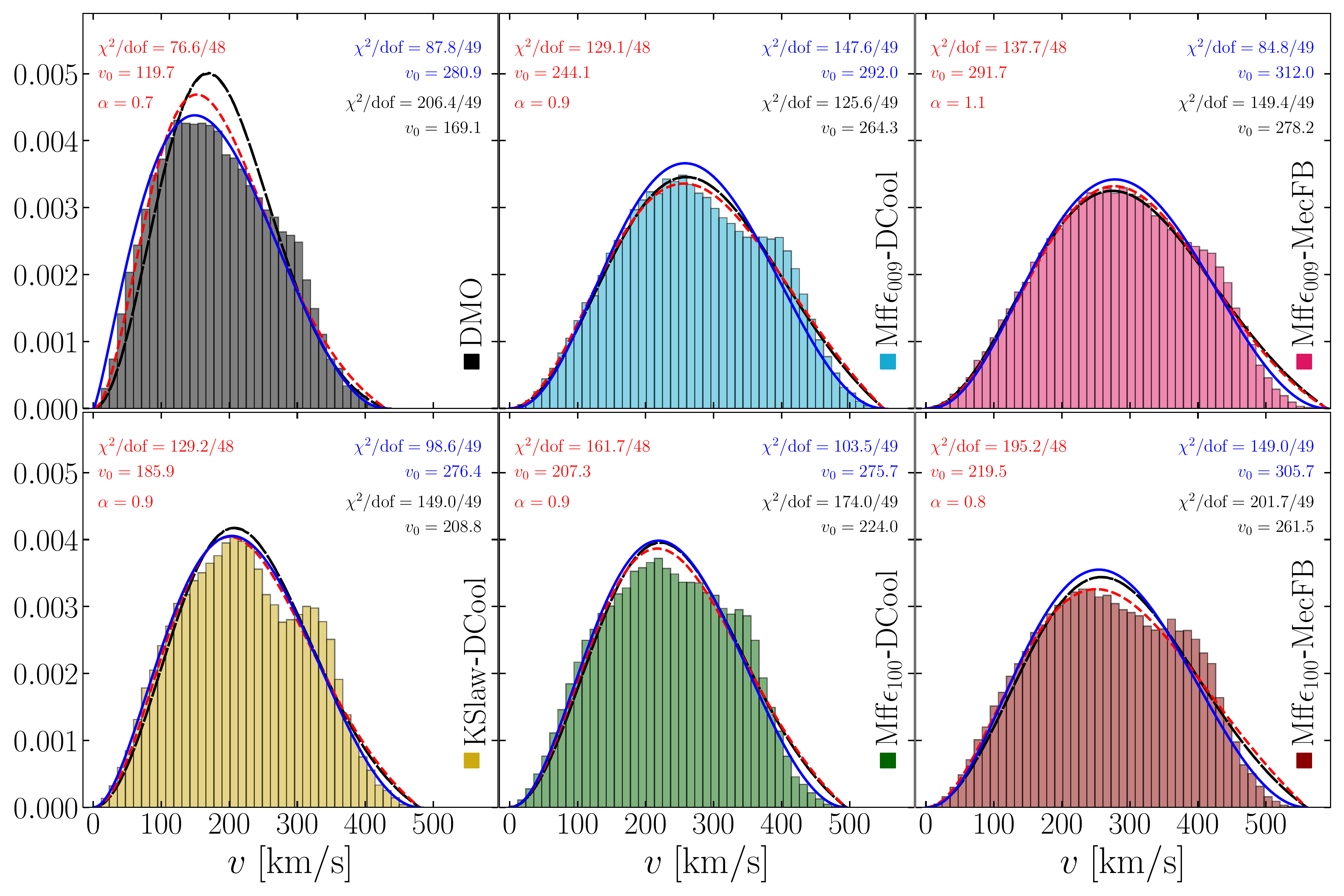}
\end{center}
  \caption{Fits of the speed distribution $f(v)$ using Maxwellian (red), generalized Maxwellian (black) and Tsallis (blue) functions. See table \ref{tab:vdffits} for details}
\label{fig:fitfdv}
\end{figure}

When it comes to the detection of DM in the solar neighbourhood one of the main assumptions to be made is the local speed distribution $f(v)$  of the DM particles. This assumption is mostly related to galactic dynamics and almost independent of the DM candidate. Cosmological simulations provide a fundamental testing ground for dynamical models of the local DM. Many different functionals have been proposed to describe the distribution of DM around the Sun and compared to simulations. The typical problematic features where such models fail to reproduce what is observed in simulations are in the hat of the distribution and the high-velocity tail. Here, three different models are fitted to $f(v)$ at 8 kpc from the centre of our simulated galaxies. The results are shown in table \ref{tab:vdffits} and figure \ref{fig:fitfdv} where the following functionals have been considered:
The Maxwellian distribution naturally results from a halo that is an isothermal sphere \cite{Ullio2001,Green2012}. This is a benchmark model used in different DM detection experiments. Where the velocity dispersion, $v_0$, can be related to the circular velocity. This functional has been extensively used even though it has several formal shortcomings \cite{Ling2010b,NunezCastineyra2019}, nevertheless, there are possible extensions that can be used, for example, the so-called generalized Maxwellian velocity distribution that has the following form:
\begin{equation}
    f(\vec{v}) = \frac{e^{-(\vec{v}^2/v_0^2)^{\alpha}}}{N(v_0,\alpha)} \;\;,
\end{equation}
\noindent and that recovers the Maxwellian distribution when $\alpha=1$. Another useful functional is the Tsallis function which is meant to describe self-gravitating structures and results from the Boltzmann-Gibbs approach, therefore, it might be more appropriate for galactic halos. It has the following form
\begin{equation}
    f(\vec{v}) = \frac{1}{N(v_0,q)} \left(1-(1-q)\frac{\vec{v}^2}{v^2 _0}\right)^{q/(1-q)}\;\;.
\end{equation} 
\noindent For the fits presented here the dynamical extension presented in \cite{NunezCastineyra2019} is used, in the extension $q=1-v_0^2/v_{\rm esc}$ and $v_{\rm esc}$ is the local escape velocity at 8 kpc. Out of all models the Tallis model seems to fit better the obtained distribution as it shows lower reduced $\chi^2$ values. Nevertheless, there are still features of the simulations data that are not reproducible by any models, notably the bumps in high-velocity tail of the distributions. These bumps are not a numerical artifact as they show in all simulations and have been proven to be independent of resolution \cite{Vogelsberger:2008qb}. It is likely that these features are built by fast remainings of disruptions that subhalos undergo after their first peri-passage. To separate both populations would require a careful detection of particles bound to subhalos throughout the evolution of the galaxy.

\begin{table}[h]

\caption[Parameters of fits]{Fits of the velocity distributions at 8 kpc for the six simulations. The normalisation are $N(v_0,\alpha) = 4\pi v_0^3 \Gamma(1+3/2\alpha)$ for the Maxwellian and $N(v_0,q)$ for the Tsallis function.}
\label{tab:vdffits}
\centering
\begin{tabular}{|c|cc|ccc|cc|}

\hline
\multirow{2}{*}{}                                                    & \multicolumn{2}{c|}{Maxwellian}             & \multicolumn{3}{c|}{Generalized Maxwellian}                                 & \multicolumn{2}{c|}{Tsallis}                \\ \cline{2-8} 
                                                                     & \multicolumn{1}{c|}{$v_0$} & $\chi^2_{red}$ & \multicolumn{1}{c|}{$v_0$} & \multicolumn{1}{c|}{$\alpha$} & $\chi^2_{red}$ & \multicolumn{1}{c|}{$v_0$} & $\chi^2_{red}$ \\ \hline
DMO                                                                  & \multicolumn{1}{c|}{206.4} & 4.2            & \multicolumn{1}{c|}{119.7} & \multicolumn{1}{c|}{0.7}      & 1.6            & \multicolumn{1}{c|}{280.9} & 1.8            \\ \hline
\textcolor{colorYELLOW}{KSlaw-DCool}              & \multicolumn{1}{c|}{208.8} & 3.0            & \multicolumn{1}{c|}{185.9} & \multicolumn{1}{c|}{0.9}      & 2.7            & \multicolumn{1}{c|}{276.4} & 2.0            \\ \hline
\textcolor{colorBLUE}{Mff$\epsilon_{009}$-DCool}  & \multicolumn{1}{c|}{264.3} & 2.6            & \multicolumn{1}{c|}{244.1} & \multicolumn{1}{c|}{0.9}      & 2.7            & \multicolumn{1}{c|}{292.0} & 3.0            \\ \hline
\textcolor{colorGREEN}{Mff$\epsilon_{100}$-DCool} & \multicolumn{1}{c|}{224.0} & 3.6            & \multicolumn{1}{c|}{207.3} & \multicolumn{1}{c|}{0.9}      & 3.4            & \multicolumn{1}{c|}{275.3} & 2.1            \\ \hline
\textcolor{colorROSE}{Mff$\epsilon_{009}$-MecFB}  & \multicolumn{1}{c|}{278.2} & 3.0            & \multicolumn{1}{c|}{291.7} & \multicolumn{1}{c|}{1.1}      & 2.9            & \multicolumn{1}{c|}{312.0} & 1.7            \\ \hline
\textcolor{colorRED}{Mff$\epsilon_{100}$-MecFB}   & \multicolumn{1}{c|}{261.5} & 4.1            & \multicolumn{1}{c|}{219.5} & \multicolumn{1}{c|}{0.8}      & 4.0            & \multicolumn{1}{c|}{305.7} & 3.0            \\ \hline

\end{tabular}

\end{table}


%

\end{document}